\documentclass[
reprint,
superscriptaddress,
showpacs,
preprintnumbers,
nofootinbib,
nobibnotes,
amsmath,
amssymb, 
aps,
prx,
floatfix
]{revtex4-2}

\pdfoutput=1 

\usepackage[utf8]{inputenc}
\usepackage[normalem]{ulem}
\usepackage{graphicx}
\usepackage{dcolumn}
\usepackage{bm}
\usepackage{color}
\usepackage{xcolor}
\usepackage[colorlinks=true,allcolors=purple]{hyperref}
\usepackage{url}
\usepackage{enumerate}
\usepackage{cleveref}
\usepackage{xspace}

\usepackage{slashed}
\usepackage{multirow}
\usepackage{mathrsfs} 
\usepackage{amsmath}

\usepackage{bbold}
\usepackage{mathrsfs}
\usepackage{braket}
\usepackage{physics}
\usepackage{multirow}

\usepackage{fontawesome} 
\definecolor{blue-violet}{rgb}{0.33, 0.17, 0.89}
\usepackage{orcidlink}

\newcommand{\refeq}[1]{Eq.~(\ref{#1})}

\newcommand{\reftab}[1]{Table~\ref{#1}}
\newcommand{\refref}[1]{Ref.~\cite{#1}}

\def\eg{\emph{e.g.}}
\def\ie{\emph{i.e.}}

\newcommand{\zprime}{\ensuremath{Z^\prime}\xspace}
\newcommand{\mzprime}{\ensuremath{m_{Z^\prime}}\xspace}
\newcommand{\mn}{\ensuremath{m_N}\xspace}
\newcommand{\pzerod}{P$\emptyset$D\xspace}
\newcommand{\epluseminus}{$e^+e^-$\xspace}
\DeclareMathOperator{\E}{\mathrm{E}}

\newcounter{CommentCount}
\definecolor{MH}{rgb}{0.0,0.6,9}
\definecolor{palatinate}{rgb}{0.494, 0.192, 0.482}

\renewcommand{\phi}{\varphi}

\renewcommand{\arraystretch}{1.2}

\begin{document}

\title{Efficiently Exploring Multi-Dimensional Parameter Spaces Beyond the Standard Model} 

\author{Carlos A. Arg{\"u}elles\,\orcidlink{0000-0003-4186-4182}}
\email{carguelles@fas.harvard.edu}
\affiliation{Department of Physics \& Laboratory for Particle Physics and Cosmology, Harvard University, Cambridge, MA 02138, USA}

\author{Nicol\`o Foppiani\,\orcidlink{0000-0001-5472-1039}}
\email{nicolofoppiani@g.harvard.edu}
\affiliation{Department of Physics \& Laboratory for Particle Physics and Cosmology, Harvard University, Cambridge, MA 02138, USA}

\author{Matheus Hostert\,\orcidlink{0000-0002-9584-8877}}
\email{mhostert@perimeterinstitute.ca}
\affiliation{Perimeter Institute for Theoretical Physics, Waterloo, ON N2J 2W9, Canada}
\affiliation{School of Physics and Astronomy, University of Minnesota, Minneapolis, MN 55455, USA}
\affiliation{William I. Fine Theoretical Physics Institute, School of Physics and Astronomy, University of Minnesota, Minneapolis, MN 55455, USA}

\date{\today}

\begin{abstract}
    We propose a method to ease the challenges of exploring multi-dimensional parameter spaces in beyond-the-Standard Model theories.
    We evaluate the model likelihood for any choice of parameters by sampling the theory parameters intelligently and building a Kernel Density Estimator.
    By reducing the number of expensive Monte-Carlo simulations, this method provides a more efficient way to test complex theories.
    We illustrate our technique to set new limits on a short-lived heavy neutrino $N$, proposed as an explanation of anomalies in neutrino experiments.
    Using a search for lepton pairs in the T2K near detector, we find exclusion limits on the model parameters in a vast region of parameter space, fully exploiting the advantages of our new method.
    With a single Monte Carlo simulation, we obtain the differential event rate for arbitrary choices of model parameters, allowing us to cast limits on any slice of the model parameter space.
    We conclude that $N$ particles with lifetimes greater than $c \tau^0 \gtrsim 3$~cm are excluded by T2K data.
    We also derive model-independent constraints in terms of the total rate, lifetime, and $N$ mass and provide an approximated analytical formula.
    This method can be applied in other branches of physics to explore the landscape of theory parameters efficiently.
\end{abstract}

\maketitle

\section{Introduction}\label{sec:introduction}

Neutrinos are a rather unique ingredient of the Standard Model (SM).
The absence of electric charge and their extremely small but non-vanishing mass implies that, contrary to all other fermions, neutrinos do not have their properties uniquely determined by the SM gauge group, $G={\rm SU}(3)\times{\rm SU}(2)\times{\rm U}(1)$.
Indeed, to uniquely determine the origin of neutrino masses, we are left to choose between the existence of new symmetries, such as $U(1)_{B-L}$, or new scales, such as the Majorana mass of their right-handed partners.
The latter is the route chosen in the canonical Type-I seesaw mechanism~\cite{Minkowski:1977sc,*Mohapatra:1979ia,*GellMann:1980vs,*Yanagida:1979as,*Lazarides:1980nt,*Mohapatra:1980yp,*Schechter:1980gr,*Cheng:1980qt,*Foot:1988aq}, which has long been the leading motivation for experimental searches for feebly-interacting Majorana particles in cosmological or laboratory settings~\cite{Abdullahi:2022jlv}.
As a complete singlet under $G$, right-handed neutrinos could also provide unique insight into the possible existence of other hypothetical particles, such as dark matter, the dark photon, or additional Higgs bosons.

While neutrino experiments have achieved remarkable success in proving that neutrinos change flavor over macroscopical distances, global data still does not point to a coherent picture. 
Short-baseline experiments, characterized by baselines and energies of $L/E \sim 1$~km/GeV, provide significant outliers in the framework of three-neutrino oscillations.
This is led mostly by the $4.8~\sigma$ excess of electron-like events at MiniBooNE~\cite{MiniBooNE:2007uho,MiniBooNE:2010idf,MiniBooNE:2013uba,MiniBooNE:2018esg,MiniBooNE:2020pnu,MiniBooNE:2022emn}, and the $3.8~\sigma$~excess of inverse-beta-decay events at LSND~\cite{LSND:1996vlr,LSND:1996ubh,LSND:1997vun,LSND:1997vqj,LSND:2001aii}.
For a long time, solutions based on a light sterile neutrino with $\Delta m_{41}^2$ of $\mathcal{O}(1)$~eV$^2$ were the leading candidates for a beyond-the-SM explanation but have yet to successfully overcome challenges with accelerator and reactor experiments, as well as cosmological observations~\cite{Planck:2018vyg}.
For a recent review on the status of this topic, see Ref.~\cite{Acero:2022wqg}.
Since then, theoretical activity in low-scale dark sectors has intensified and, unsurprisingly, new dark-sector solutions to the short-baseline puzzle have been brought to light~\cite{Gninenko:2009ks,Gninenko:2010pr,Gninenko:2012rw,Masip:2012ke,Radionov:2013mca,Magill:2018jla,Vergani:2021tgc,Alvarez-Ruso:2021dna,Denton:2018dqq,Bertuzzo:2018itn,Bertuzzo:2018ftf,Ballett:2018ynz,Arguelles:2018mtc,Ballett:2019pyw,Ballett:2019cqp,Abdullahi:2020nyr,Datta:2020auq,Dutta:2020scq,Abdallah:2020biq,Abdallah:2020vgg,Hammad:2021mpl,Dutta:2021cip,Fischer:2019fbw,Chang:2021myh,Abdallah:2022grs}.

Among them is the possibility that light dark particles are produced in the scattering of neutrinos with matter and misidentified as electron-neutrinos due to their electromagnetic decays.
These models have been popularized by their connection to neutrino masses in low-scale seesaw models, the possibility to explain other low-energy anomalies, and, chiefly, due to their falsifiability.

Often, these dark sector solutions involve several new particles and, with them, several independent parameters. 
While the experimental signatures are straightforward to identify, thoroughly exploring the model parameter space can quickly become unmanageable.
This curse of dimensionality is especially burdensome due to the expensive Monte Carlo simulations of the experiments.
This article provides a solution to this problem using a new method based on re-weighing Monte Carlo samples.
Thoroughly exploring parameter spaces is essential both when placing constraints, since parameters can often be correlated, and when searching for degenerate new-physics solutions since they can be missed in coarse scanning procedures.
Our method is a generic way to explore any model with a complex parameter space; thus, it applies to several branches of physics.

The novelty of our technique lies in the usage of a single Monte Carlo simulation that simultaneously samples physical quantities, like phase space variables, as well as model parameters, such as the masses of heavy neutrinos and the dark photon.
With sufficient statistics, these samples can be used to construct a Kernel Density Estimator (KDE), which computes the model prediction and corresponding Likelihood for any choice of model parameters within the boundaries of the simulation.
The result is a fast interpolation of the posterior probability of the model and greater flexibility in determining confidence intervals in various slices of parameter space.
This method does away with the usual methods used in the phenomenology community of repeating the full experimental simulation on a $n$-dimensional grid, which can be a much more inefficient method to build the posterior.
By adaptively sampling the model parameter space, we maximize the sampling around the regions of interest.

We illustrate our method in the context of a dark sector model that can explain the MiniBooNE excess.
The model will contain heavy neutrinos and a dark photon, whose properties are defined by parameters embedded in a multi-dimensional parameter space.
We apply our method to a search for \epluseminus in the multi-component near detector of T2K, ND280~\cite{Abe:2019kgx}.
The signatures arise from the upscattering of neutrinos inside the high-density region of the detector, followed by the decay of the heavy neutrino into \epluseminus pairs inside the gaseous Argon (GAr) Time Projection Chambers (TPC) of ND280.
By leveraging the power of our technique, we can take advantage of our detailed detector simulation throughout a much broader parameter space.

This article is divided as follows.
In \Cref{sec:model} we introduce the model we study in this work, providing a phenomenology-friendly parametrization.
\Cref{sec:analysis} provides details on the ND280 detector and the analyses we use to set constraints.
In \Cref{sec:simulation_and_MC} we discuss our dedicated Monte Carlo simulation and introduce our re-weighing scheme and KDE techniques. 
Finally, in \Cref{sec:results}, we present the resulting limits in slices of parameter space and conclude in \Cref{sec:conclusions}.

\section{Dark Neutrinos}\label{sec:model}
\begin{figure}[t]
    \centering
    \includegraphics[width=0.45\textwidth]{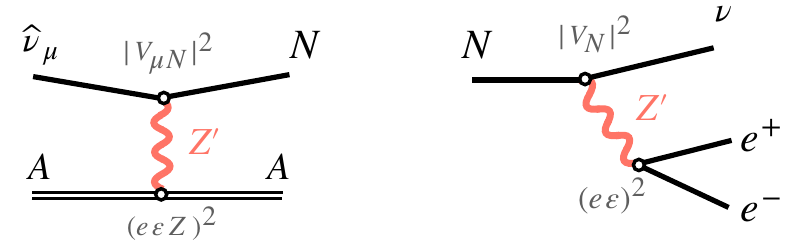}
    \caption{The diagrams for coherent neutrino-nucleus upscattering ($\widehat{\nu}_\mu A \to N A$) and heavy neutrino decays ($N\to \nu e^+e^-$) are considered in this work. We indicate the relevant parametrization of each interaction vertex.}
    \label{fig:feyn_diagrams}
\end{figure}
The idea that low-scale seesaw extensions of the SM can co-exist with new gauge symmetries, most famously with $B-L$~\cite{Buchmuller:1991ce,Khalil:2006yi,Perez:2009mu,Khalil:2010iu,Harnik:2012ni,Batell:2016zod,Dib:2014fua,DeRomeri:2017oxa,Camargo:2019ukv,Arguelles:2019ziu},
has been discussed throughout the literature also in the context of baryonic~\cite{Pospelov:2011ha}, 
leptonic ~\cite{Baek:2015mna,Kamada:2015era},
or completely hidden gauge symmetries~\cite{Baek:2013qwa,Okada:2014nsa,Ko:2014bka,Diaz:2017edh,Nomura:2018ibs,Hagedorn:2018spx,Shakya:2018qzg}. These models present a complicated mass spectrum and a self-interacting dark sector that can be challenging to identify experimentally. Nevertheless, they remain viable and testable examples of low-scale neutrino mass mechanisms and deserve experimental scrutiny.

In this article, we focus on a low-scale dark sector containing heavy neutrino states and a broken $U(1)_D$ gauge symmetry~\cite{Bertuzzo:2018itn, Bertuzzo:2018ftf, Ballett:2018ynz, Arguelles:2018mtc, Ballett:2019pyw, Ballett:2019cqp, Abdullahi:2020nyr}. 
The heavy neutrinos interact with the mediator of the dark gauge group, the dark photon, and with SM neutrinos via mixing.
Through a combination of neutrino mixing and kinetic mixing between the SM photon and the dark photon, this dark sector leads to new interactions of neutrinos with charged particles. 
It contains heavy neutrino states that decay primarily through the new force. 
This model is especially interesting in the context of short-baseline anomalies as it predicts the production of heavy neutrinos inside detectors, which subsequently decay to electromagnetic final states, mimicking $\nu_\mu \to \nu_e$ appearance signatures.

We start with the definition of a simplified, low-energy Lagrangian that will be used throughout this work. 
The minimal particle content we consider contains a single mediator, corresponding to a kinetically-mixed dark photon $Z^\prime$, and heavy neutrino states $\nu_{i \ge 4}$.
We provide further details on the UV completions of the model at the end of this section.
In terms of the physical fields, our Lagrangian reads
\begin{align}\label{eq:Lagrangian}
    \mathcal{L} \supset \mathcal{L}_{\nu\text{-mass}} +\frac{m_{Z^\prime}^2}{2} Z^{\prime \mu} Z^\prime_{\mu} + Z^\prime_\mu \left(  e \epsilon {J}_{\rm EM}^\mu +  g_D {J}_{D}^\mu\right),
\end{align}
where $\mathcal{L}_{\nu\text{-mass}}$ contains all the mass terms for the neutrino fields after proper diagonalization.
The dark photon interacts with the SM electromagnetic current, $J_{\rm EM}^\mu$, proportionally to the electric charge $e$ and kinetic mixing $\epsilon$.
It also interacts with the neutral leptons dark current, $J_D^\mu$, proportionally to the gauge coupling $g_D$.
The above Lagrangian includes all interactions of interest in the limit of small $\epsilon$ and $(m_{Z^\prime}/m_Z)^2$. 

In terms of an interaction matrix $V$, the dark current in the mass basis is given by
\begin{align}\label{eq:dark_current}
     {J}^\mu_D = \sum_{i,j}^{n+3} V_{i j} \overline{\nu_i} \gamma^\mu \nu_j,
\end{align}
where $n$ is the number of heavy neutrino states. 
Here, $\nu_{1,2,3}$ are the mostly-SM-flavor light neutrinos, and $\nu_{i\ge 4}$ are the heavy neutrinos that contain small admixtures of SM flavors. 
We can express $V_{ij}$ in terms of the mixing matrix $U_{di}$ between the mass eigenstates $i$ and dark flavor states $d$ as $V_{ij} = \sum_d Q_d U_{di}^*U_{dj}$, where the index $d$ runs over all dark-neutrino flavors $\nu_d$ of $U(1)_D$ charge $Q_d$. 
Assuming $Q_d=1$ for all flavors, $|V_{ij}| \leq 1$ for all $i$ and $j$ due to the unitarity of the full neutrino mixing matrix. 
Since we are not interested in the specifics of the flavor structure of the full model, we stick with the generic notation of~\Cref{eq:dark_current}, noting that experimental constraints require $V_{ij}\ll 1$ if either $i$ or $j$ are in $\{1,2,3\}$.

We now consider the upscattering of light neutrinos into one of the $n$ heavy neutrinos states $\nu_N$, hereafter referred to as $N$ for brevity, and its subsequent decay into lighter neutrinos and an $e^+e^-$ pair. Specifically, 
\begin{align}
    \widehat{\nu}_\mu + A \to (N \to \nu e^+e^-) + A,
\end{align}
where $\widehat{\nu}_\mu$ stands for the coherent superposition of $\nu_{1,2,3}$ produced in the neutrino beam, and $\nu$ for all possible daughter neutrinos.
Throughout this work, we consider only coherent scattering on the nucleus $A$.

\subsection{Upscattering and decay}
\begin{figure}[t]
    \centering
    \includegraphics[width=0.49\textwidth]{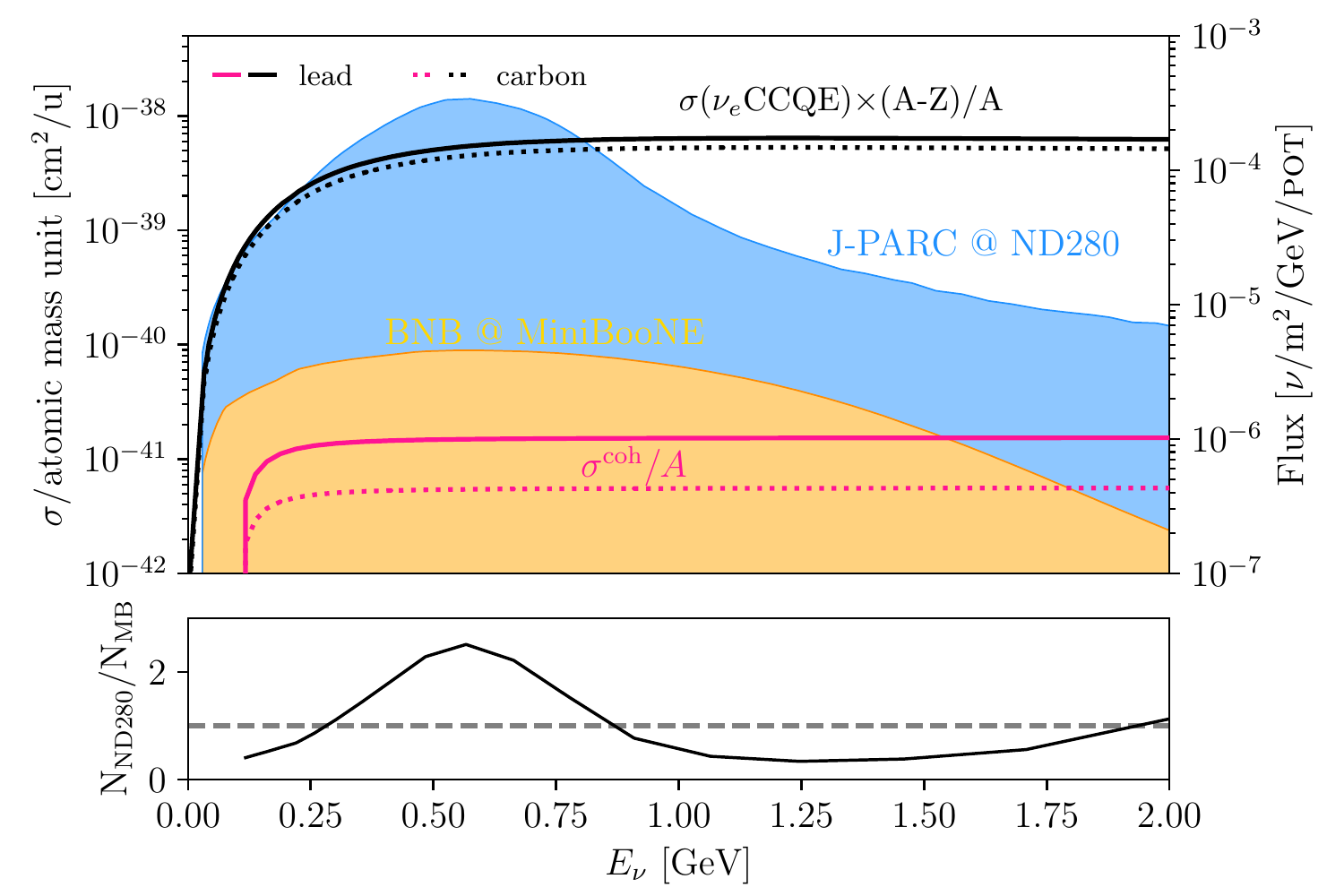}
    \caption{Comparison between relevant neutrino fluxes and cross sections. Although MiniBooNE has a much larger mass, ND280 benefits from a significantly larger neutrino flux and higher Z materials.
    The solid and dotted lines show the cross section per atomic mass unit for lead and carbon, respectively.
    The $\nu_e$CCQE cross sections are shown as black lines and the coherent upscattering cross sections for our heavy dark photon benchmark (B) as pink lines.
    When considering all the active material in ND280 and MiniBooNE we found that the ratio of upscattering between the two experiments is $\mathcal{O}(1)$ across the energy spectrum.
    \label{fig:xsec_flux_comparison}}
\end{figure}

The coherent neutrino-nucleus scattering is mediated by the dark photon with amplitude
\begin{align}
\label{eqn:upscattering_amplitude}
\mathcal{M}_{\rm ups}^{Z^\prime} &= \frac{e\epsilon g_D}{q^2 - m_{Z^\prime}^2} \, \ell_\mu h^\mu,
\end{align}
where $q^2$ is the momentum exchange with the nucleus, $h^{\mu} = \bra{A} {J}_{\rm EM}^\mu \ket{A}$ is the elastic electromagnetic transition amplitude for the nuclear ground state of $A$, and $\ell^\mu$ is the leptonic current
\begin{align}
\ell^\mu &= \bra{N} {J}_{D}^\mu \ket{\hat{\nu}_\mu}  = \frac{\sum_{i\le 3}U_{\mu i}^* V_{i N}   \bra{N} \overline{N} \gamma^\mu \nu_i  \ket{\nu_i}}{\left(\sum_{k\le 3} |U_{ki}|^2\right)^{1/2}},
\\\nonumber
&= V_{\mu N}\bra{N} \overline{N} \gamma^\mu \nu_i \ket{\nu_i},
\end{align}
where we defined the vertex factor
\begin{equation}
    V_{\alpha N} \equiv \frac{\sum_{i\le3} U_{\alpha i}^* V_{i N}}{{\left(\sum_{k\le 3} |U_{ki}|^2\right)^{1/2}}}.
\end{equation} 
In a model with a single dark flavor $d=D$ and one heavy neutrino $N=\nu_4$, one can show that $V_{\alpha N} =  U_{\alpha 4} |U_{D4}|^2 \simeq U_{\alpha 4}$, which is small and directly constrained by laboratory experiments.
The full cross-section is then computed in the usual fashion. 
We have implemented a data-driven Fourier-Bessel parametrization for the nuclear form factors~\cite{DeJager:1974liz}. 
An approximate analytic formula for the full cross-section is provided in \Cref{app:cross_sec_app}. 

The decay process can be computed similarly, now summing over the daughter neutrinos incoherently,
\begin{align}
    |\mathcal{M}_{\rm dec}^{Z^\prime}|^2 &\equiv \sum_{i<N} |V_{Ni} \mathcal{M}(m_i)|^2 \simeq |\mathcal{M}(0)|^2 \sum_{i<N} |V_{Ni}|^2,
\end{align}
where we factorize the matrix elements assuming that all daughter neutrinos have a negligible mass with respect to $m_N$.
We define the remaining vertex factor as
\begin{align}
    |V_N|^2 &= \sum_{i<N} |V_{iN}|^2.
\end{align}
As before, for a model with a single dark flavor and one heavy neutrino, $|V_N|^2 = |U_{D4}|^2 (1 - |U_{D4}|^2) \simeq |U_{e 4}|^2+|U_{\mu 4}|^2+|U_{\tau 4}|^2$.
Clearly, $|V_N|^2$ may be similar in size or much larger than $|V_{\mu N}|^2$, depending on the flavor structure of the model.\footnote{This was the idea proposed in \refref{Ballett:2018ynz}, where by virtue of $|U_{\tau 4}|^2 \gg |U_{\mu 4}|^2$, the daughter neutrino produced had a significant admixture of the tau flavor.}
This way, the production cross section is effectively decoupled from the lifetime of $N$. 
This statement is irrelevant for light dark photons ($m_{Z^\prime} < m_N$), as the decay is always prompt, rendering most signatures independent of $|V_N|^2$. 
However, in MiniBooNE explanations where $N$ decays via an off-shell dark photon, the requirement $|V_{N}| > |V_{\mu N}|$ helps ensure that the production of $N$, as well as its decays, happen inside the detector.

\renewcommand{\arraystretch}{1.3}
\begin{table}[t]
    \centering
    \begin{tabular}{|c|c|c|}
    \hline
    Benchmark & (A) Light $Z^\prime$ & (B) Heavy $Z^\prime$
    \\
    \hline\hline
    $m_N$ (MeV) &  $100$ & $100$
    \\\hline
    $m_{Z^\prime}$ (GeV) & $0.03$ & $1.25$
    \\\hline
    $|V_{\mu N}|^2$ & $8 \times 10^{-9}$ & $2.2 \times 10^{-7}$
    \\\hline
    $\alpha_D$ & $1/4$ & $0.4$
    \\\hline
    $\epsilon$ & $1.7\times10^{-4}$ & $2\times10^{-2}$
    \\\hline
    $|V_{N}|^2$ & $|V_{\mu N}|^2$ & $1$
    \\
    \hline \hline
    $c\tau^0_N$ (cm) & $7 \times 10^{-5}$ & $0.54$
    \\
    $N_{\rm upscattering}^{\rm T2K}$ & $15.5$ & $560$
    \\
    $N_{\rm decay}^{\rm T2K}$ & $15.5$ & $4.6$
    \\
\hline\hline
    \end{tabular}
    \caption{Parameters of our two benchmark points. These choices are compatible with the excess of events observed at MiniBooNE. 
    \label{tab:my_label}}
\end{table}

We also keep the number of daughter neutrinos unspecified to effectively cover models where $N$ does not decay only into $\nu_{1,2,3}$, but also into other heavy neutrinos $\nu_j$ with $3<j<N$. 
In this case, $V_{jN}$ is mainly insensitive to the direct limits on the mixing of active and heavy neutrinos, $|U_{\alpha 4}|^2$, and can be of order one.
The properties of these new states are rather model-dependent, so we conservatively consider them invisible and not observable.
For simplicity, we require that $\nu_j$ be light enough such that $(m_N - m_{j})/m_N \ll 1$. Therefore, we do not consider scenarios with small mass splittings between the upscattered and the daughter neutrinos.

The relevant decay rate for a Dirac $N$ with off-shell $Z^\prime$ is
\begin{equation}
\label{eqn:gamma_heavy}
\Gamma_{N\to \nu e^+e^-} =  \frac{\alpha\alpha_D\epsilon^2 |V_{N}|^2 }{48\pi} \frac{m_{N}^5}{m_{Z^\prime}^4} L(m_N^2/m_{Z^\prime}^2),
\end{equation}
where $L(x) = \frac{12}{x^4} \left(x -\frac{x^2}{2} - \frac{x^3}{6} - (1-x)\log{\frac{1}{1-x}}\right)$, with $L(0) = 1$.
For a light, on-shell $Z^\prime$ we need only compute $N \to \nu Z^\prime$ since $Z^\prime \to e^+e^-$ is always prompt,
\begin{equation}
\label{eqn:gamma_light}
\Gamma_{N \to \nu Z^\prime} =  \frac{ \alpha_D |V_{N}|^2 }{4} \frac{m_{N}^3}{m_{Z^\prime}^2} \left(1-\frac{m_{Z^\prime}^2}{m_{N}^2}\right)^2\left(\frac{1}{2}+\frac{m_{Z^\prime}^2}{m_{N}^2}\right).
\end{equation}
Note that the decay rate is bounded from above and below by
\begin{equation}\label{eq:decay_range}
\Gamma_N(|V_N|= |V_{\mu N}|)<\Gamma_N<\Gamma_N(|V_N|=1),
\end{equation} 
where MiniBooNE explanations prefer to saturate the right-most inequality whenever $ m_N < m_{Z^\prime}$.
While for the light $Z^\prime$ case the lifetimes of both $N$ and $Z^\prime$ are always prompt, for the heavy case they are much longer. For instance, taking $|V_N|=1$ and $m_{Z^\prime} = 1.25$~GeV, we find
\begin{equation}
c\tau^0_{\rm min} \simeq 1 \text{ cm} \times \left( \frac{10^{-2}}{\epsilon} \right)^2\left( \frac{100 \text{ MeV}}{m_N} \right)^5 \left( \frac{m_{Z^\prime}}{1.25 \text{ GeV}} \right)^4.
\end{equation}

\subsection{Benchmark points and MiniBooNE}

We now comment on the broader context of the MiniBooNE anomaly and present two choices of model parameters that will help us benchmark the MiniBooNE explanation in the context of dark photon models.
Despite being around for over two decades, the MiniBooNE and LSND anomalies have remained unsolved.
Most recently, the origin of the MiniBooNE excess was recently searched for in the MicroBooNE experiment~\cite{MicroBooNE:2021rmx, MicroBooNE:2021nxr, MicroBooNE:2021jwr, MicroBooNE:2021sne}. 
As argued in \refref{Arguelles:2021meu}, the sterile neutrino interpretation, although disfavored, is not entirely ruled out, and several hypotheses behind the excess remain untested, including those involving large $\nu_e$ disappearance~\cite{Denton:2021czb}. 
This observation is also true for explanations that rely on particle misidentification, such as models with $e^\pm \leftrightarrow \gamma$ or $e^\pm \leftrightarrow e^+e^-$ misidentification.
Several models exploring this mechanism have been put forward. 
However, only a small subset of those can also explain LSND~\cite{Abdallah:2022grs} due to the much harder-to-fake inverse-beta-decay signature.
In that case, a neutral mediator in neutrino-nucleus scattering processes should kick out a neutron from inside the Carbon nucleus.
Not only is this a negligible effect for a dark photon mediator, but it also requires larger neutrino energies to produce the heavy neutrinos and remain above the $E_e>20$~MeV analysis threshold.
Therefore, we proceed to present benchmark points compatible with the MiniBooNE observation only.

\paragraph{Benchmark A, light $Z^\prime$.---} Previously, in Ref.~\cite{Arguelles:2018mtc}, we picked the same benchmark as in Ref.~\cite{Bertuzzo:2018itn} to present bounds set by the neutrino-electron elastic scattering measurement performed by MINER$\nu$A. 
Here we would like to target the low $N$ mass region of the parameter space. 
This part of the parameter space is not constrained very effectively by MINER$\nu$A due to systematic uncertainties in the background estimate.

\paragraph{Benchmark B, heavy $Z^\prime$.---} 
This benchmark is inspired by the benchmarks provided in Refs.~\cite{Ballett:2019pyw} and \cite{Abdullahi:2020nyr}. It illustrates the case of a heavy dark photon, where the coherent upscattering contribution is not dominant but still significant.

Both choices of parameters above can explain the MiniBooNE energy spectrum but not the angular spectrum. 
The exchange of a dark photon with the nucleus gives rise to very low-$Q^2$ processes and, therefore, very forward $e^+e^-$ final states, making the model in apparent contradiction with the MiniBooNE observation. 
Quantifying this tension, however, is not currently possible due to the lack of public information on the background systematics in $\cos{\theta}$. 
Since systematic uncertainties in the background prediction dominate the significance of the MiniBooNE excess, a proper fit should include the systematic uncertainties in the angular bins and, most importantly, their correlations.

Models with helicity-flipping interactions and heavy mediators, such as scalar mediator models~\cite{Abdallah:2022grs}, have a better chance of describing the angular spectrum.
These also have the advantage of having larger cross sections with neutrons and being interesting in the context of LSND.
Due to their broader angular spectrum, we expect these models to have smaller selection efficiencies at T2K than the ones we find for the dark photon model. 
We leave the exploration of these models for future work after proper fits to the MiniBooNE excess have been performed.

It should also be noted that there have been a series of constraints posed on the model above, from accelerator neutrino experiments to kaon decays.
We note the study of Ref.~\cite{Brdar:2020tle}, where the authors point out a large set of experimental observables that can be used to constrain MiniBooNE explanations, among which the ND280 data that we make use of.
In our analysis, we properly take into account detector effects and systematics, carefully describing the geometry of the detector, which is an essential ingredient to correctly determine the bounds on the heavy \zprime case, which are extremely sensitive to the lifetime of the heavy neutrino.

\subsection{UV completions}
Possible UV completions of \refeq{eq:Lagrangian} have been discussed in Refs.~\cite{Bertuzzo:2018ftf,Ballett:2019pyw,Abdullahi:2020nyr}.
The general idea is to consider new fermions $\nu_D$ charged under the new gauge symmetry, which mix with SM neutrinos upon symmetry breaking.
Two main categories can be identified, depending on the pattern of the $U(1)_D$ breaking.
Schematically, they make use of the following operators,
\begin{align}
    \label{eq:opI}
    \text{(I): }& (\overline{L} \tilde{H}_D)\nu_D,\\
    \label{eq:opII}
    \text{(II): }& (\overline{L} \tilde{H})(\Phi \nu_D).
\end{align}
The first route requires new $SU(2)_L$ scalar doublets, $H_D$, also charged under the dark symmetry.
The mixing between SM neutrinos and the dark leptons $\nu_D$ is then generated by the expectation value of $H_D$, which breaks the $U(1)_D$, together with the SM Higgs, also the electroweak symmetry.
The second method considers instead an SM-singlet dark scalar $\Phi$, whose expectation value breaks only the dark symmetry. The main idea is illustrated by the dimension-six operator in \Cref{eq:opII}, which induces a mixing term between $\nu_D$ and the SM neutrinos after symmetry breaking.
This effective operator can be easily generated by the exchange of a singlet (sterile) neutrino $\nu_s$, which serves as a bridge between the SM and the dark sector via $(\overline{L}\tilde{H})\nu_s$ and $\overline{\nu_s}(\nu_D \Phi)$. 

Note that in both cases, the dark photon gets a mass from breaking $U(1)_D$, while the masses of the neutral leptons and the additional scalar degrees of freedom will depend on the specifics of the model.
The interplay between the expectation values of the new scalars, Yukawa couplings, and new arbitrary Majorana mass terms will determine the coupling vertices $V_{ij}$ in \Cref{eq:dark_current} and should also generate the correct value for the light neutrino masses.
In particular, both types of model are flexible enough for $N$ to be a (pseudo-)Dirac or Majorana particle. 
However, pseudo-Dirac states are preferred to generate small neutrino masses due to the approximate conservation of the lepton number.
Any additional fermion in the model can be heavier than a few GeV, where their interactions with the SM would be poorly constrained at the values of neutrino mixing we consider.

\section{ND280 analyses}\label{sec:analysis}
\begin{figure*}[t]
    \centering
    \includegraphics[width=\textwidth]{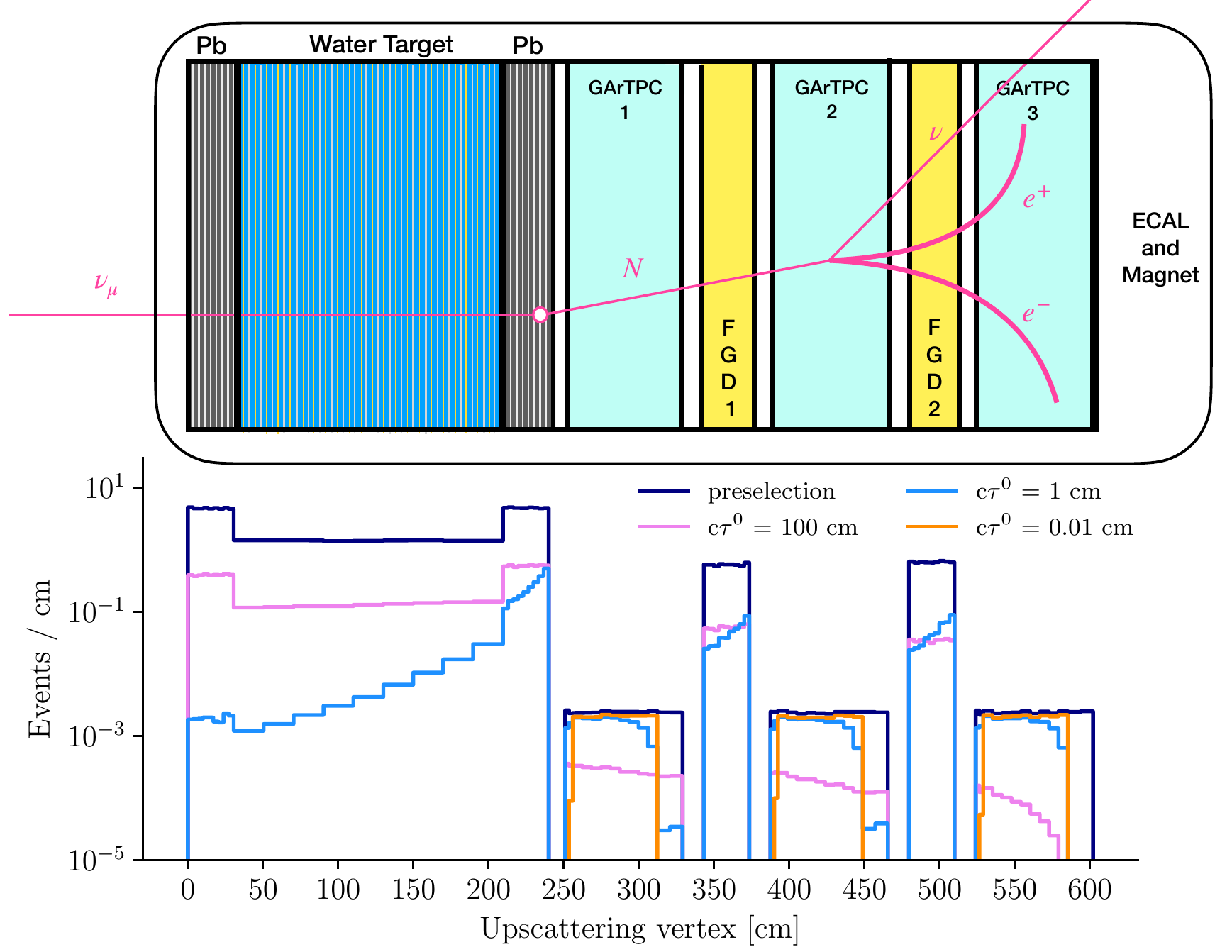}
    \caption{Diagram of the T2K near detector, ND280, showing all the active components of the detector and the new-physics signature we are interested in. Below, we show the event rate distribution as a function of the upscattering position $z$ before and after geometrical and analysis selection. For long lifetimes ($c\tau^0\gtrsim1$~cm), the event rate is dominated by upscattering on lead, while for much smaller values, only upscattering on the gaseous argon modules contributes. \label{fig:ND280_diagram}}
\end{figure*}

ND280 is the off-axis near detector of T2K located at 280~m from the target at an angle of $\sim 2.042^\circ$ with respect to the beam~\cite{T2K:2011qtm}. 
The mean neutrino energy at this location is very similar to that of the Booster Neutrino Beam, where MiniBooNE is located. 
The comparison of the two fluxes in \Cref{fig:xsec_flux_comparison} clearly shows that the flux seen by ND280 is significantly larger than that seen by MiniBooNE for the same exposure.
The active mass, however, is much smaller -- MiniBooNE contains a total of $818$~t of mineral liquid scintillator (CH$_2$) compared with $18$~t of total active mass.
Considering these two elements, we expect a similar number of upscattering events to happen in the two detectors, enabling T2K to directly test the dark neutrino interpretation of the MiniBooNE excess.

Our analysis reinterprets two public results by T2K: the search for the in-flight decays of long-lived heavy neutrinos~\cite{Abe:2019kgx} and the $\nu_e$CCQE cross-section measurement~\cite{T2K:2020lrr}. 
The former directly searched for appearing $e^+e^-$ pairs inside the low-density region of the detector, while the latter measured the rate of single photons that convert into $e^+e^-$ pairs inside one of the tracking components of the detector. 
We discuss the detector components below to then discuss the two analyses.

\subsection{The ND280 detector}

ND280 is a highly segmented and magnetized detector. 
The detector modules used in our work are shown in \Cref{fig:ND280_diagram} and constitute most of the active volume of the detector. 
The first three modules constitute the \pzerod detector, a layered arrangement of high-Z material such as lead and brass, intertwined with plastic scintillators and water bags that serves both as an electromagnetic calorimeter (ECAL) and an active water target. 
This subdetector is specifically designed to study $\pi^0$ production and neutrino cross sections in water, both critical inputs for the oscillation analyses using the Super-Kamiokande far detector.
The first and third ECALs contain only layers of lead and scintillator plates.
The module in between contains the water bags as well as layers of brass and scintillator plates.
Downstream we have the tracking modules composed of three gaseous Argon time projection chambers (GArTPC), separated by fine-grained scintillator detectors (FGD).
These components have fewer neutrino interactions and provide a better environment for particle identification. 
The whole detector is in a 0.2~T magnetic field and is surrounded by additional side and back ECAL as well as side muon detectors.
The dimensions and composition of each region are summarized in \reftab{tab:detector_info} (see Refs.~\cite{Assylbekov:2011sh,Gilje:2014cwd,Lamoureux:2018owo} for more information on ND280).
In our simulation, we assume that the lead, brass, water, and scintillator layers are distributed uniformly inside each module.

Each GArTPC is enclosed in a plastic and aluminum cage.
The cage, in turn, is composed of an external wall ($10.1$~kg), an internal volume of CO$_2$ gas, and an internal wall ($6.9$~kg).
Inside the TPC, on the Y-Z plane, one can find the cathode ($6$~kg, made of $34\%$~C, $25\%$~O, $17\%$~Cu, $17\%$~Si).
Even though these materials are the closest to the active volume of the TPC, we neglect them as the total number of neutrino interactions recorded in them of order thousand times smaller than in the other targets.

\subsection{Analysis I -- heavy neutrino searches in the GAr TPC}
\label{sec:analysis_tpc}

\begin{figure}
    \centering
    \includegraphics[width=0.45\textwidth]{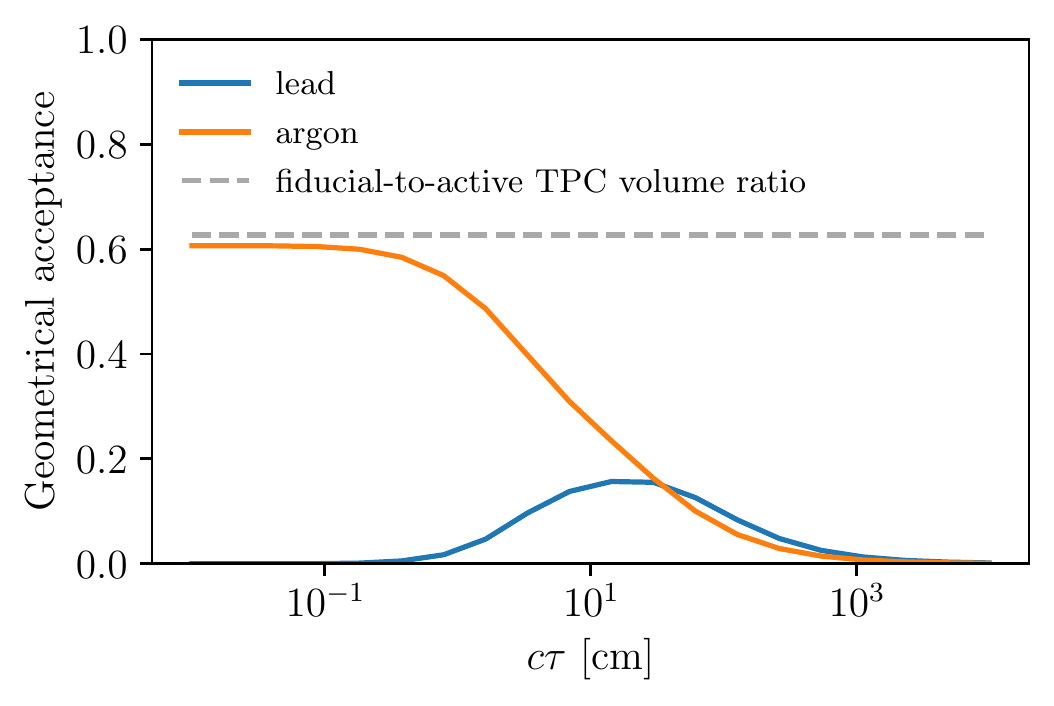}
    \caption{The geometrical acceptance of ND280 as a function of the $N$ proper lifetime. For the smallest lifetimes, only argon can pick up the decays.}
    \label{fig:geometrical_acceptance}
\end{figure}

The analysis in~\cite{Abe:2019kgx} looked for the decay in flight (DIF) of heavy neutrinos inside the three GAr TPCs.
Heavy neutrinos are produced in the target through mixing between active and sterile neutrinos, and, after propagating from the target to the detector, they decay in the detector TPCs.
They look for multiple final states, although the relevant one for our analysis is $N\to \nu e^+e^-$, which also extends to the lowest masses, with a threshold of $m_N \sim 1$ MeV.
This analysis benefits from a clear signature with zero background, as the total Argon mass is so tiny that neutrino interactions inside the TPC do not produce a relevant background.
In order to achieve the zero background, the original selection imposes a tight fiducial volume cut in the TPC, with a requirement of no additional visible energy deposition in the detector in addition to the charged particles produced in the TPC.
Our model can be tested with this analysis because it predicts a large coherent cross-section, resulting in a very low-energy nuclear recoil that is invisible in these detectors.

This analysis is a counting experiment performed over 12.34 $\times 10^{20}$ proton on target (POT) in neutrino mode and 6.29 $\times 10^{20}$ POT in anti-neutrino mode. 
Having observed zero $e^+e^-$ events over the neutrino background expectation of 0.563 in neutrino mode and 0.015 in anti-neutrino mode, the analysis sets strong limits on new physics.
Limits on long-lived heavy neutrinos produced in kaon decays at the target have been discussed in Ref.~\cite{Abe:2019kgx,Arguelles:2021dqn}.

\begin{figure*}[t]
    \centering
    \includegraphics[width=0.32\textwidth]{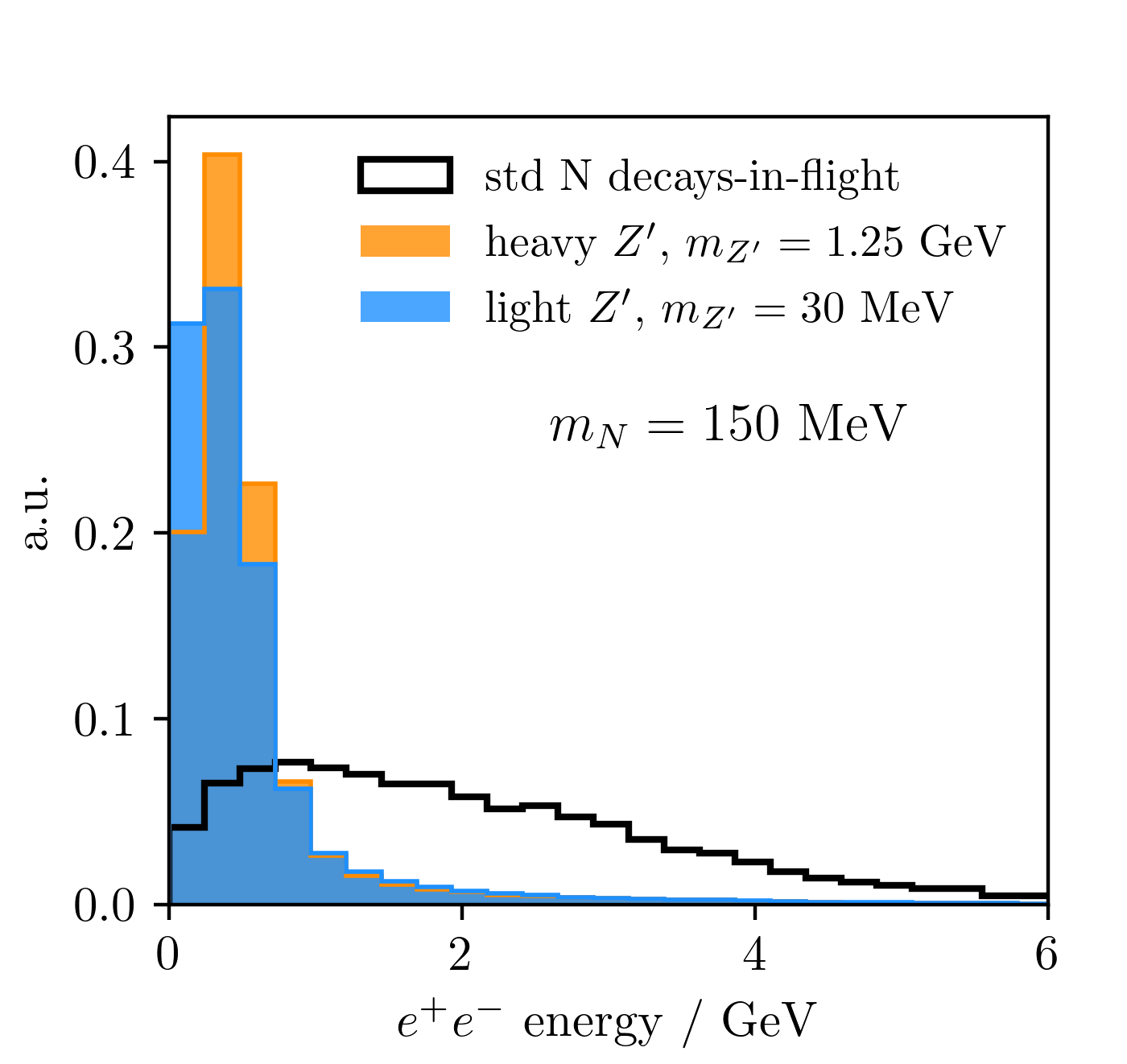}
    \includegraphics[width=0.32\textwidth]{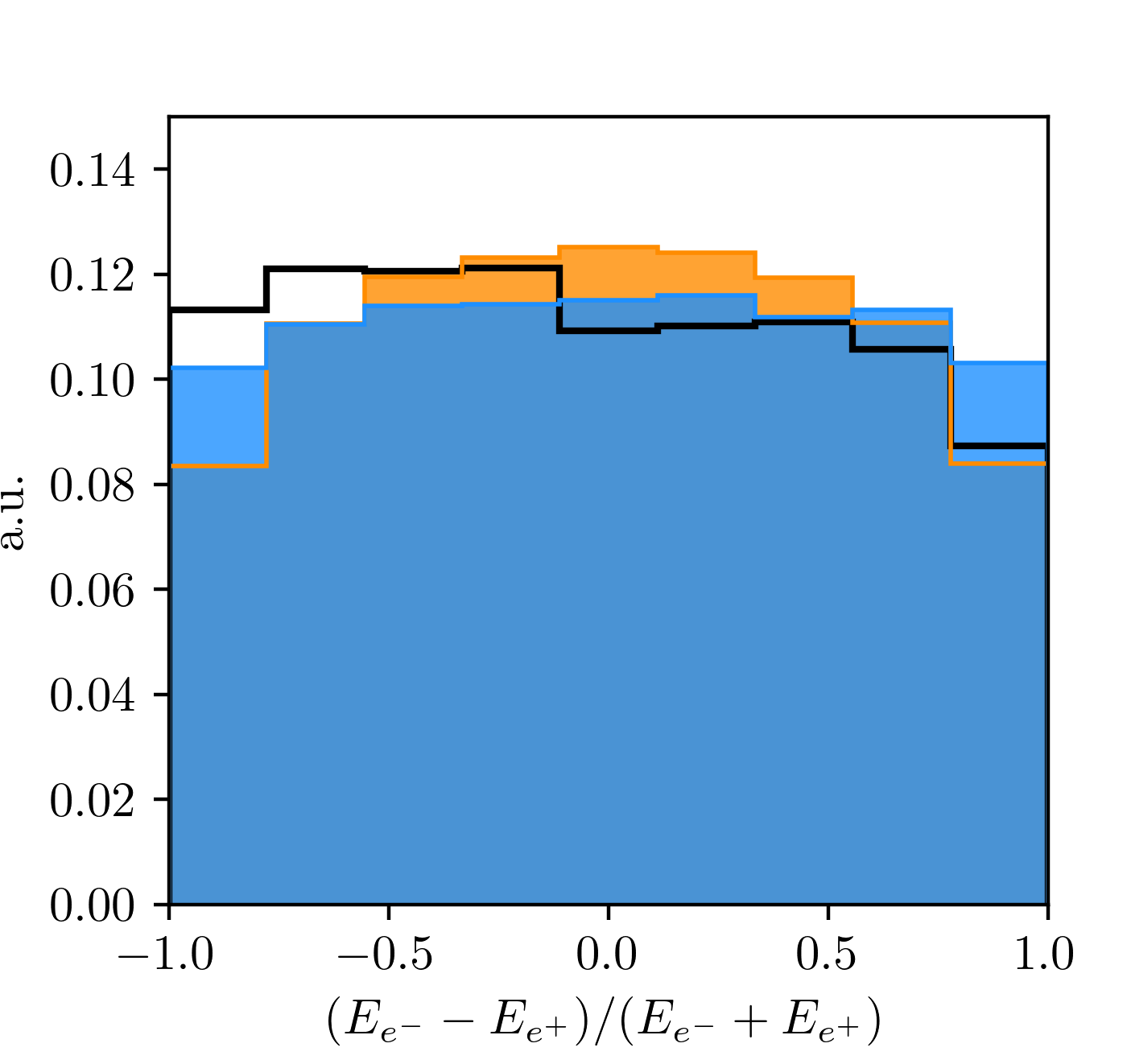}
    \includegraphics[width=0.32\textwidth]{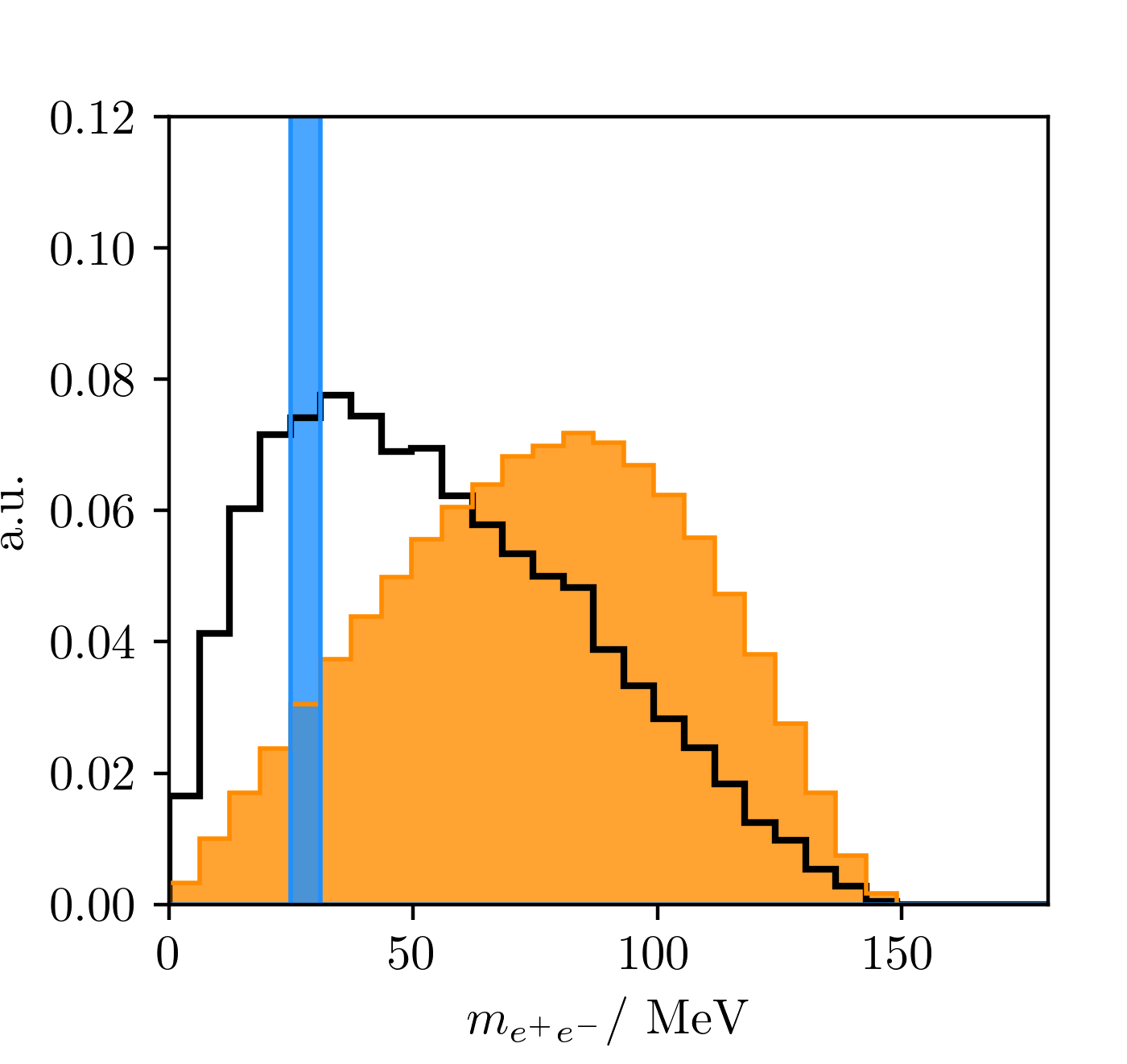}    \includegraphics[width=0.32\textwidth]{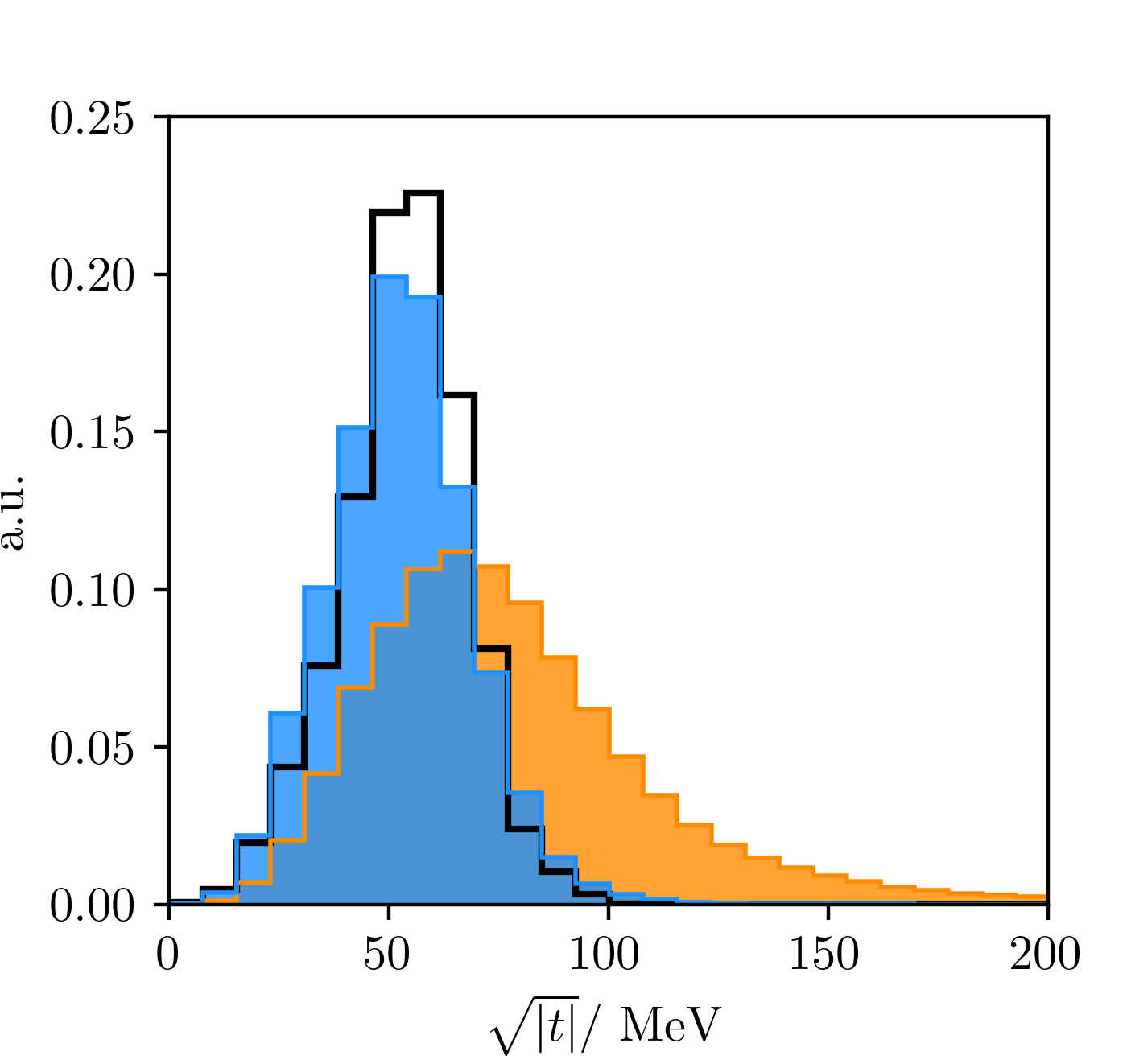}
    \includegraphics[width=0.32\textwidth]{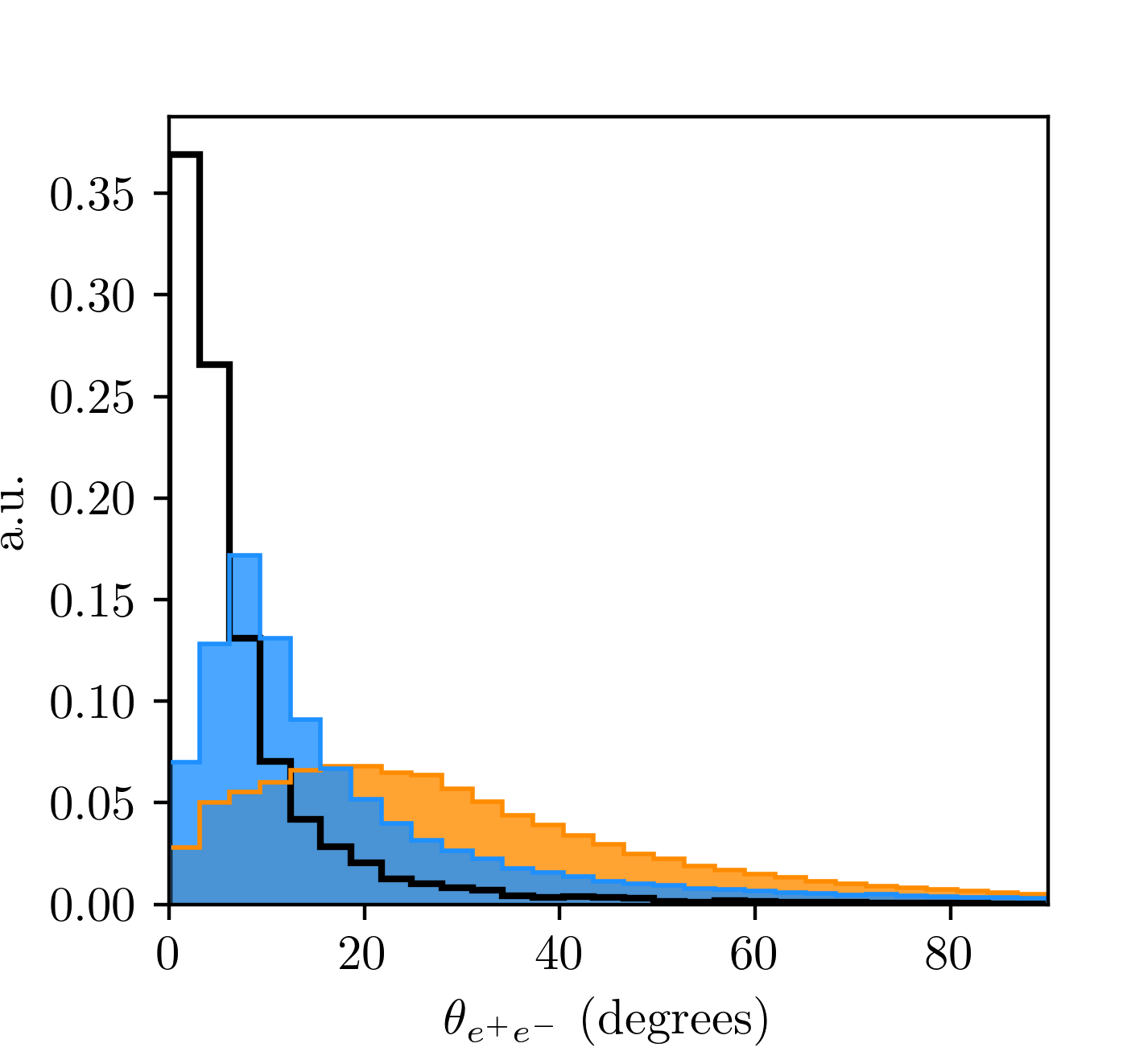}
    \includegraphics[width=0.32\textwidth]{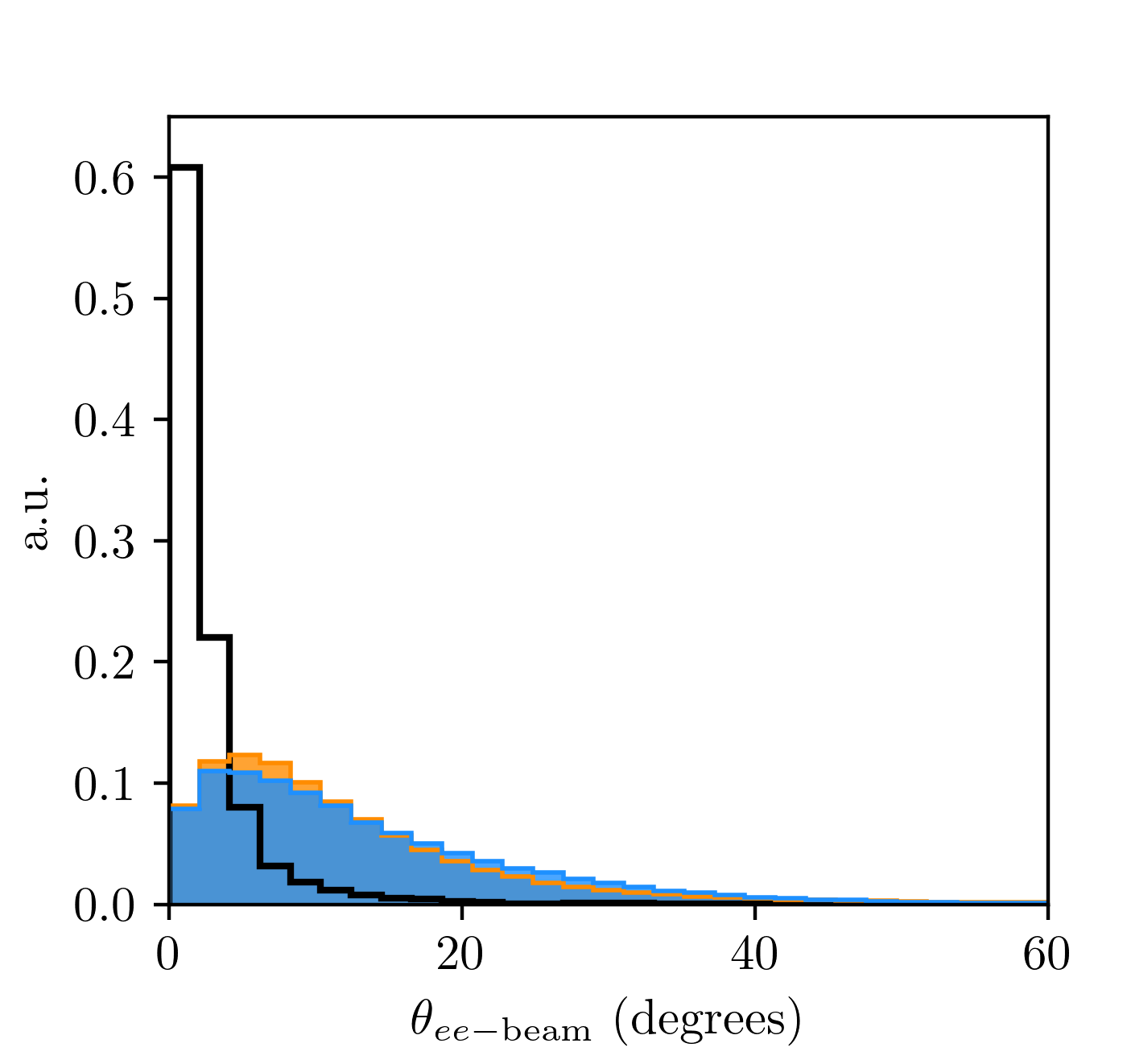}
    \caption{Comparison between standard decay-in-flight $N\to \nu e^+e^-$ signatures of heavy neutral leptons produced at the target (solid black) and that of heavy neutrino decays initiated by coherent scattering in a light dark photon (filled blue) and heavy dark photon (filled orange) model. Histograms are area-normalized. For variable definitions, see \refeq{eq:cuts}.}
    \label{fig:comparison_openingangle}
\end{figure*}

For dark neutrino models, those can be recast by considering the production of $N$ via upscattering inside the detector.
In particular, we will focus on parameters such that the lifetimes of $N$ are not much larger than $\mathcal{O}(10)$~m, as otherwise, these particles would not provide a good fit to MiniBooNE as well. 
If $N$ propagates more than $\mathcal{O}(15)$~cm, it can be produced via upscattering in the dense material of the \pzerod, where the cross-section is significantly enhanced due to the coherent scaling with proton number, $Z^2$.
It would then decay into a visible \epluseminus pair inside one of the TPCs.
This particular signature is present in most of the interesting parameter space for the heavy mediator case.
When the lab-frame lifetime becomes shorter than $\mathcal{O}(15)$~cm, the heavy neutrinos produced in the \pzerod decay before entering the TPCs, and therefore the \epluseminus are rejected by the selection to avoid large neutrino-induced backgrounds. Nevertheless, the upscattering can happen inside the TPCs, where they would be visible.
Despite the relatively small number of targets in the TPC fiducial volume, a handful of events is enough to constrain the model due to the absence of backgrounds.
Such fast decays always happen in the light dark photon case ($m_{Z^\prime} < m_N$), but it can also happen in regions of large $|V_N|^2$ values of the heavy dark photon parameter space.
\Cref{fig:geometrical_acceptance} shows the fraction of heavy neutrino decaying in one of the three TPCs as a function of the proper lifetime, for the case the upscattering happens in the lead or argon.

To estimate them, we developed a simplified detector simulation and implemented the analysis selection criteria on upscattering events generated by our own modified version of the \textsc{d}ark\textsc{n}ews generator~\cite{Abdullahi:2022cdw}.

We expect differences in the reconstruction and selection efficiencies for upscattering with respect to the decay-in-flight signatures considered in Refs.~\cite{Abe:2019kgx, Arguelles:2021dqn}. 
\Cref{fig:comparison_openingangle} shows the comparison between standard heavy neutrino signatures and our scattering-induced signatures for both the heavy and the light mediator case.
The reconstruction efficiency depends on the kinematics of the heavy neutrino decay products, including the detector capability to separate the \epluseminus tracks as a function of their opening angle and their distance of closest proximity.
While the selection efficiency relies on the following cuts (see~\Cref{eq:cuts}), which we implemented in our analysis.
No smearing of the kinematics of the electron and positron has been applied at this stage.

\begin{subequations}
\label{eq:cuts}
\begin{align}
    E_{e^+e^-} \equiv E_{e^+} + E_{e^-} &> 0.150 \text{ GeV},
    \\
    m_{e^+e^-} \equiv \sqrt{(p_{e^+} + p_{e^-})^2} & < 0.7 \text{ GeV},
    \\
    |t| \equiv (E_{e^+e^-} - p_{e^+e^-}^z)^2 - |\vec{p}_{e^+e^-}^{\,T}|^2 &< 0.03 \text{ GeV}^2,
\\
    \cos{\theta_{e^+e^-}} \equiv \frac{\vec{p}_{e^+}\dot \vec{p}_{e^-}}{|\vec{p}_{e^+}| |\vec{p}_{e^-}|} & > 0,
\\
    \cos{\theta_{ee-{\rm beam}}} \equiv p_{e^+e^-}^z/p_{e^+e^-} &> 0.99.
\end{align}
\end{subequations}
The efficiency for these cuts is of order 50\% for our dark neutrino BPs.
Given that the efficiency in the original standard heavy neutrino analysis is of the order 10-15 \%, we applied an additional 10\% efficiency factor to account for reconstruction effects in a conservative way.

We also perform a sensitivity study by projecting the status of this analysis by the end of the T2K data taking.
The current analysis could be extended to about 4 $\times 10^{21}$ POT which have already been collected by the experiment.
Moreover, ND280 is currently being upgraded to a new configuration~\cite{T2K:2019bbb}: the \pzerod is being replaced by two new GArTPCs and a Super FGD module.
A future search post-upgrade, looking only at upscattering inside the argon, could be performed on the 3 TPCs, plus the two new TPCs, on a forecast of $16\times 10^{21}$ POT~\cite{Abe:2016tii}.
This conservative estimate neglects improvements to reconstruction and background rejection and a benefit of a tailored analysis for this model.

\subsection{Analysis II -- photons in the FGD}
\label{sec:analysis_fgd}

\begin{figure}
    \centering
    \includegraphics[width=0.45\textwidth]{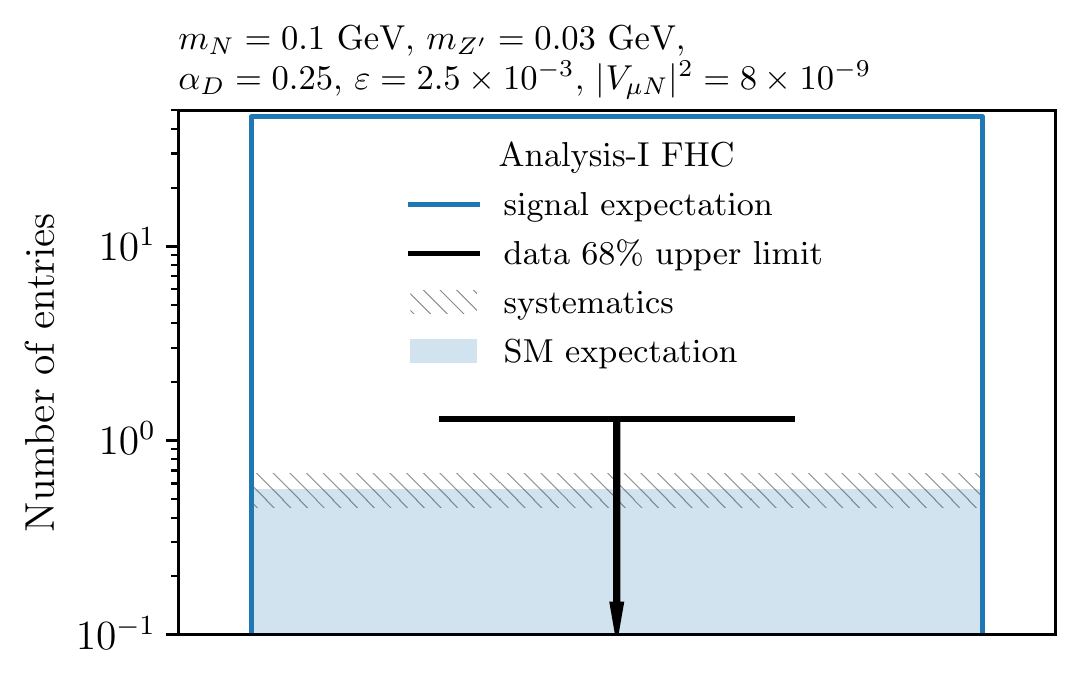}
    \includegraphics[width=0.45\textwidth]{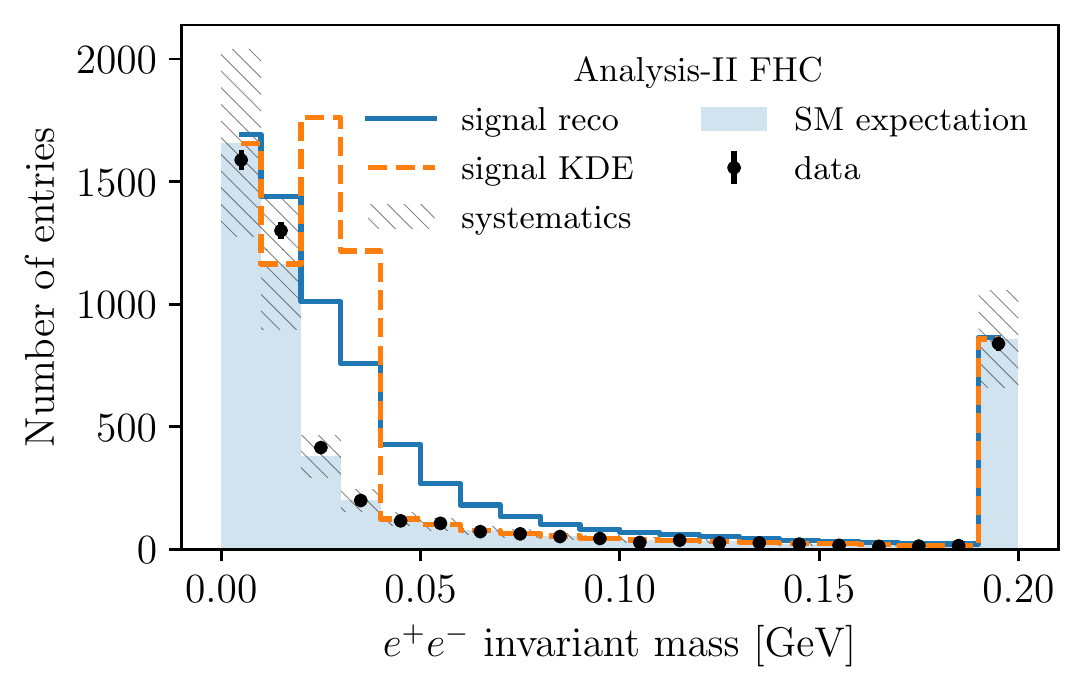}
    \caption{
    The data we are using for Analysis-I (first panel) and Analysis-II (second panel) in neutrino mode (the plots for the antineutrino mode case are analogous).
    Analysis-I is a one-bin experiment, while Analysis-II is a search for a resonance on the \epluseminus invariant mass spectrum.
    The shaded blue region represents the expected SM background, while the black points show the observed data.
    No event was observed in Analysis-I. 
    Therefore we display the upper limit at 68\% CL.
    The blue lines show the signal we expect to observe for the light \zprime benchmark point but for a larger $\varepsilon$.
    The orange line illustrates the signal at the true level, smeared only by our KDE interpolation.
    It cannot be compared with the data but gives a sense of the effect of the experimental resolution in measuring $e^+e^-$ pairs in the FGD.
    \label{fig:ee_inv_mass_spectrum}
    }
\end{figure}

The second analysis uses the photon-like control sample of the $\nu_e$CCQE cross section measurement in the first FGD.
Even though it focuses on a very different measurement, it can provide an important constraint for our model.
The largest background for this analysis comes from photons that convert inside of the FGD and for which one of the two particles has not been reconstructed.
In order to better measure this background, they look at a specific sideband, selecting \epluseminus in the FGD in the same way they select single electrons or positrons for the primary measurement.
The \epluseminus invariant mass is a helpful quantity for them to select real photons, and it can be used to constrain the dark neutrino signal in the case of a light mediator.
The $Z^{\prime}$ is produced on-shell and decays promptly to an \epluseminus pair, which, if reconstructed correctly, shows a peak in the invariant mass spectrum at $m_{ee} = m_{Z^{\prime}}$.
\Cref{fig:ee_inv_mass_spectrum} shows an example of the measured $m_{ee}$ spectrum 
We implement the smearing of momenta and zenith angles using the matrices provided in Ref.~\cite{king_sophie_2021_5543856}.
We consider a flat 10\% reconstruction and selection efficiency, which considers the 30\% efficiency for the $\nu_e CCQE$ analysis, squared to account for the two leptons.

We also estimate the sensitivity of a projection of this analysis by expanding to two FGDs, with a larger dataset and including the SuperFGD after the upgrade.

\section{Simulation and Monte Carlo Techniques}\label{sec:simulation_and_MC}

Extended dark sectors like the ones considered in this work typically involve a large number of independent parameters.
This observation poses a challenge from the phenomenology point of view. 
Performing inference in this ample parameter space requires predicting distributions of observables for all possible choices of parameters.
It can be costly and quickly become unfeasible for more detailed simulations. 
For this reason, promptly predicting our signal is crucial to improving our coverage of the model parameter space.
We apply existing statistical methods to interpolate model predictions across the parameter space, allowing the computation of the model prediction from a single simulated sample.
Fast interpolations of physical predictions across the parameter space have been discussed in the context of event generators for colliders~\cite{Buckley:2009bj, Krishnamoorthy:2021nwv} as well as in fast generators of dark matter direct detection signatures~\cite{Cerdeno:2018bty}.
These methods, however, rely on the parametrization of the prediction in terms of analytic functions.
Our technique complements these methods by deriving a non-parametric estimate of the observables.
A similar approach to the one discussed here has been proposed for treating nuisance parameters and systematic uncertainties in IceCube~\cite{IceCube:2019lxi}.
The IceCube scheme overcomes the curse of dimensionality of the production of many distinct Monte Carlo samples, sometimes described as the "multiple Universes" approach.
While the IceCube method derives a non-parametric estimate of the observables as a function of the nuisance parameters in the neighborhood of the central value, our method applies to the entire parameter space.

\subsection{General idea}

We now present the general idea behind our method. 
We can think of a model as a family of probability density functions (PDF) $p(x|\theta)$, where $\theta$ are the physical parameters of the theory over which we want to perform inference, like masses and couplings, while $x$ are observables, like particle momenta.
The model also predicts a normalization factor $\mathcal{N}(\theta)$: not just the observable distribution depends on the parameter, but also the total rate.
Both $\theta$ and $x$ are multi-dimensional, varying from several to $\mathcal{O}(10)$ dimensions.

Inference is performed by computing the expectation value $\E_{\theta}[T(x)]$ of a test statistic $T$ for each value of $\theta$.
The typical approach proceeds as follows:
\emph{i)} start from an initial definition of a multi-dimensional grid of a total of $m$ points in the parameter space $\theta_{j=1,...,m}$, 
\emph{ii)} run a simulation for each $\theta_j$, \ie{}, draw $n_j$ samples $x_{i=1,...,n^j} \sim p(x|\theta_j)$, 
\emph{iii)} compute the expectation value for each $\theta_j$ as:
\begin{equation}
    \E_{\theta_j}[T(x)] = \sum_{i=1}^{n^j} w_i^j T(x_i^j),
\end{equation}
where $w_i^j$ are weights associated with the sampling, such as importance-sampling weights. 
Finally, \emph{iv)} one eventually interpolate these values across the parameter space, in order to predict $\E_{\bar{\theta}}[T(x)]$ for a $\bar{\theta}$ which has not been simulated.

While the method above works, our procedure described next provides a more efficient way to interpolate the expectation.
It allows us to rapidly compute multiple and more complex test statistics, like histograms, using a single set of samples, \ie{}, running only one simulation.
We promote $\theta$ to a random variable by considering $p(x, \theta) = p(x|\theta)\mathcal{N}(\theta)q(\theta)$ where $q(\theta)$ is a prior over $\theta$, and we sample $x_i, \theta_i \sim p(x, \theta)$ with weights $w_i$, for $i=1,...,n$.
Using these samples, we obtain $\E_{\bar{\theta}}[T(x)]$ by interpolating across the parameter space using Kernel Density Estimation:
\begin{align}
    \E_{\bar{\theta}}[T(x)] &= \sum_i^n w_i T(x_i) \frac{w(\bar{\theta}, \theta_i)}{q(\theta_i)} \nonumber \\ 
    &= \sum_i w_i T(x_i) \frac{K(d(\bar{\theta}, \theta_i), \delta)}{q(\theta_i)},
\end{align}
where $K(d, \delta)$ is a Kernel function, $d(\bar{\theta}, \theta_i)$ is a distance in parameter space, and $\delta$ is the bandwidth or smoothing parameter.
By sampling over parameter space, we exploit the fact that neighbor parameters will produce similar observable distributions.
Using some sampling techniques, like importance adaptive sampling or Markov Chain Monte Carlo, this method will guarantee to sample observables with significant contribution to any test statistic, \ie{}, large weights.
However, the adaptation over the parameter space will also make to sample parameters where they result in larger weights.
The function $q(\theta)$ allows us to control this effect and skew the distribution of samples towards our preferences.
For example, if performing inference by conditioning on the posterior, \eg{}, when setting limits on a slice of the parameter space, fixing a subset of the total parameters, and varying the other ones, we might be in a region where there are no samples, as that slides contains a small probability with respect to the total model.
\begin{figure}[t]
    \centering
    \includegraphics[width=\columnwidth]{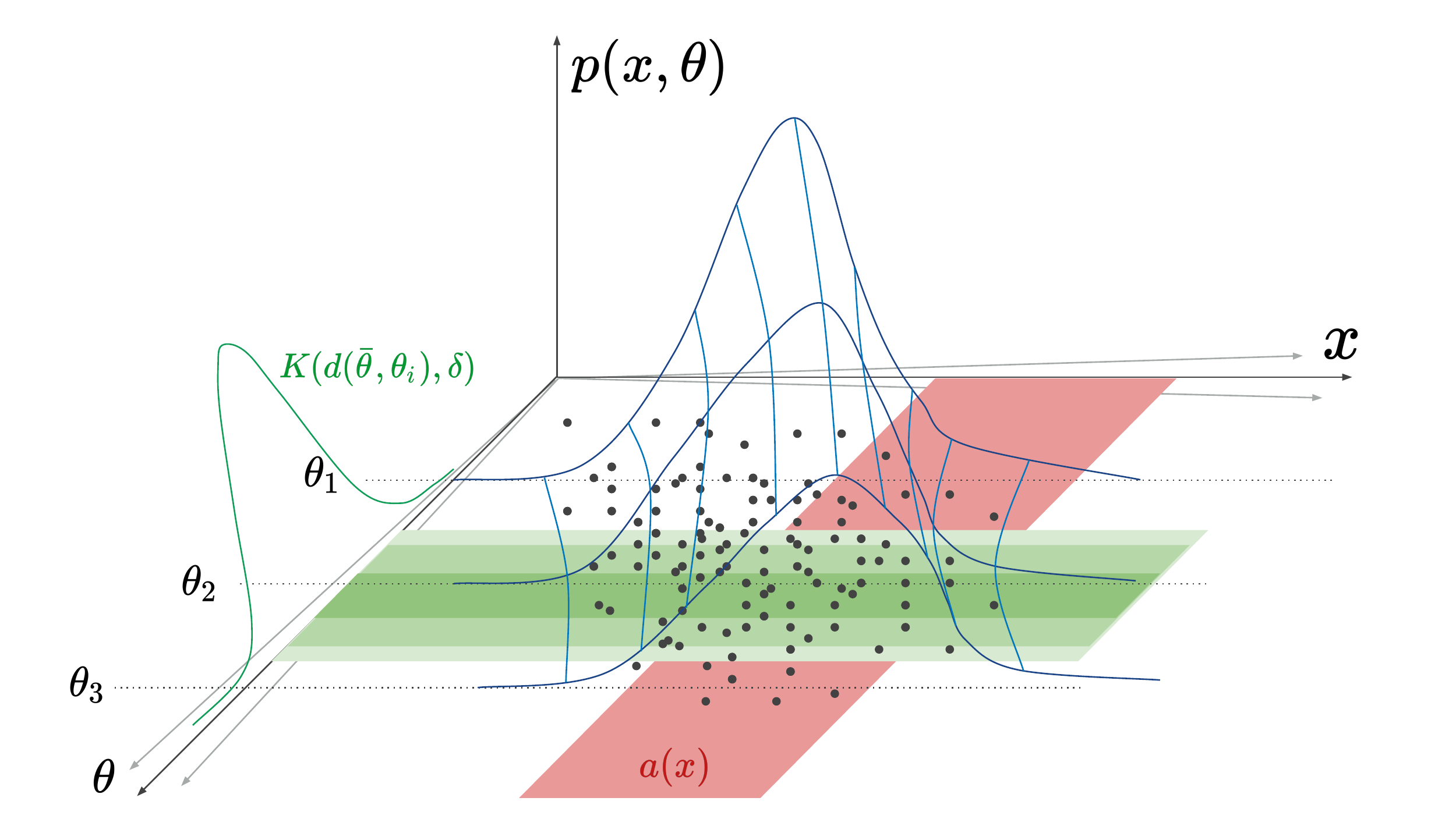}
    \caption{An illustration of the construction of the posterior distribution using the Kernel Density Estimator (KDE). 
    In one set of axes, we show the model parameters, $\theta_i$, and in another, we show the observable variables, $x_i$.
    In the vertical axis, we have the posterior distribution, $p(x,\theta)$.
    The black dots represent the samples thrown in the hyperspace of the model and phase space variables for which we calculate the posterior $p(x,\theta)$.
    The green and red planes are the support of the KDE, and the lines represent the posterior as interpolated by the KDE.\label{fig:kde_cartoon}}
\end{figure}

Finally, despite the power of this interpolation, it introduces some statistical uncertainty related to the finite sample size.
Whenever a close formula for $w(\bar{\theta}, \theta_i)$ is present, it is more effective to use that.
For this reason, we split the parameter space into $\theta^\alpha$, for which a KDE weight is computed, and $\theta^\beta$, for which the weight is computed using an analytical formula.
This procedure is illustrated in \Cref{fig:kde_cartoon}.

\subsection{Application to dark neutrino sectors}

We use this framework in the context of the dark neutrino model discussed in the previous sections.
In this model, $\theta$ is a 5-dimensional parameter space for the light mediator case and 6-dimensional for the heavy case since $|V_N|$ is only a relevant variable for the latter.
More precisely, $\theta^\alpha = \{m_N, \mzprime \}$, while $\theta^{\beta, \rm light} = \{|V_{\mu N}|^2, \alpha_D, \varepsilon \}$, while $\theta^{\beta, \rm heavy} = \theta^{\beta, \rm light} \cup \{|V_N|^2\}$.
The differential cross section in the observable space $x = \{\Omega_N, \Omega_{ee} \}$ is our $p(x|\theta)$, as
\begin{equation}
    p(x|\theta) = \frac{d\sigma}{d\Omega} = \frac{d\sigma}{d\Omega_N}(\Omega_N|\theta) 
    \frac{1}{\Gamma(\theta)} \frac{d\Gamma}{d\Omega_{ee}}(\Omega_{ee}|\Omega_N, \theta),
\end{equation}
where $\Omega_N$ describes the kinematics of the heavy neutrino and the nuclear recoil, while $\Omega_{ee}$ describes the kinematics of the \epluseminus pair and the final state neutrino.
However, since we are not interested in the incoming neutrino energy, sampled according to the flux and the degrees of freedom of the recoil and the final state neutrino, we implicitly integrate over those variables.
Here, $\Gamma$ is the total decay width of the heavy neutrino and is given by
\begin{equation}
    \Gamma(\theta) = \int d\Omega_{ee} \frac{d\Gamma}{d\Omega_{ee}}(\Omega_{ee}| \Omega_N, \theta),
\end{equation}
and can be computed analytically using \Cref{eqn:gamma_heavy} and \Cref{eqn:gamma_light} for the heavy and light scenarios, respectively.
Finally, an essential parameter for the simulation of the events in the detector is the heavy neutrino lifetime in its rest frame~$\tau^0(\theta) = \hbar/\Gamma(\theta)$.

\subsection{Monte Carlo Event Generator}
We implemented the physics matrix elements in a Monte Carlo event generator, and we sample events using the Vegas Monte Carlo algorithm~\cite{PETERLEPAGE1978192, Lepage:2020tgj} with its Python implementation~\cite{peter_lepage_2022_5893494}.

In a typical simulation, we would sample the following integral
\begin{equation}
\label{eqn:cross_section_integral}
    \sigma(\theta) = \int d\Omega_N \frac{d\sigma}{d\Omega_N}(\Omega_N|\theta)
    \int d\Omega_{ee} \frac{1}{\Gamma(\theta)} \frac{d\Gamma}{d\Omega_{ee}}(\Omega_{ee}|\Omega_N),
\end{equation}
that we now extend to
\begin{equation}
\label{eqn:cross_section_integral_extended}
    \int d\theta^{\alpha} q(\theta^{\alpha}) \int d\Omega_N \frac{d\sigma}{d\Omega_N}(\Omega_N|\theta)
    \int d\Omega_{ee} \frac{1}{\Gamma(\theta)} \frac{d\Gamma}{d\Omega_{ee}}(\Omega_{ee}|\Omega_N),
\end{equation}
where $\theta^{\alpha} = \{m_N, \mzprime\}$.
We use $q(\theta^{\alpha})^{\rm light} = \mzprime^2/m_N^{3.5}$ and $q(\theta^{\alpha})^{\rm heavy} = \mzprime^8/m_N^5$, designed to provide samples distributed as uniformly as possible in the $\{m_N, \mzprime\}$ plane.

We employ the total number of selected events in a single-bin analysis as test statistics,
\begin{align}
\label{eqn:mu_intermsof_xsec}
    \mu(\theta) = &\sum_{k} n_t^{k} \times \rm POT \times\nonumber \\
    &\times d\Omega_N \frac{d\sigma}{d\Omega_N}(\Omega_N|\theta) \times\nonumber\\
    &\times \frac{1}{\Gamma(\theta)} d\Omega_{ee}(\Omega_{ee}|\Omega_N) \epsilon(\Omega_{ee}) a(\Omega_N, \Gamma(\theta)),
\end{align}
where the cross-section has been multiplied by $n_t^{k}$, the number of targets for each material, indexed by $k$, and by the collected beam exposure in terms of protons on target (POT).
We also folded in the selection efficiency $\epsilon(\Omega_{ee})$ and detector acceptance $a(\Omega_N, \Gamma(\theta))$, which depends on the kinematics as well as on the lifetime of the heavy neutrino.
Both functions can be computed as multidimensional cuts on the observables.

By introducing weights for parameters $\theta^\alpha$ and $\theta^\beta$ such that:
\begin{align}
\label{eqn:weights_theta_alpha_beta}
    \E_{\bar{\theta}}[T] &= \int d\Omega_N d\Omega_{ee} T(x) \frac{d\sigma}{d\Omega_N}(\Omega_N|\bar{\theta}) 
    \frac{1}{\Gamma(\bar{\theta})} (\Omega_{ee}|\Omega_N) \nonumber \\
    &= \sum_i^n w_i T(x_i) w_i^{\rm KDE}(\bar{\theta}^\alpha, \theta_i^\alpha) w_i^{\rm \sigma}(\bar{\theta}^\beta, \theta_i^\beta),
\end{align}
and by defining $\epsilon(\Omega_{ee, i}) = w^{\varepsilon}_i(\Omega_{ee, i})$, $a(\Omega_{N, i}, \Gamma(\theta_i)) = w_i^{\tau_0}(\theta_i, \Omega_{N, i})$, and $n_t^{k_i} \times \rm POT = w^{\rm n_t, POT}_i$ if event $i$ is generated on material $k_i$, we can rewrite \Cref{eqn:mu_intermsof_xsec} as a product of weights:
\begin{align}
\label{eqn:weights_product}
    \mu(\theta) \simeq \sum_i^n &w_i w_i^{\rm KDE}(\bar{\theta}^\alpha, \theta_i^\alpha) w_i^{\rm \sigma}(\bar{\theta}^\beta, \theta_i^\beta)\nonumber\\ &w^{\varepsilon}_i(\Omega_{ee, i}) w_i^{\tau_0}(\theta_i, \Omega_{N, i}) w^{\rm n_t, POT}_i.
\end{align}

For simplicity, we will discuss the method by writing the expectation for a single-bin analysis, which is the case for Analysis-I.
Binned analyses, like Analysis-II, represent a trivial extension.

\subsection{Multidimensional re-weighting scheme}
\begin{figure}[!ht]
    \centering
    \includegraphics[width=0.45\textwidth]{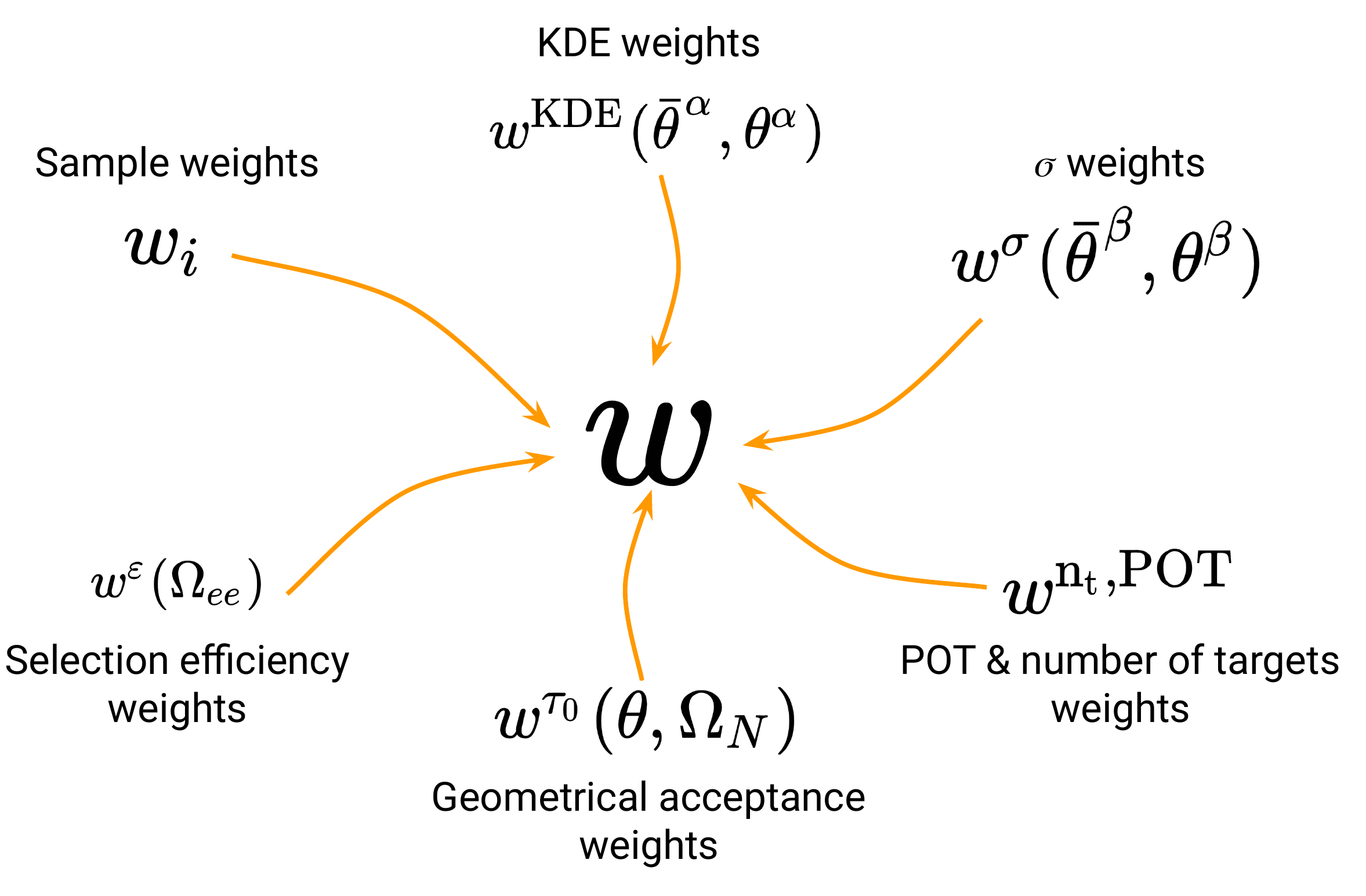}
    \caption{In order to quickly compute the model prediction across the parameter space, we implement a set of weights that accounts for the simulation of the cross-section, the kinematics, and detector effects.
    By taking the product of all these weights, we obtain a single weight that can be used to compute the final model prediction for any observable.
    \label{fig:reweighting_scheme}
    }
\end{figure}

We now detail the various weights appearing in \Cref{eqn:weights_product}.
\Cref{fig:reweighting_scheme} summarizes the different aspects of the re-weighting scheme.

\paragraph{Cross section KDE weights.---}
The cross-section has a non-trivial dependence on $\theta^\alpha = \{m_N, \mzprime \}$.
We define the KDE weight as
\begin{equation}
    w_i^{\rm KDE}(\bar{\theta}^\alpha, \theta_i^\alpha) = K(d(\bar{\theta}^\alpha, \theta_i^\alpha), \delta)/q(\theta_i^\alpha).
\end{equation}
We studied the accuracy of different kernels, distance functions, and smoothing parameters by comparing the interpolation on a benchmark grid with a dedicated, high-statistics sample for different values of $\theta$.
In the rest of the work, we used the Epanechnikov kernel, a logarithmic distance, and $\delta = 0.005$ along the direction of both parameters, in an uncorrelated way.

\begin{figure*}[t]
    \centering
    \includegraphics[width=0.48\textwidth]{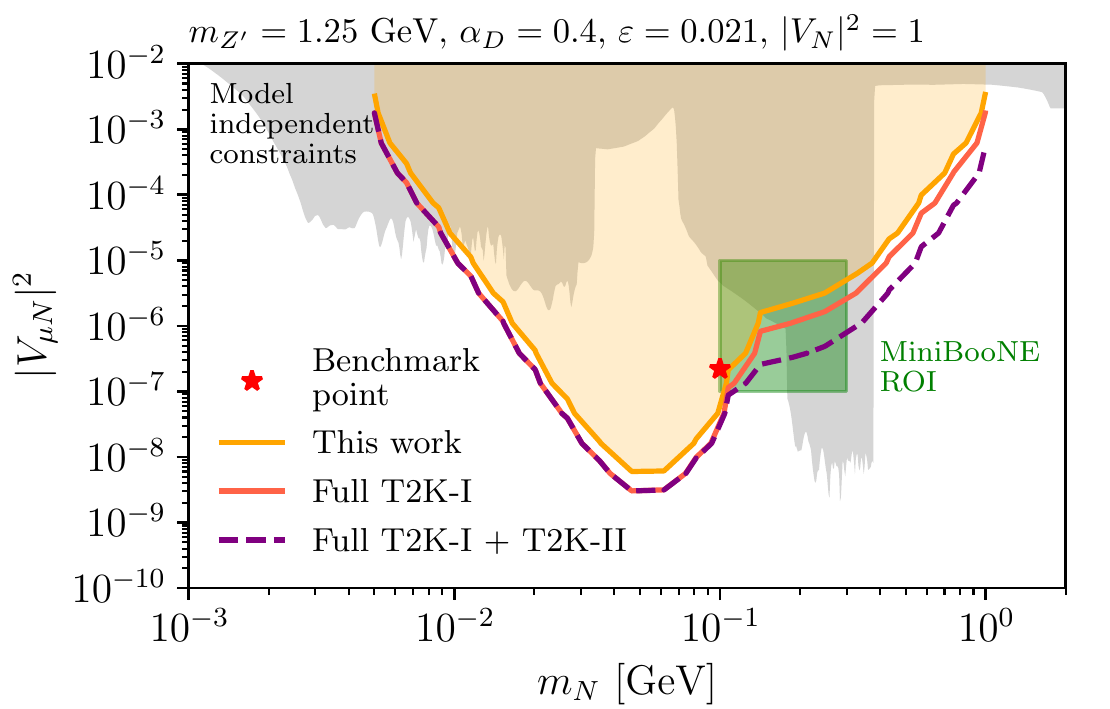}
    \includegraphics[width=0.48\textwidth]{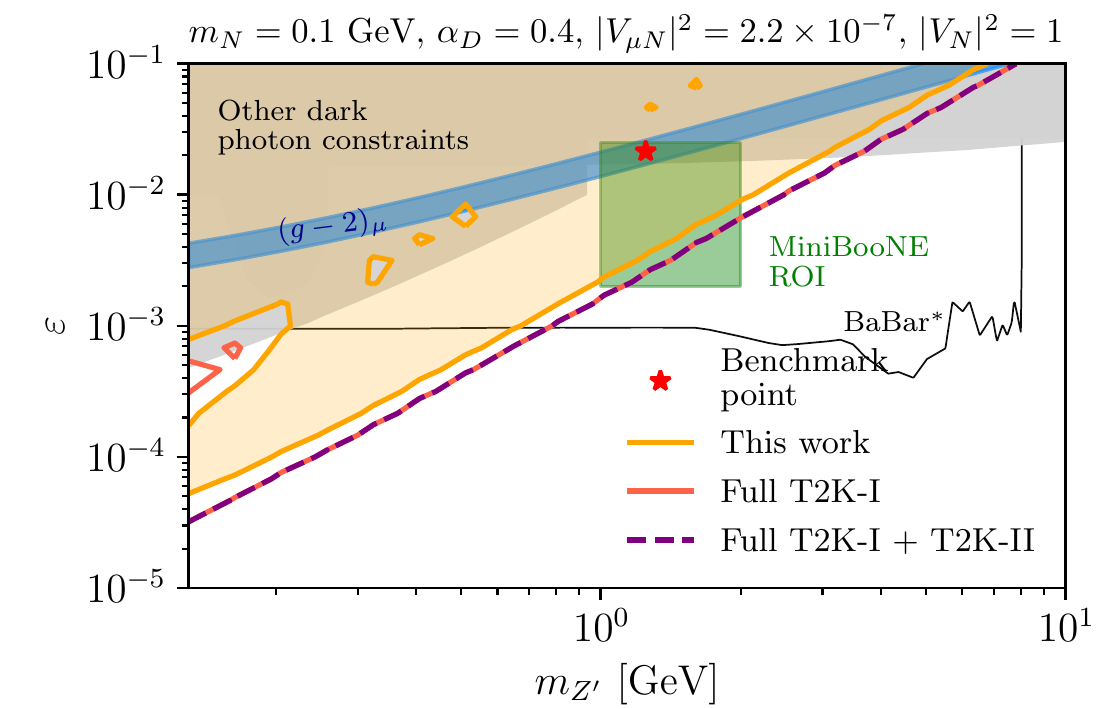}
    \caption{Limits on the dark neutrino model for a scenario with a heavy dark photon, where $c\tau_N^0$ is typically greater than centimeters. 
    On the left, we show the limits on $|V_{\mu N}|^2$ as a function of $m_N$ and on the right on $\epsilon$ as a function of $m_{Z^\prime}$, choosing the remaining parameters according to benchmark (B). The MiniBoonE region of interest (ROI) is shown as a large green area surrounding the benchmark point in~\refref{Ballett:2019pyw}. \label{fig:heavy_exclusion}}
\end{figure*}
\begin{figure*}[t]
    \centering
    \includegraphics[width=0.48\textwidth]{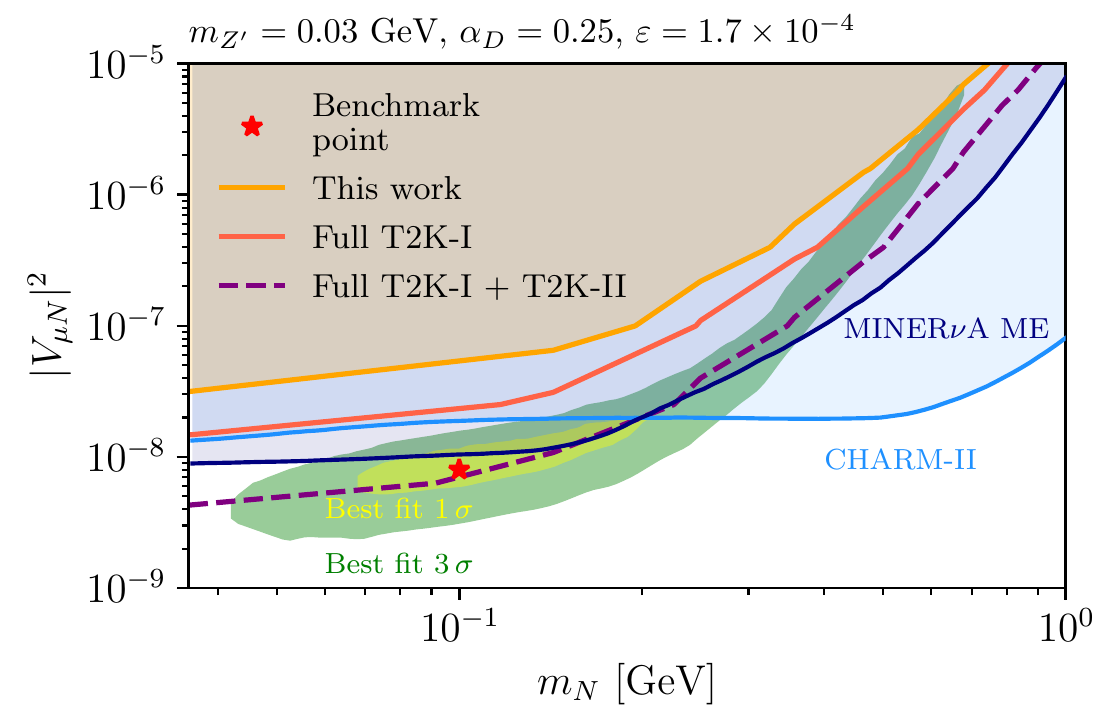}
    \includegraphics[width=0.48\textwidth]{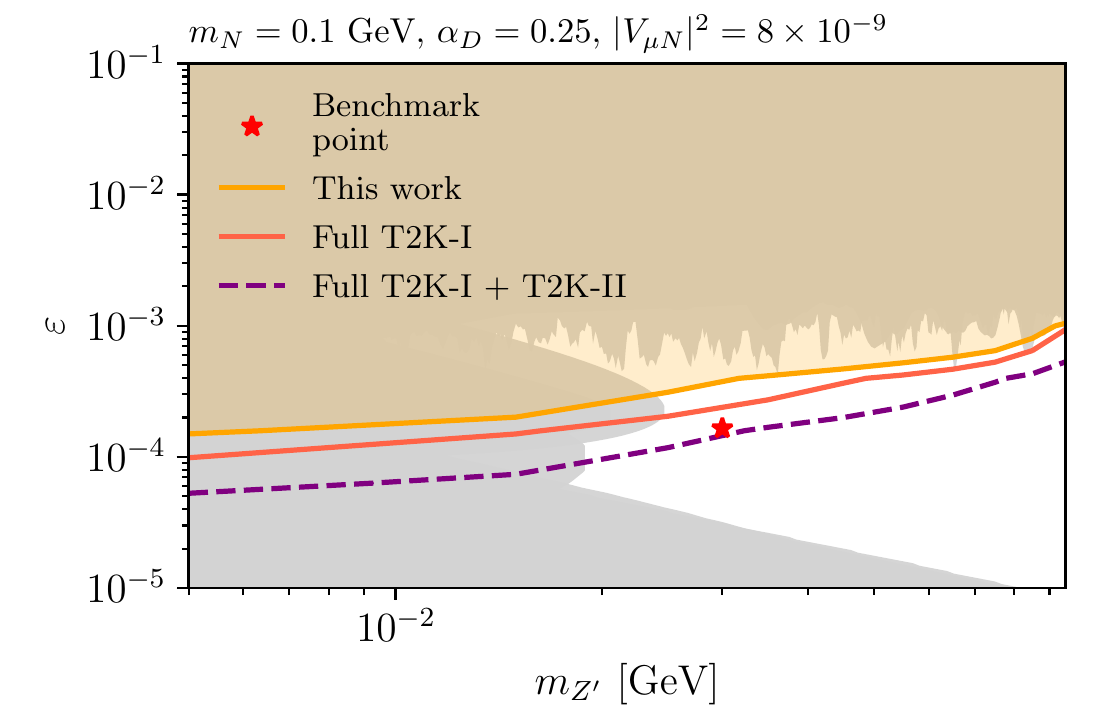}
    \caption{Limits on the dark neutrino model for a scenario with a light dark photon, where $N$ decays are prompt. On the left we show the limits on $|V_{\mu N}|^2$ as a function of $m_N$ for $m_{Z^\prime} = 30$~MeV and on the right on $\epsilon$ as a function of $m_{Z^\prime}$ for $m_N = 100$~MeV. We show the allowed region from Ref.~\cite{Bertuzzo:2018itn} on the left. Limits on the light and visible dark photon have been obtained from \refref{Ilten:2018crw}. \label{fig:light_exclusion}}
\end{figure*}

\paragraph{Cross section analytical weights.---}
The up-scattering cross section is proportional to $|V_{\mu N}|^2 \alpha_D (e \epsilon)^2$, as seen by squaring the amplitude in \Cref{eqn:upscattering_amplitude}.
We implement a trivial scaling along with the parameters $\theta^{\beta}$ allowing a quick re-weight of the events in this parameter space.
In this case, we define:
\begin{equation}
    w_i^{\rm \sigma}(\bar{\theta}^\beta, \theta_i^\beta) = \frac{|V_{\mu N}|^2 \alpha_D (e \epsilon)^2}{(|V_{\mu N}|^2 \alpha_D (e \epsilon)^2)_i},
\end{equation}
where the parameters $\bar{\theta}^\beta$ are fixed in the entire simulation and so independent of $i$, but could, in principle, be varied as well.

\paragraph{Reconstruction and selection efficiency weights.---}
We implemented $\epsilon(\Omega_{ee})$ as a function that is 0 for the events which are not selected and $\bar{\epsilon}$ for the events that are selected.
For Analysis-I, this weight is $\bar{\epsilon} = 10\%$ and the selection follows \Cref{eq:cuts}.
For Analysis-II, this weight is a flat $\bar{\epsilon} = 10\%$ for every event.

\paragraph{Lifetime re-weighting.---}
This weight applies only to the heavy case, where lifetimes span multiple orders of magnitude, while the light case always leads to a prompt decay ($c\tau^0 \le$ 0.1 cm). 
In this latter case, we simulate interactions directly in the fiducial volume.
The easiest way to compute the acceptance for different lifetimes is to sample a number from the exponential distribution with a scale parameter equal to the $N$ lab-frame lifetime, propagate $N$ to the detector, and accept or reject the event if the decay point happens within the TPC fiducial volume.
However, this method has an important drawback as it produces small effective sample sizes, especially at short lifetimes, where most interactions from the \pzerod will not make it to the detector.
To avoid this issue, we instead account for the geometrical acceptance by multiplying by a lifetime weight, which is equal to the integral of the trajectory within the TPC weighted by the exponential distribution.
The trajectory of the heavy neutrino in the lab frame in the event $i$ enters and exits each of the three different TPC at points $(a^j_i, b^j_i)$, where $j = 0, 1, 2$ is the TPC index.
If the heavy neutrino never enters a given TPC, we can take both numbers as infinity.
For each event, we can compute $(\beta \gamma)_i = p_i/m_{N}$, and given a value of the lifetime in the proper frame $c \tau_0$, the lifetime weight is computed as:
\begin{align}
    w_i^{\tau_0}(\theta, \Omega_N) &= \sum_j \int_{a^j_i}^{b^j_i} \frac{ds}{(\beta \gamma)_i c \tau_0} e^{-s/(\beta \gamma)_i c \tau_0} \nonumber \\
    &= \sum_j (e^{-a^j_i/(\beta \gamma)_i c \tau_0} - e^{-b^j_i/(\beta \gamma)_i c \tau_0}).
\end{align}

\paragraph{POT and number of targets weights.---}
Re-weighting for the POT and number of targets is trivial, as it depends only on these multiplicative factors
\begin{equation}
    w^{\rm n_t, POT}_i = \rm n_t \times \rm POT,
\end{equation}
and can be computed on the fly to change the beam exposure and target material and mass easily. 

\subsection{Likelihood evaluation}
We compute a Poisson likelihood of the observed data ($N_{obs}$) given the expectation, summing up the expected background ($b$ and the signal ($\mu(\theta)$) across the parameter space.
We account for systematic uncertainties by using the \textit{effective likelihood} framework \cite{Arguelles:2019izp}, which provides an analytic formula to marginalize over systematic uncertainties:
\begin{equation}
\label{eqn:likelihood_def}
    \mathcal{L}(\theta) = \mathcal{L}^{eff}(\theta | N_{obs}, b + \mu(\theta), \sigma^2(b, \mu(\theta))),
\end{equation}
where 
\begin{equation}
\label{eqn:sigma_def}
    \sigma^2(b, \mu(\theta)) = \sum_i w_i^2 + (b + \mu(\theta))^2 * \eta^2,
\end{equation}
accounts for systematic uncertainties. 
The first addend accounts for the finite sample size, while the second includes the analysis systematics (\eg{}, flux and cross section), using the fractional systematic uncertainties published with the analysis, which are typically close to a flat 20\%.
This formula can be easily extended to a multi-bin analysis by taking the product of the likelihood for each bin.
When combining different analyses, like the TPC search and the FGD sideband, we simply sum the likelihood together.
When computing projections, we scale the signal and the  background proportional to the number of targets and the POT, and we assume $N_{obs} = int(b + \mu(\theta))$, where $int()$ is just approximating to an integer number.

\begin{figure*}[t]
    \centering
    \includegraphics[width=0.48\textwidth]{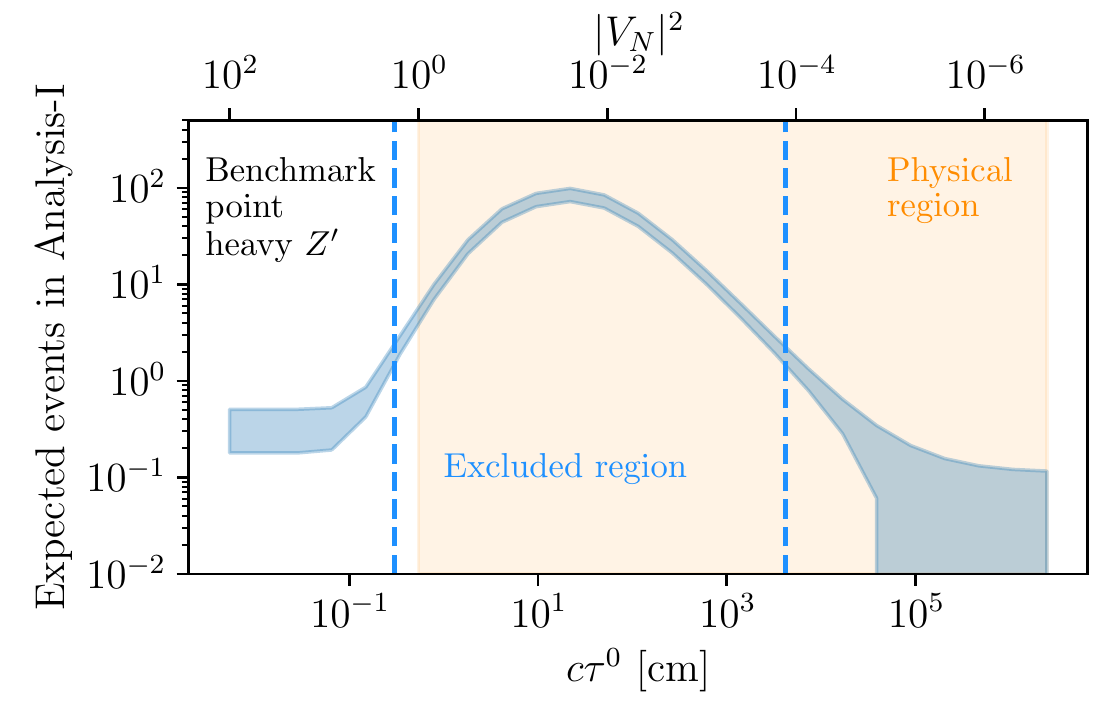}
    \includegraphics[width=0.48\textwidth]{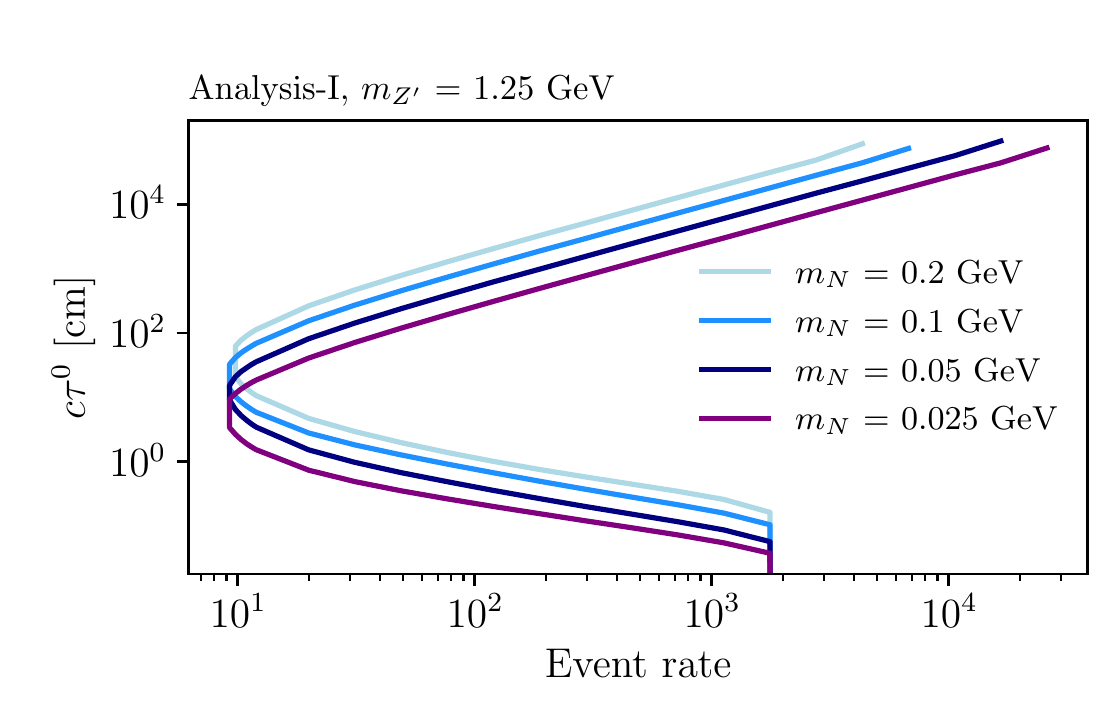}
    \includegraphics[width=0.8\textwidth]{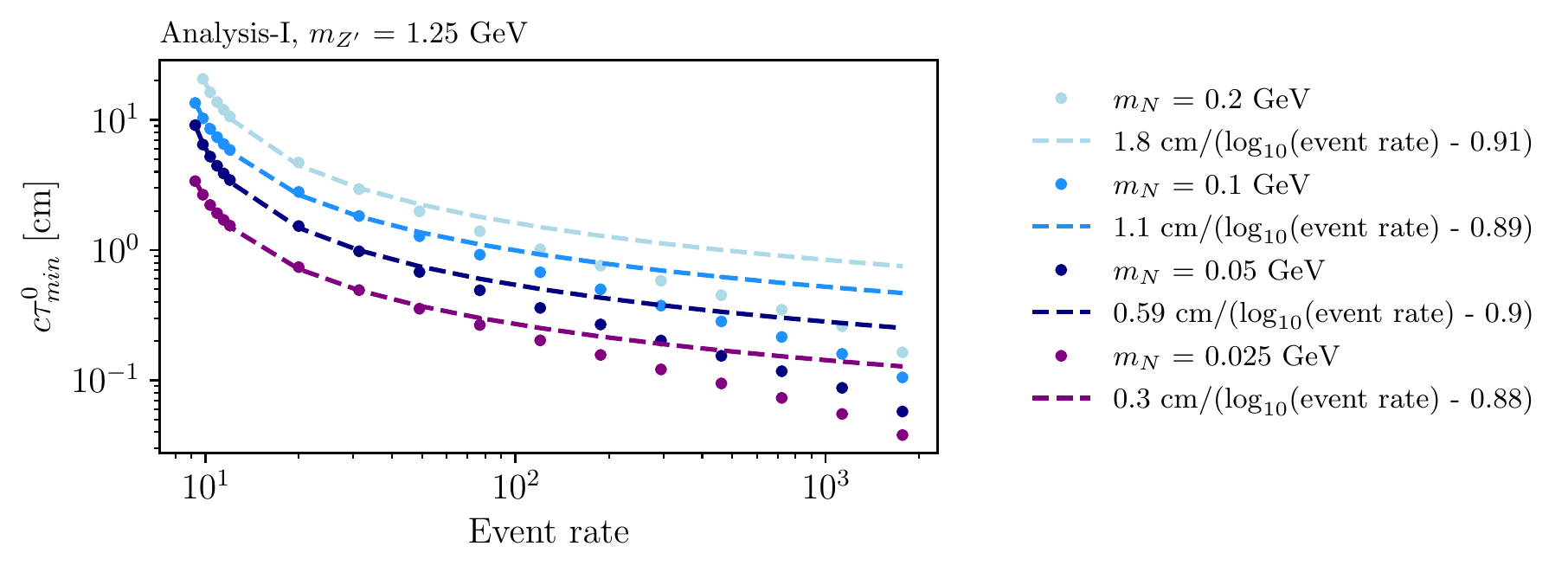}
    \caption{Top left: The expected number of events in Analysis-I after the full selection as a function of the proper decay length of $N$, $c\tau^0$, for the heavy \zprime benchmark point.
    The blue band represents the uncertainty as obtained using \Cref{eqn:sigma_def}. 
    Once we fix all the other parameters in the model, there is a bijection between $c\tau^0$ and $|V_N|^2$, as shown on the top x-axis.
    However, $|V_N|^2$ is physical only in the region allowed by \Cref{eq:decay_range}, shaded in orange.
    Besides very long-lived heavy neutrinos, we exclude most of the physical region.
    The minimum decay length we exclude is 0.3 cm, although our model cannot generate such short lifetimes for this combination of parameters.
    Top right: The limits on the $N$ decay length as a function of the total number of events after signal selection, but before requiring that $N$ decays in the TPC volume.
    The region enclosed by the solid lines is excluded at 90\% C.L.
    At first order approximation, different \mzprime values result in different event rates without affecting the kinematics of the decay, while different $m_N$ values result in different production thresholds and lab-frame lifetimes, explaining the difference between the different curves.
    Bottom: Zoom of the previous plot in the lower-left part of the graph. 
    The dots represent the curves obtained with the analysis, while the dashed lines are the best-fit values obtained with a simplified model.
    Although the functional form cannot perfectly fit all the points, it describes the curves reasonably accurately, providing model-independent constraints which can be applied to models predicting a similar phenomenology. 
    \label{fig:heavy_ctau}}
\end{figure*}

\section{Results and Discussion}\label{sec:results}

Given that neither analysis observed any excess of events with respect to the background prediction, they can be used to constrain the parameter space of the dark neutrino model.
We show our limits for two particular projections in parameter space: the $m_N - |U_{\mu N}|^2$ plane, describing the heavy neutrino properties, and the $m_{Z^{\prime}} - \epsilon$ plane, describing the dark photon properties.
We compute the likelihood on the plane by summing up the negative log-likelihoods of the relevant analyses, subtracting the minimum, and tracing the contour at constant 2.3, which produces regions of exclusion at 90\% C.L.

Given that no fit of this model to the MiniBooNE data has been performed for the heavy case, we consider a region of interest around the benchmark point, while for the light case, we consider the best-fit region from~\cite{Bertuzzo:2018itn}.
For the heavy case, we also determine model-independent bounds on the proper decay length $c\tau^0$ as a function of the upscattering rate after selection but before requiring that the decay happened in the TPC, as shown in \Cref{fig:heavy_ctau}.
We derive approximated functional forms for the bounds, which could easily be applied to other models predicting similar signatures.

As previously discussed in \refref{Asaka:2012bb, Abe:2019kgx, Arguelles:2021dqn}, the gaseous Argon (GAr) Time-Projection-Chambers (TPCs) of the T2K near detector, ND280, provide a powerful probe of long-lived particles.
Visible decays inside the low-density volume of the TPCs, where neutrino-induced backgrounds are negligible, are clearly identified. 
In this work, we showed that the combination of the high-density material in the \pzerod detector with the low-density TPCs downstream is even more powerful.
The former enhances the production of new particles in neutrino-nucleus scattering due to the large mass of lead. At the same time, the latter provides a desirable volume to search for charged final states.
In addition, the magnetic field allows for improved identification of $\ell^+\ell^-$ pairs even at the smallest opening angles and energies. 

In Analysis-I, we use an existing, background-free search for $e^+e^-$ pairs inside the volume of the TPCs~\cite{Abe:2019kgx} to set limits on a short-lived heavy neutrino $N$, produced in neutrino-nucleus upscattering upstream in the detector.
Having observed zero events, T2K data strongly constrains these particles as explanations of the MiniBooNE excess but does not entirely rule them out for sufficiently short lifetimes, $c\tau_N^0/m_N \lesssim 2$~cm/GeV.
In Analysis-II, we explored the measurement of single-photon events in the Fine-Grained Detector~\cite{T2K:2020lrr} to set limits on the parameter space with the shortest lifetimes of $N$. 
Due to the larger backgrounds and lower event rate, this analysis is not as sensitive as Analysis-I and therefore does not exclude allowed regions of parameter space.
These limits can be significantly improved using all T2K-I run data taken up to 2021, as well as with future data that will be taken with the upgraded ND280 detector and the more intense neutrino beam at J-PARC.

We note that we have not exhausted the list of models, having not covered cases with small mass splittings between $N$ and the daughter neutrinos, scalar mediator models, and other $2\to3$ scattering signatures involving emission of on-shell dark bosons~\cite{deGouvea:2018cfv,Datta:2020auq,Dutta:2020scq,Abdallah:2020biq,Abdallah:2020vgg,Hammad:2021mpl,Dutta:2021cip}. 
We expect different signal selection efficiencies for these models, especially those that better fit the MiniBooNE angular spectrum.
We encourage the T2K collaboration to pursue a dedicated search for all such upscattering signatures, including the one discussed in this paper, leveraging the full power of their detector simulation.
In particular, we expect that a complete reconstruction simulation by the collaboration overcomes the simplifying assumption in this work of energy-independent signal efficiencies. 
In addition, further public data on the reconstruction efficiencies as a function of physical observables, like energies and angles, rather than model parameters, would be incredibly beneficial to the phenomenology community.

\section{Conclusions}\label{sec:conclusions}

We have proposed a method to cover multi-dimensional model parameter spaces efficiently. 
We used importance sampling to sample model parameters and physical observables to find model predictions for the
number of events in a particular experimental setup.
Because both types of parameters are sampled intelligently, the construction of the likelihood is more efficient.
We also built a Kernel Density Estimator (KDE) to interpolate the likelihood inside the sampled region of parameter space.
The KDE gives us greater flexibility in hypothesis testing and offers a valuable product that can be distributed to other users interested in the model likelihood.

The interpolation is achieved with a single batch of simulated events and can provide limits in arbitrary slices of the parameter space.
This tool enabled us to set the limits shown in the four planes of \Cref{fig:heavy_exclusion} and \Cref{fig:heavy_exclusion}.
In the context of particle physics experiments, this method will constitute a useful tool for experimental collaborations and phenomenologists to explore more complex physical models.
In neutrino physics, with the latest progress in building new phenomenology-friendly Monte Carlo generators for new physics processes~\cite{Isaacson:2020wlx, Isaacson:2021xty, Campbell:2022qmc}, 
improving our ability to cover the large-dimensional parameter spaces of dark sectors will be all the more relevant.

Future directions include applying our methodology to searches for new physics outside the context of short-baseline anomalies.
Among models of interest are higgsed low-scale $U(1)$ symmetries, co-annihilating dark matter models, light axion-like particles with multiple interactions, as well as flavor-rich models of heavy neutrinos.
In these theories, decay-in-flight signatures of multiple light dark particles are ubiquitous and require more efficient exploration methods to be thoroughly tested.
We believe our method can increase the theory reach of experiments like the Large Hadron Collider, the Short-Baseline Neutrino program at Fermilab~\cite{MicroBooNE:2015bmn, Machado:2019oxb}, atmospheric neutrino experiments like IceCube and KM3NET~\cite{KM3Net:2016zxf}, as well as future high-intensity long-baseline experiments like DUNE~\cite{DUNE:2020ypp} and Hyper-Kamiokande~\cite{Hyper-Kamiokande:2018ofw}.

\acknowledgments
We would like to thank
Mathieu Lamoureux, Teppei Katori, Sophie King, and Tianlu Yuan for useful discussion on ND280.
C.A.A. is supported by the Faculty of Arts and Sciences of Harvard University and the Alfred P. Sloan Foundation. 
N.F.'s work is supported by the Department of Energy grant award number DE-SC0007881.
The research of M.H. was supported in part by Perimeter Institute for Theoretical Physics. 
Research at Perimeter Institute is supported by the Government of Canada through the Department of Innovation, Science and Economic Development and by the Province of Ontario through the Ministry of Research, Innovation, and Science. 
Part of M.H.'s work was performed at the Aspen Center for Physics, which is supported by the National Science Foundation grant PHY-1607611.

\appendix

\section{Analytical approximation for upscattering cross section}\label{app:cross_sec_app}
For convenience, a crude approximation for the upscattering cross sections above $E_\nu>1$ GeV is given below. These have been obtained assuming a box function for the coherent and dipole form factors with cut-offs around the QCD scale and vector mass, respectively. For upscattering on nuclei, 
\begin{align}
    \sigma^{ \nu_\alpha \to N}_{\rm coh} &\simeq \frac{|V_{\alpha h}|^2(Z e \,\epsilon)^2 \alpha_D}{4 E_\nu^2m_{Z^\prime}^4} \left[ 2(M^4+s^2)-sM^2(x_A^2+4)\right],
\end{align}
where $Z$ is the atomic number of the nucleus with mass $M$, $x_A=2\Lambda_{\rm QCD}/A^{1/3}$ with $\Lambda_{\rm QCD}$ from $100$ to $200$~MeV, and $A$ the atomic mass number. The dependence of the total cross section on $\Lambda_{\rm QCD}$ is stronger for lower energies. 

\section{Detector description}
We simulate the three subdetectors of ND280: the \pzerod, the two FGDs, and the three GAr TPCs.
In \Cref{tab:detector_info}, we report the sizes of the active volume, where upscaterring occurs, and the fiducial volume, where \epluseminus pairs are detected.
We also report total active and fiducial mass, as well as the material composition in mass.
We account for gaps between the detector volumes and report the $z$ coordinate along the beam axis, where the active or fiducial volume begins.

\begin{table*}[!ht]
\begin{tabular}{c|c|c|c|c|c|c|c}
ND280 module & \begin{tabular}[c]{@{}c@{}}Active volume\\ $X \times Y \times Z$ {[}cm{]}\end{tabular} & $z_{\rm begin}^{\rm active}$ & $M_{\rm tot}^{\rm active}$ & \begin{tabular}[c]{@{}c@{}}Fiducial volume\\ $X \times Y \times Z$ {[}cm{]}\end{tabular} & $z_{\rm begin}^{\rm fiducial}$ & $M_{\rm tot}^{\rm fiducial}$ & composition (in mass) \\
\hline\hline
P0D-ECAL1    & 210$\times$224$\times$30.5                                                             & 0                                                                               & 2.9                                                                            &                                                                                          &                                                                                   &                                                                                  & 6.5\% H, 40\% C, 53.5\% Pb              \\
P0D-water    & 210$\times$224$\times$179                                                              & 30.5                                                                            & 10                                                                             &                                                                                          &                                                                                   &                                                                                  & 10\% H, 43\% C, 22\% O, 16\% Cu, 9\% Zn \\
P0D-ECAL2    & 210$\times$224$\times$30.4                                                             & 209.6                                                                           & 2.9                                                                            &                                                                                          &                                                                                   &                                                                                  & 6.5\% H, 40\% C, 53.5\% Pb              \\
GArTPC1      & 186$\times$206$\times$78                                                               & 251                          & 0.016                      & 170$\times$196$\times$56                                                                & 256                            & 0.010                        & 100\% Ar              \\
FGD1         & 186$\times$186$\times$30                                                               & 343                          & 1.1                        & 175$\times$175$\times$29                                                                 & 344                            & 0.92                         & 8\% H, 88\% C, 4\% O  \\
GArTPC2      & 186$\times$206$\times$78                                                               & 387                          & 0.016                      & 170$\times$196$\times$56                                                                 & 256                            & 0.010                        & 100\% Ar              \\
FGD2         & 186$\times$186$\times$30                                                               & 480                          & 1.1                        & 175$\times$175$\times$29                                                                 & 481                            & 0.92                         & 9\% H, 50\% C, 41\% O \\
GArTPC3      & 186$\times$206$\times$78                                                               & 524                          & 0.016                      & 170$\times$196$\times$56                                                                 & 256                            & 0.010                        & 100\% Ar\\
\hline\hline
\end{tabular}
    \caption{The active and fiducial volume dimensions, mass, and composition of each ND280 module simulated in our analysis. 
    No fiducial volume is shown for the \pzerod because we do not employ this detector for measuring heavy neutrino decays but only for its production.
    For the GAr modules, we show the fiducial volume and mass of each TPC, taking $\rho_{\rm GAr} = 1.78$~g/cm$^3$.}
    \label{tab:detector_info}
\end{table*}

\begin{figure*}[!ht]
    \centering
    \includegraphics[width=0.8\textwidth]{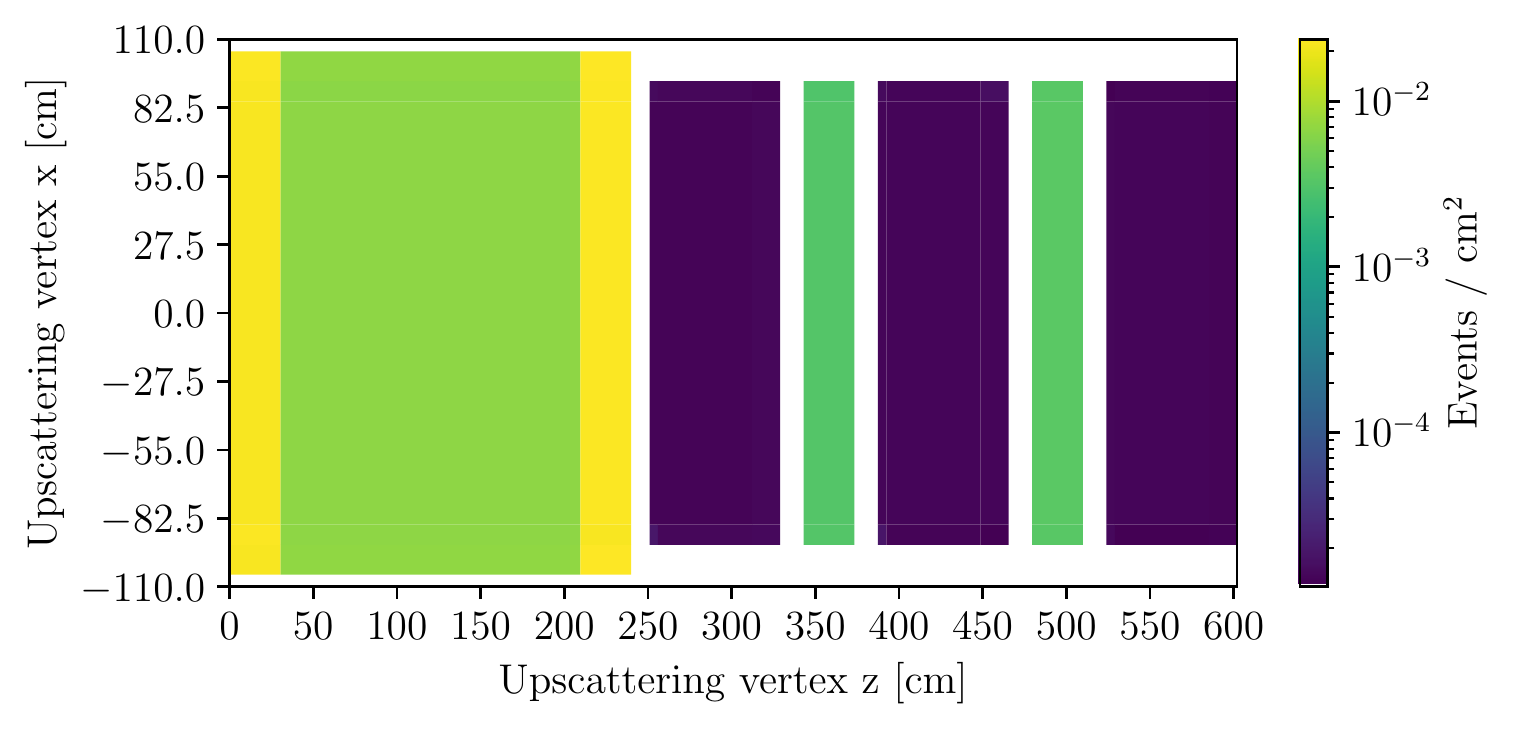}
    \caption{Distribution of the upscattering vertex across the z-x plane, for the heavy \zprime case, using our benchmark point.
    The color bar shows the number of upscattering events per bin
    \label{fig:detector_2d}
    }
\end{figure*}

\Cref{fig:detector_2d} shows the 2d distribution density of upscattering vertices along the z and x axes for the heavy \zprime case, using our benchmark point in parameter space.
The three different sections of the \pzerod, the three different TPCs, and the two FGDs are clearly distinguishable, together with the gaps between volumes.

\section{Kinematical selection efficiency in Analysis-I}

\begin{figure*}[!ht]
    \centering
    \includegraphics[width=0.58\textwidth]{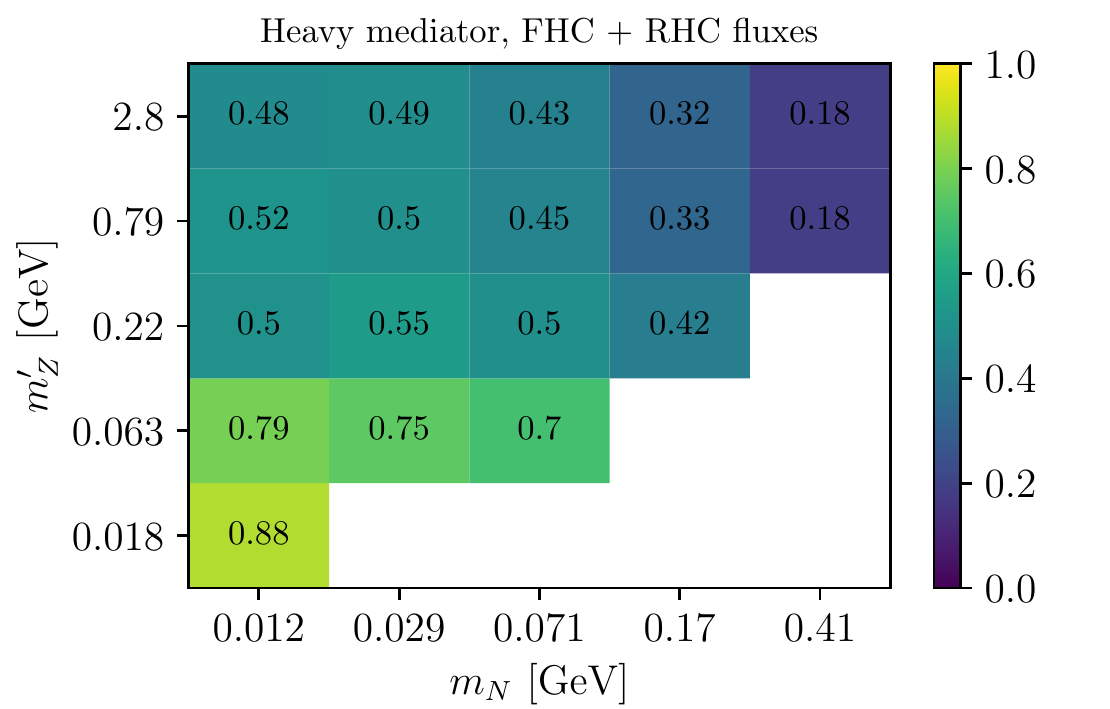}
    \caption{Selection efficiency resulting from the kinematical cuts in \cref{eq:cuts} in Analysis-I for the heavy mediator case, as a function of the parameter space \mzprime - $m_N$.
    \label{fig:selection_efficiency_vs_mz_mn}}
\end{figure*}

\Cref{fig:selection_efficiency_vs_mz_mn} shows a map of the efficiency for the kinematical cuts in \cref{eq:cuts} as a function of $m_N$ and \mzprime, for the heavy \zprime case in analysis-I.
The distributions of the kinematical variables and, therefore, the selection efficiency depends only on these two parameters because all the others affect only the total upscattering rate or the lifetime.
The efficiency is about 50\% for our benchmark point, and it is independent of \mzprime for large values of \mzprime when \mzprime only determines the total rate, similar to the $W$ mass in the muon decay.
However, it grows to almost 90\% at lower values of \mzprime, while it decreases to $\sim 20\%$ at larger $m_N$.
While automatically considered when scanning the parameter space through our procedure, this variation does not significantly impact the final result.

\section{Complementarity of the ND280 sub-detectors}

\subsection{Upscattering in the \texorpdfstring{\pzerod}{pzerod} and in the GArTPCs}
\begin{figure*}[!ht]
    \centering
    \includegraphics[width=0.48\textwidth]{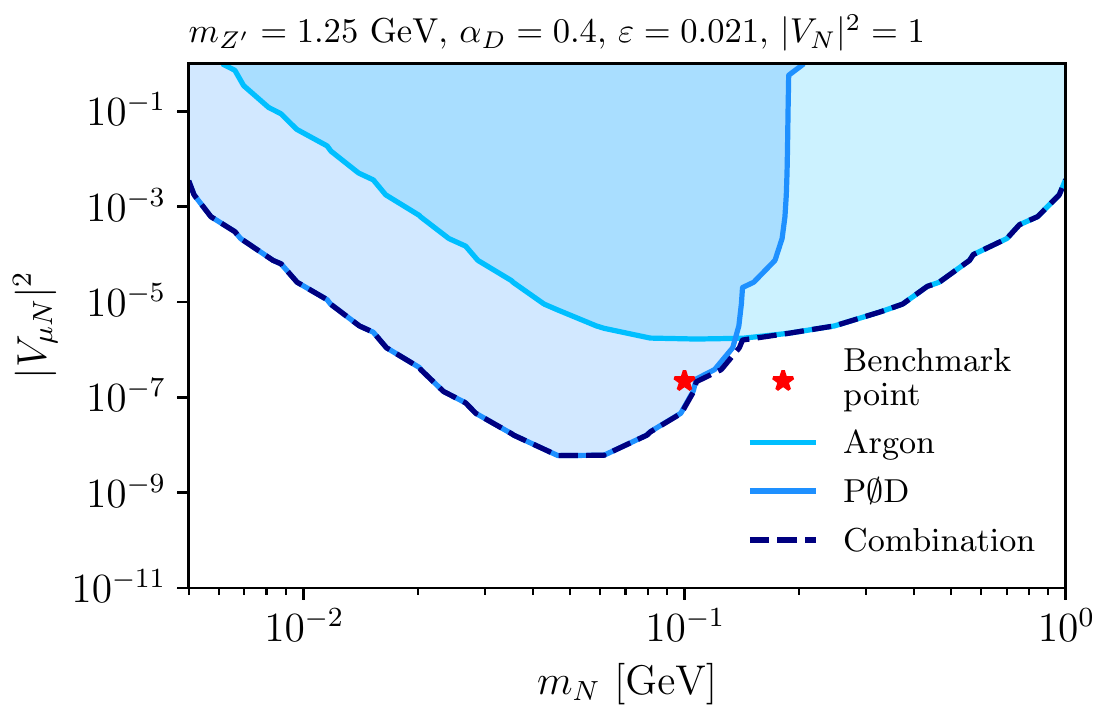}
    \includegraphics[width=0.48\textwidth]{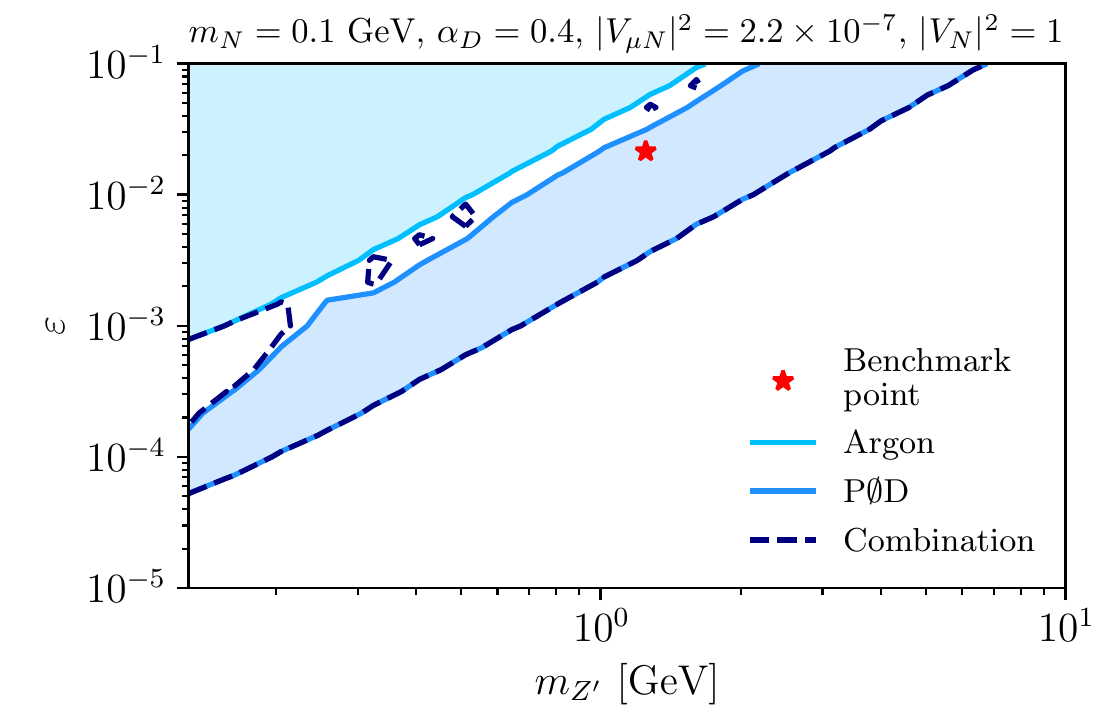}
    \caption{Limits for the heavy \zprime case, as shown in \Cref{fig:heavy_exclusion}, splitting into the two regions where upscattering happens in the GArTPCs or the \pzerod.
    \label{fig:pod_vs_argon}}
\end{figure*}

In the heavy mediator case, both the scattering in the \pzerod and the argon contribute significantly to the constraints, however, in different regions of the parameter space.
Scattering in the gaseous argon is rare because of the low density, but it is the most powerful component in constraining the shortest lifetimes since it is where the fiducial volume of the analysis is contained.
\Cref{fig:pod_vs_argon} shows our constraints, as in \Cref{fig:heavy_exclusion}, splitting the limits into the contribution from the GArTPCs and from the \pzerod.
Between $m_N = 0.1$ GeV and $m_N = 0.2$ GeV, the model is very short lived, and all heavy neutrinos produced in the \pzerod decay before reaching the TPCs.
This region is constrained only by prompt decays of heavy neutrinos produced inside the argon and is, therefore, less constrained.
In the right plot, we show the dark photon parameter space, where, despite the larger upscattering rate at smaller \zprime masses, the model cannot be constrained by the \pzerod events due to the short lifetimes.
However, in several regions of parameter space, the model predicts a significant number of events in the argon, which allows for a robust exclusion of the largest values of $\varepsilon$.

\subsection{Upscattering in the FGD and the GArTPCs}

\begin{figure*}[!ht]
    \centering
    \includegraphics[width=0.48\textwidth]{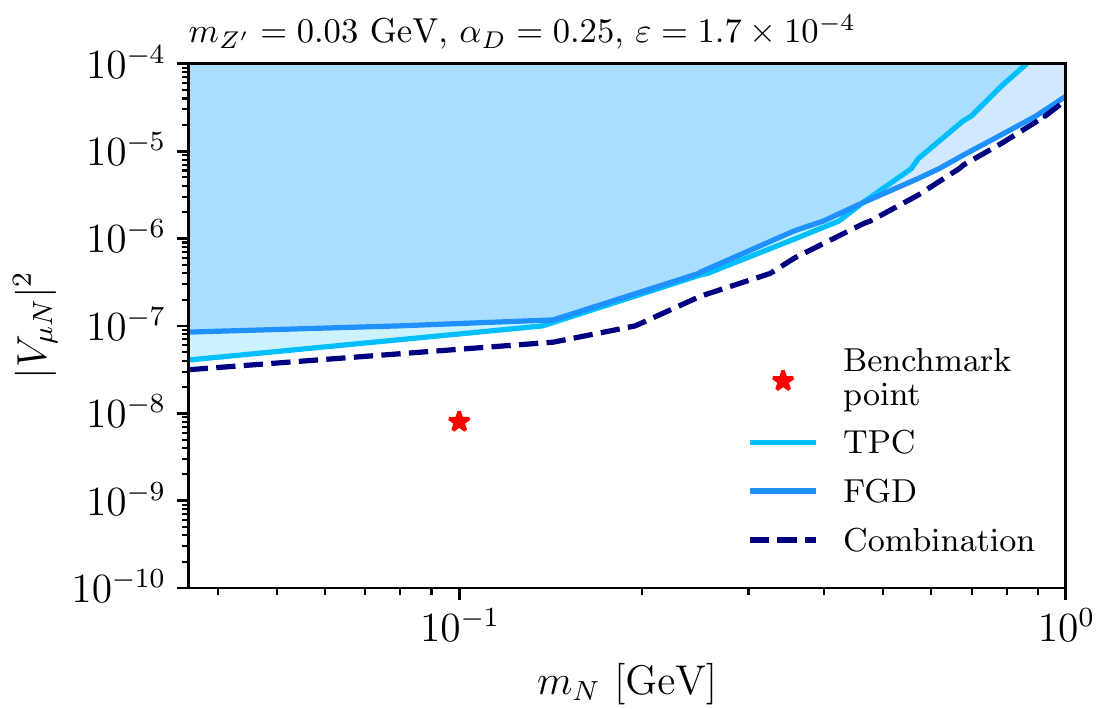}
    \includegraphics[width=0.48\textwidth]{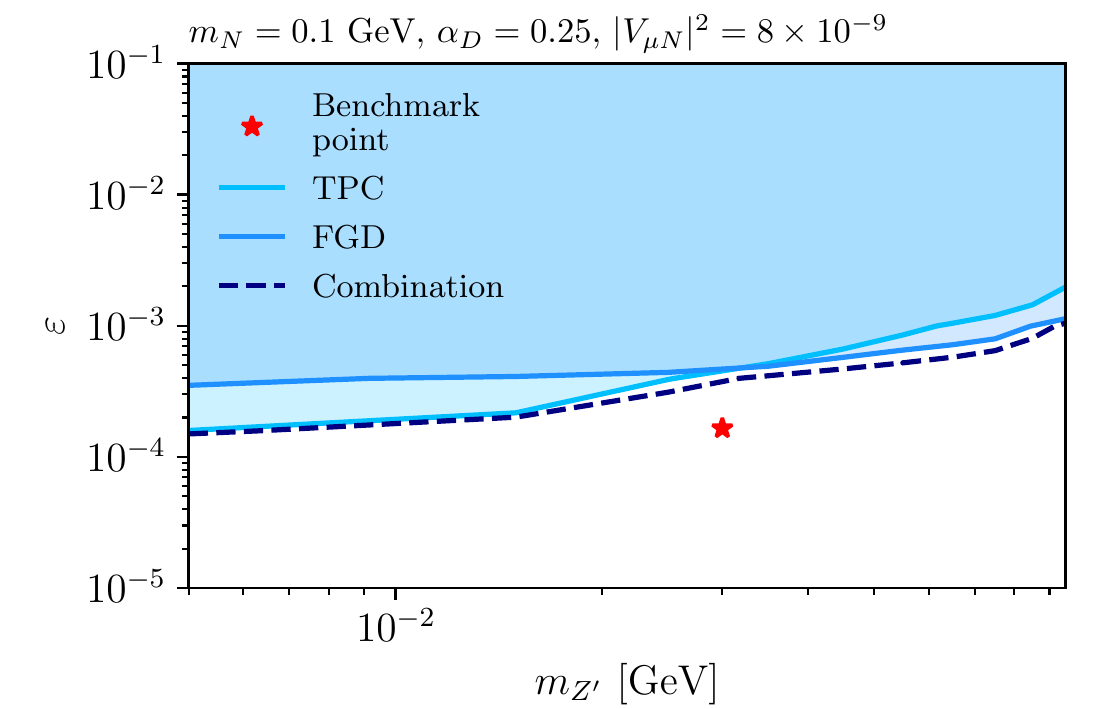}
    \caption{Limits for the heavy \zprime case, as shown in \Cref{fig:light_exclusion}, splitting into the two regions where upscattering happens in the FGD or the GArTPCs.
    \label{fig:fgd_vs_argon}}
\end{figure*}

In the light mediator case, we combine Analysis-I and Analysis-II, considering upscattering happening in the GArTPCs, for the first case, and in the FGDs, in the second case.
The two analyses contribute similarly to the limit, as shown in \Cref{fig:fgd_vs_argon}.

\section{Functional forms of the model-independent constraints}

The bottom plot of \cref{fig:heavy_ctau} shows two limits: the dots represent the limits derived through the full analysis, corresponding to lines of the top right figure, while the dashed lines show best-fit curves using the well-motivated functional form:
\begin{equation}
\label{eqn:fit_ctau_dark_nus}
    c \tau^0_{min} = \frac{\alpha [\text{cm}]}{\log_{10}({\text{event rate}}) - \beta},
\end{equation}
where $\alpha$ and $\beta$ are free parameters.
It can be derived by assuming a point-like source of heavy neutrinos decaying in a detector of size $d$ at a distance $z$ and assuming that the process happens in one dimension only.
The number of decays in the detector is given by $\text{event rate} \times (e^{-z/\beta\gamma c \tau^0} - e^{-(z+d)/\beta\gamma c \tau^0})$.
We are considering the region of the smallest lifetimes we exclude, so $c \tau^0 \ll z$, and the second term of the exponential is negligible.
Finally, the limit is set when the event rate is equal to a constant value $C$, so the equation defining the limit can be re-written as $C = e^{-z/\beta\gamma c \tau^0}$, with $\alpha = z/\beta\gamma$ and $\beta = f(C)$ as free parameters, with $f$ a certain functional form.
It can now be inverted to obtain \cref{eqn:fit_ctau_dark_nus}.

The best fit to the different curves accurately describes the points at small event rates, underestimating the limit at smaller lifetimes.
As expected, the values of $\beta$ are roughly constant, while $\alpha$ varies with \mn, roughly in a proportional way.
The reason is that the momentum available in the heavy neutrino production is constant, so $\mn \beta \gamma$ is roughly constant.
Combining these two observations, we deduce that we exclude 
\begin{equation}
\frac{c\tau^0}{m_N} \gtrsim \frac{1.1~\text{cm}}{0.1~\text{GeV}} \frac{1}{\log_{10}(\text{event rate}) - 0.9}.
\end{equation}
While this functional form is limited and cannot replace the complete study, it provides a model-independent benchmark, which can easily show the size of the constraints resulting from this analysis for an arbitrary model without needing a complete analysis.

\newpage
\bibliographystyle{apsrev4-1}
\bibliography{lib}{}

\begin{thebibliography}{106}%
\makeatletter
\providecommand \@ifxundefined [1]{%
 \@ifx{#1\undefined}
}%
\providecommand \@ifnum [1]{%
 \ifnum #1\expandafter \@firstoftwo
 \else \expandafter \@secondoftwo
 \fi
}%
\providecommand \@ifx [1]{%
 \ifx #1\expandafter \@firstoftwo
 \else \expandafter \@secondoftwo
 \fi
}%
\providecommand \natexlab [1]{#1}%
\providecommand \enquote  [1]{``#1''}%
\providecommand \bibnamefont  [1]{#1}%
\providecommand \bibfnamefont [1]{#1}%
\providecommand \citenamefont [1]{#1}%
\providecommand \href@noop [0]{\@secondoftwo}%
\providecommand \href [0]{\begingroup \@sanitize@url \@href}%
\providecommand \@href[1]{\@@startlink{#1}\@@href}%
\providecommand \@@href[1]{\endgroup#1\@@endlink}%
\providecommand \@sanitize@url [0]{\catcode `\\12\catcode `\$12\catcode
  `\&12\catcode `\#12\catcode `\^12\catcode `\_12\catcode `\%12\relax}%
\providecommand \@@startlink[1]{}%
\providecommand \@@endlink[0]{}%
\providecommand \url  [0]{\begingroup\@sanitize@url \@url }%
\providecommand \@url [1]{\endgroup\@href {#1}{\urlprefix }}%
\providecommand \urlprefix  [0]{URL }%
\providecommand \Eprint [0]{\href }%
\providecommand \doibase [0]{http://dx.doi.org/}%
\providecommand \selectlanguage [0]{\@gobble}%
\providecommand \bibinfo  [0]{\@secondoftwo}%
\providecommand \bibfield  [0]{\@secondoftwo}%
\providecommand \translation [1]{[#1]}%
\providecommand \BibitemOpen [0]{}%
\providecommand \bibitemStop [0]{}%
\providecommand \bibitemNoStop [0]{.\EOS\space}%
\providecommand \EOS [0]{\spacefactor3000\relax}%
\providecommand \BibitemShut  [1]{\csname bibitem#1\endcsname}%
\let\auto@bib@innerbib\@empty
\bibitem [{\citenamefont {Minkowski}(1977)}]{Minkowski:1977sc}%
  \BibitemOpen
  \bibfield  {author} {\bibinfo {author} {\bibfnamefont {P.}~\bibnamefont
  {Minkowski}},\ }\href {\doibase 10.1016/0370-2693(77)90435-X} {\bibfield
  {journal} {\bibinfo  {journal} {Phys. Lett.}\ }\textbf {\bibinfo {volume}
  {67B}},\ \bibinfo {pages} {421} (\bibinfo {year} {1977})}\BibitemShut
  {NoStop}%
\bibitem [{\citenamefont {Mohapatra}\ and\ \citenamefont
  {Senjanovic}(1980)}]{Mohapatra:1979ia}%
  \BibitemOpen
  \bibfield  {author} {\bibinfo {author} {\bibfnamefont {R.~N.}\ \bibnamefont
  {Mohapatra}}\ and\ \bibinfo {author} {\bibfnamefont {G.}~\bibnamefont
  {Senjanovic}},\ }\href {\doibase 10.1103/PhysRevLett.44.912} {\bibfield
  {journal} {\bibinfo  {journal} {Phys. Rev. Lett.}\ }\textbf {\bibinfo
  {volume} {44}},\ \bibinfo {pages} {912} (\bibinfo {year} {1980})},\ \bibinfo
  {note} {[,231(1979)]}\BibitemShut {NoStop}%
\bibitem [{\citenamefont {Gell-Mann}\ \emph {et~al.}(1979)\citenamefont
  {Gell-Mann}, \citenamefont {Ramond},\ and\ \citenamefont
  {Slansky}}]{GellMann:1980vs}%
  \BibitemOpen
  \bibfield  {author} {\bibinfo {author} {\bibfnamefont {M.}~\bibnamefont
  {Gell-Mann}}, \bibinfo {author} {\bibfnamefont {P.}~\bibnamefont {Ramond}}, \
  and\ \bibinfo {author} {\bibfnamefont {R.}~\bibnamefont {Slansky}},\
  }\bibfield  {booktitle} {\emph {\bibinfo {booktitle} {{Supergravity Workshop
  Stony Brook, New York, September 27-28, 1979}}},\ }\href@noop {} {\bibfield
  {journal} {\bibinfo  {journal} {Conf. Proc.}\ }\textbf {\bibinfo {volume}
  {C790927}},\ \bibinfo {pages} {315} (\bibinfo {year} {1979})},\ \Eprint
  {http://arxiv.org/abs/1306.4669} {arXiv:1306.4669 [hep-th]} \BibitemShut
  {NoStop}%
\bibitem [{\citenamefont {Yanagida}(1979)}]{Yanagida:1979as}%
  \BibitemOpen
  \bibfield  {author} {\bibinfo {author} {\bibfnamefont {T.}~\bibnamefont
  {Yanagida}},\ }\bibfield  {booktitle} {\emph {\bibinfo {booktitle}
  {{Proceedings: Workshop on the Unified Theories and the Baryon Number in the
  Universe: Tsukuba, Japan, February 13-14, 1979}}},\ }\href@noop {} {\bibfield
   {journal} {\bibinfo  {journal} {Conf. Proc.}\ }\textbf {\bibinfo {volume}
  {C7902131}},\ \bibinfo {pages} {95} (\bibinfo {year} {1979})}\BibitemShut
  {NoStop}%
\bibitem [{\citenamefont {Lazarides}\ \emph {et~al.}(1981)\citenamefont
  {Lazarides}, \citenamefont {Shafi},\ and\ \citenamefont
  {Wetterich}}]{Lazarides:1980nt}%
  \BibitemOpen
  \bibfield  {author} {\bibinfo {author} {\bibfnamefont {G.}~\bibnamefont
  {Lazarides}}, \bibinfo {author} {\bibfnamefont {Q.}~\bibnamefont {Shafi}}, \
  and\ \bibinfo {author} {\bibfnamefont {C.}~\bibnamefont {Wetterich}},\ }\href
  {\doibase 10.1016/0550-3213(81)90354-0} {\bibfield  {journal} {\bibinfo
  {journal} {Nucl. Phys.}\ }\textbf {\bibinfo {volume} {B181}},\ \bibinfo
  {pages} {287} (\bibinfo {year} {1981})}\BibitemShut {NoStop}%
\bibitem [{\citenamefont {Mohapatra}\ and\ \citenamefont
  {Senjanovic}(1981)}]{Mohapatra:1980yp}%
  \BibitemOpen
  \bibfield  {author} {\bibinfo {author} {\bibfnamefont {R.~N.}\ \bibnamefont
  {Mohapatra}}\ and\ \bibinfo {author} {\bibfnamefont {G.}~\bibnamefont
  {Senjanovic}},\ }\href {\doibase 10.1103/PhysRevD.23.165} {\bibfield
  {journal} {\bibinfo  {journal} {Phys. Rev.}\ }\textbf {\bibinfo {volume}
  {D23}},\ \bibinfo {pages} {165} (\bibinfo {year} {1981})}\BibitemShut
  {NoStop}%
\bibitem [{\citenamefont {Schechter}\ and\ \citenamefont
  {Valle}(1980)}]{Schechter:1980gr}%
  \BibitemOpen
  \bibfield  {author} {\bibinfo {author} {\bibfnamefont {J.}~\bibnamefont
  {Schechter}}\ and\ \bibinfo {author} {\bibfnamefont {J.~W.~F.}\ \bibnamefont
  {Valle}},\ }\href {\doibase 10.1103/PhysRevD.22.2227} {\bibfield  {journal}
  {\bibinfo  {journal} {Phys. Rev.}\ }\textbf {\bibinfo {volume} {D22}},\
  \bibinfo {pages} {2227} (\bibinfo {year} {1980})}\BibitemShut {NoStop}%
\bibitem [{\citenamefont {Cheng}\ and\ \citenamefont
  {Li}(1980)}]{Cheng:1980qt}%
  \BibitemOpen
  \bibfield  {author} {\bibinfo {author} {\bibfnamefont {T.~P.}\ \bibnamefont
  {Cheng}}\ and\ \bibinfo {author} {\bibfnamefont {L.-F.}\ \bibnamefont {Li}},\
  }\href {\doibase 10.1103/PhysRevD.22.2860} {\bibfield  {journal} {\bibinfo
  {journal} {Phys. Rev.}\ }\textbf {\bibinfo {volume} {D22}},\ \bibinfo {pages}
  {2860} (\bibinfo {year} {1980})}\BibitemShut {NoStop}%
\bibitem [{\citenamefont {Foot}\ \emph {et~al.}(1989)\citenamefont {Foot},
  \citenamefont {Lew}, \citenamefont {He},\ and\ \citenamefont
  {Joshi}}]{Foot:1988aq}%
  \BibitemOpen
  \bibfield  {author} {\bibinfo {author} {\bibfnamefont {R.}~\bibnamefont
  {Foot}}, \bibinfo {author} {\bibfnamefont {H.}~\bibnamefont {Lew}}, \bibinfo
  {author} {\bibfnamefont {X.~G.}\ \bibnamefont {He}}, \ and\ \bibinfo {author}
  {\bibfnamefont {G.~C.}\ \bibnamefont {Joshi}},\ }\href {\doibase
  10.1007/BF01415558} {\bibfield  {journal} {\bibinfo  {journal} {Z. Phys.}\
  }\textbf {\bibinfo {volume} {C44}},\ \bibinfo {pages} {441} (\bibinfo {year}
  {1989})}\BibitemShut {NoStop}%
\bibitem [{\citenamefont {Abdullahi}\ \emph
  {et~al.}(2022{\natexlab{a}})\citenamefont {Abdullahi} \emph
  {et~al.}}]{Abdullahi:2022jlv}%
  \BibitemOpen
  \bibfield  {author} {\bibinfo {author} {\bibfnamefont {A.~M.}\ \bibnamefont
  {Abdullahi}} \emph {et~al.},\ }in\ \href@noop {} {\emph {\bibinfo {booktitle}
  {{2022 Snowmass Summer Study}}}}\ (\bibinfo {year} {2022})\ \Eprint
  {http://arxiv.org/abs/2203.08039} {arXiv:2203.08039 [hep-ph]} \BibitemShut
  {NoStop}%
\bibitem [{\citenamefont {Aguilar-Arevalo}\ \emph {et~al.}(2007)\citenamefont
  {Aguilar-Arevalo} \emph {et~al.}}]{MiniBooNE:2007uho}%
  \BibitemOpen
  \bibfield  {author} {\bibinfo {author} {\bibfnamefont {A.~A.}\ \bibnamefont
  {Aguilar-Arevalo}} \emph {et~al.} (\bibinfo {collaboration} {MiniBooNE}),\
  }\href {\doibase 10.1103/PhysRevLett.98.231801} {\bibfield  {journal}
  {\bibinfo  {journal} {Phys. Rev. Lett.}\ }\textbf {\bibinfo {volume} {98}},\
  \bibinfo {pages} {231801} (\bibinfo {year} {2007})},\ \Eprint
  {http://arxiv.org/abs/0704.1500} {arXiv:0704.1500 [hep-ex]} \BibitemShut
  {NoStop}%
\bibitem [{\citenamefont {Aguilar-Arevalo}\ \emph {et~al.}(2010)\citenamefont
  {Aguilar-Arevalo} \emph {et~al.}}]{MiniBooNE:2010idf}%
  \BibitemOpen
  \bibfield  {author} {\bibinfo {author} {\bibfnamefont {A.~A.}\ \bibnamefont
  {Aguilar-Arevalo}} \emph {et~al.} (\bibinfo {collaboration} {MiniBooNE}),\
  }\href {\doibase 10.1103/PhysRevLett.105.181801} {\bibfield  {journal}
  {\bibinfo  {journal} {Phys. Rev. Lett.}\ }\textbf {\bibinfo {volume} {105}},\
  \bibinfo {pages} {181801} (\bibinfo {year} {2010})},\ \Eprint
  {http://arxiv.org/abs/1007.1150} {arXiv:1007.1150 [hep-ex]} \BibitemShut
  {NoStop}%
\bibitem [{\citenamefont {Aguilar-Arevalo}\ \emph {et~al.}(2013)\citenamefont
  {Aguilar-Arevalo} \emph {et~al.}}]{MiniBooNE:2013uba}%
  \BibitemOpen
  \bibfield  {author} {\bibinfo {author} {\bibfnamefont {A.~A.}\ \bibnamefont
  {Aguilar-Arevalo}} \emph {et~al.} (\bibinfo {collaboration} {MiniBooNE}),\
  }\href {\doibase 10.1103/PhysRevLett.110.161801} {\bibfield  {journal}
  {\bibinfo  {journal} {Phys. Rev. Lett.}\ }\textbf {\bibinfo {volume} {110}},\
  \bibinfo {pages} {161801} (\bibinfo {year} {2013})},\ \Eprint
  {http://arxiv.org/abs/1303.2588} {arXiv:1303.2588 [hep-ex]} \BibitemShut
  {NoStop}%
\bibitem [{\citenamefont {Aguilar-Arevalo}\ \emph {et~al.}(2018)\citenamefont
  {Aguilar-Arevalo} \emph {et~al.}}]{MiniBooNE:2018esg}%
  \BibitemOpen
  \bibfield  {author} {\bibinfo {author} {\bibfnamefont {A.~A.}\ \bibnamefont
  {Aguilar-Arevalo}} \emph {et~al.} (\bibinfo {collaboration} {MiniBooNE}),\
  }\href {\doibase 10.1103/PhysRevLett.121.221801} {\bibfield  {journal}
  {\bibinfo  {journal} {Phys. Rev. Lett.}\ }\textbf {\bibinfo {volume} {121}},\
  \bibinfo {pages} {221801} (\bibinfo {year} {2018})},\ \Eprint
  {http://arxiv.org/abs/1805.12028} {arXiv:1805.12028 [hep-ex]} \BibitemShut
  {NoStop}%
\bibitem [{\citenamefont {Aguilar-Arevalo}\ \emph {et~al.}(2021)\citenamefont
  {Aguilar-Arevalo} \emph {et~al.}}]{MiniBooNE:2020pnu}%
  \BibitemOpen
  \bibfield  {author} {\bibinfo {author} {\bibfnamefont {A.~A.}\ \bibnamefont
  {Aguilar-Arevalo}} \emph {et~al.} (\bibinfo {collaboration} {MiniBooNE}),\
  }\href {\doibase 10.1103/PhysRevD.103.052002} {\bibfield  {journal} {\bibinfo
   {journal} {Phys. Rev. D}\ }\textbf {\bibinfo {volume} {103}},\ \bibinfo
  {pages} {052002} (\bibinfo {year} {2021})},\ \Eprint
  {http://arxiv.org/abs/2006.16883} {arXiv:2006.16883 [hep-ex]} \BibitemShut
  {NoStop}%
\bibitem [{\citenamefont {Aguilar-Arevalo}\ \emph {et~al.}(2022)\citenamefont
  {Aguilar-Arevalo} \emph {et~al.}}]{MiniBooNE:2022emn}%
  \BibitemOpen
  \bibfield  {author} {\bibinfo {author} {\bibfnamefont {A.~A.}\ \bibnamefont
  {Aguilar-Arevalo}} \emph {et~al.} (\bibinfo {collaboration} {MiniBooNE}),\
  }\href@noop {} {\  (\bibinfo {year} {2022})},\ \Eprint
  {http://arxiv.org/abs/2201.01724} {arXiv:2201.01724 [hep-ex]} \BibitemShut
  {NoStop}%
\bibitem [{\citenamefont {Athanassopoulos}\ \emph
  {et~al.}(1996{\natexlab{a}})\citenamefont {Athanassopoulos} \emph
  {et~al.}}]{LSND:1996vlr}%
  \BibitemOpen
  \bibfield  {author} {\bibinfo {author} {\bibfnamefont {C.}~\bibnamefont
  {Athanassopoulos}} \emph {et~al.} (\bibinfo {collaboration} {LSND}),\ }\href
  {\doibase 10.1103/PhysRevC.54.2685} {\bibfield  {journal} {\bibinfo
  {journal} {Phys. Rev. C}\ }\textbf {\bibinfo {volume} {54}},\ \bibinfo
  {pages} {2685} (\bibinfo {year} {1996}{\natexlab{a}})},\ \Eprint
  {http://arxiv.org/abs/nucl-ex/9605001} {arXiv:nucl-ex/9605001} \BibitemShut
  {NoStop}%
\bibitem [{\citenamefont {Athanassopoulos}\ \emph
  {et~al.}(1996{\natexlab{b}})\citenamefont {Athanassopoulos} \emph
  {et~al.}}]{LSND:1996ubh}%
  \BibitemOpen
  \bibfield  {author} {\bibinfo {author} {\bibfnamefont {C.}~\bibnamefont
  {Athanassopoulos}} \emph {et~al.} (\bibinfo {collaboration} {LSND}),\ }\href
  {\doibase 10.1103/PhysRevLett.77.3082} {\bibfield  {journal} {\bibinfo
  {journal} {Phys. Rev. Lett.}\ }\textbf {\bibinfo {volume} {77}},\ \bibinfo
  {pages} {3082} (\bibinfo {year} {1996}{\natexlab{b}})},\ \Eprint
  {http://arxiv.org/abs/nucl-ex/9605003} {arXiv:nucl-ex/9605003} \BibitemShut
  {NoStop}%
\bibitem [{\citenamefont {Athanassopoulos}\ \emph
  {et~al.}(1998{\natexlab{a}})\citenamefont {Athanassopoulos} \emph
  {et~al.}}]{LSND:1997vun}%
  \BibitemOpen
  \bibfield  {author} {\bibinfo {author} {\bibfnamefont {C.}~\bibnamefont
  {Athanassopoulos}} \emph {et~al.} (\bibinfo {collaboration} {LSND}),\ }\href
  {\doibase 10.1103/PhysRevLett.81.1774} {\bibfield  {journal} {\bibinfo
  {journal} {Phys. Rev. Lett.}\ }\textbf {\bibinfo {volume} {81}},\ \bibinfo
  {pages} {1774} (\bibinfo {year} {1998}{\natexlab{a}})},\ \Eprint
  {http://arxiv.org/abs/nucl-ex/9709006} {arXiv:nucl-ex/9709006} \BibitemShut
  {NoStop}%
\bibitem [{\citenamefont {Athanassopoulos}\ \emph
  {et~al.}(1998{\natexlab{b}})\citenamefont {Athanassopoulos} \emph
  {et~al.}}]{LSND:1997vqj}%
  \BibitemOpen
  \bibfield  {author} {\bibinfo {author} {\bibfnamefont {C.}~\bibnamefont
  {Athanassopoulos}} \emph {et~al.} (\bibinfo {collaboration} {LSND}),\ }\href
  {\doibase 10.1103/PhysRevC.58.2489} {\bibfield  {journal} {\bibinfo
  {journal} {Phys. Rev. C}\ }\textbf {\bibinfo {volume} {58}},\ \bibinfo
  {pages} {2489} (\bibinfo {year} {1998}{\natexlab{b}})},\ \Eprint
  {http://arxiv.org/abs/nucl-ex/9706006} {arXiv:nucl-ex/9706006} \BibitemShut
  {NoStop}%
\bibitem [{\citenamefont {Aguilar-Arevalo}\ \emph {et~al.}(2001)\citenamefont
  {Aguilar-Arevalo} \emph {et~al.}}]{LSND:2001aii}%
  \BibitemOpen
  \bibfield  {author} {\bibinfo {author} {\bibfnamefont {A.}~\bibnamefont
  {Aguilar-Arevalo}} \emph {et~al.} (\bibinfo {collaboration} {LSND}),\ }\href
  {\doibase 10.1103/PhysRevD.64.112007} {\bibfield  {journal} {\bibinfo
  {journal} {Phys. Rev. D}\ }\textbf {\bibinfo {volume} {64}},\ \bibinfo
  {pages} {112007} (\bibinfo {year} {2001})},\ \Eprint
  {http://arxiv.org/abs/hep-ex/0104049} {arXiv:hep-ex/0104049} \BibitemShut
  {NoStop}%
\bibitem [{\citenamefont {Aghanim}\ \emph {et~al.}(2020)\citenamefont {Aghanim}
  \emph {et~al.}}]{Planck:2018vyg}%
  \BibitemOpen
  \bibfield  {author} {\bibinfo {author} {\bibfnamefont {N.}~\bibnamefont
  {Aghanim}} \emph {et~al.} (\bibinfo {collaboration} {Planck}),\ }\href
  {\doibase 10.1051/0004-6361/201833910} {\bibfield  {journal} {\bibinfo
  {journal} {Astron. Astrophys.}\ }\textbf {\bibinfo {volume} {641}},\ \bibinfo
  {pages} {A6} (\bibinfo {year} {2020})},\ \bibinfo {note} {[Erratum:
  Astron.Astrophys. 652, C4 (2021)]},\ \Eprint
  {http://arxiv.org/abs/1807.06209} {arXiv:1807.06209 [astro-ph.CO]}
  \BibitemShut {NoStop}%
\bibitem [{\citenamefont {Acero}\ \emph {et~al.}(2022)\citenamefont {Acero}
  \emph {et~al.}}]{Acero:2022wqg}%
  \BibitemOpen
  \bibfield  {author} {\bibinfo {author} {\bibfnamefont {M.~A.}\ \bibnamefont
  {Acero}} \emph {et~al.},\ }\href@noop {} {\  (\bibinfo {year} {2022})},\
  \Eprint {http://arxiv.org/abs/2203.07323} {arXiv:2203.07323 [hep-ex]}
  \BibitemShut {NoStop}%
\bibitem [{\citenamefont {Gninenko}(2009)}]{Gninenko:2009ks}%
  \BibitemOpen
  \bibfield  {author} {\bibinfo {author} {\bibfnamefont {S.~N.}\ \bibnamefont
  {Gninenko}},\ }\href {\doibase 10.1103/PhysRevLett.103.241802} {\bibfield
  {journal} {\bibinfo  {journal} {Phys. Rev. Lett.}\ }\textbf {\bibinfo
  {volume} {103}},\ \bibinfo {pages} {241802} (\bibinfo {year} {2009})},\
  \Eprint {http://arxiv.org/abs/0902.3802} {arXiv:0902.3802 [hep-ph]}
  \BibitemShut {NoStop}%
\bibitem [{\citenamefont {Gninenko}(2011)}]{Gninenko:2010pr}%
  \BibitemOpen
  \bibfield  {author} {\bibinfo {author} {\bibfnamefont {S.~N.}\ \bibnamefont
  {Gninenko}},\ }\href {\doibase 10.1103/PhysRevD.83.015015} {\bibfield
  {journal} {\bibinfo  {journal} {Phys. Rev. D}\ }\textbf {\bibinfo {volume}
  {83}},\ \bibinfo {pages} {015015} (\bibinfo {year} {2011})},\ \Eprint
  {http://arxiv.org/abs/1009.5536} {arXiv:1009.5536 [hep-ph]} \BibitemShut
  {NoStop}%
\bibitem [{\citenamefont {Gninenko}(2012)}]{Gninenko:2012rw}%
  \BibitemOpen
  \bibfield  {author} {\bibinfo {author} {\bibfnamefont {S.~N.}\ \bibnamefont
  {Gninenko}},\ }\href {\doibase 10.1016/j.physletb.2012.02.071} {\bibfield
  {journal} {\bibinfo  {journal} {Phys. Lett. B}\ }\textbf {\bibinfo {volume}
  {710}},\ \bibinfo {pages} {86} (\bibinfo {year} {2012})},\ \Eprint
  {http://arxiv.org/abs/1201.5194} {arXiv:1201.5194 [hep-ph]} \BibitemShut
  {NoStop}%
\bibitem [{\citenamefont {Masip}\ \emph {et~al.}(2013)\citenamefont {Masip},
  \citenamefont {Masjuan},\ and\ \citenamefont {Meloni}}]{Masip:2012ke}%
  \BibitemOpen
  \bibfield  {author} {\bibinfo {author} {\bibfnamefont {M.}~\bibnamefont
  {Masip}}, \bibinfo {author} {\bibfnamefont {P.}~\bibnamefont {Masjuan}}, \
  and\ \bibinfo {author} {\bibfnamefont {D.}~\bibnamefont {Meloni}},\ }\href
  {\doibase 10.1007/JHEP01(2013)106} {\bibfield  {journal} {\bibinfo  {journal}
  {JHEP}\ }\textbf {\bibinfo {volume} {01}},\ \bibinfo {pages} {106} (\bibinfo
  {year} {2013})},\ \Eprint {http://arxiv.org/abs/1210.1519} {arXiv:1210.1519
  [hep-ph]} \BibitemShut {NoStop}%
\bibitem [{\citenamefont {Radionov}(2013)}]{Radionov:2013mca}%
  \BibitemOpen
  \bibfield  {author} {\bibinfo {author} {\bibfnamefont {A.}~\bibnamefont
  {Radionov}},\ }\href {\doibase 10.1103/PhysRevD.88.015016} {\bibfield
  {journal} {\bibinfo  {journal} {Phys. Rev. D}\ }\textbf {\bibinfo {volume}
  {88}},\ \bibinfo {pages} {015016} (\bibinfo {year} {2013})},\ \Eprint
  {http://arxiv.org/abs/1303.4587} {arXiv:1303.4587 [hep-ph]} \BibitemShut
  {NoStop}%
\bibitem [{\citenamefont {Magill}\ \emph {et~al.}(2018)\citenamefont {Magill},
  \citenamefont {Plestid}, \citenamefont {Pospelov},\ and\ \citenamefont
  {Tsai}}]{Magill:2018jla}%
  \BibitemOpen
  \bibfield  {author} {\bibinfo {author} {\bibfnamefont {G.}~\bibnamefont
  {Magill}}, \bibinfo {author} {\bibfnamefont {R.}~\bibnamefont {Plestid}},
  \bibinfo {author} {\bibfnamefont {M.}~\bibnamefont {Pospelov}}, \ and\
  \bibinfo {author} {\bibfnamefont {Y.-D.}\ \bibnamefont {Tsai}},\ }\href
  {\doibase 10.1103/PhysRevD.98.115015} {\bibfield  {journal} {\bibinfo
  {journal} {Phys. Rev. D}\ }\textbf {\bibinfo {volume} {98}},\ \bibinfo
  {pages} {115015} (\bibinfo {year} {2018})},\ \Eprint
  {http://arxiv.org/abs/1803.03262} {arXiv:1803.03262 [hep-ph]} \BibitemShut
  {NoStop}%
\bibitem [{\citenamefont {Vergani}\ \emph {et~al.}(2021)\citenamefont
  {Vergani}, \citenamefont {Kamp}, \citenamefont {Diaz}, \citenamefont
  {Arg\"uelles}, \citenamefont {Conrad}, \citenamefont {Shaevitz},\ and\
  \citenamefont {Uchida}}]{Vergani:2021tgc}%
  \BibitemOpen
  \bibfield  {author} {\bibinfo {author} {\bibfnamefont {S.}~\bibnamefont
  {Vergani}}, \bibinfo {author} {\bibfnamefont {N.~W.}\ \bibnamefont {Kamp}},
  \bibinfo {author} {\bibfnamefont {A.}~\bibnamefont {Diaz}}, \bibinfo {author}
  {\bibfnamefont {C.~A.}\ \bibnamefont {Arg\"uelles}}, \bibinfo {author}
  {\bibfnamefont {J.~M.}\ \bibnamefont {Conrad}}, \bibinfo {author}
  {\bibfnamefont {M.~H.}\ \bibnamefont {Shaevitz}}, \ and\ \bibinfo {author}
  {\bibfnamefont {M.~A.}\ \bibnamefont {Uchida}},\ }\href {\doibase
  10.1103/PhysRevD.104.095005} {\bibfield  {journal} {\bibinfo  {journal}
  {Phys. Rev. D}\ }\textbf {\bibinfo {volume} {104}},\ \bibinfo {pages}
  {095005} (\bibinfo {year} {2021})},\ \Eprint
  {http://arxiv.org/abs/2105.06470} {arXiv:2105.06470 [hep-ph]} \BibitemShut
  {NoStop}%
\bibitem [{\citenamefont {Alvarez-Ruso}\ and\ \citenamefont
  {Saul-Sala}(2021)}]{Alvarez-Ruso:2021dna}%
  \BibitemOpen
  \bibfield  {author} {\bibinfo {author} {\bibfnamefont {L.}~\bibnamefont
  {Alvarez-Ruso}}\ and\ \bibinfo {author} {\bibfnamefont {E.}~\bibnamefont
  {Saul-Sala}},\ }\href {\doibase 10.1140/epjs/s11734-021-00293-9} {\bibfield
  {journal} {\bibinfo  {journal} {Eur. Phys. J. ST}\ }\textbf {\bibinfo
  {volume} {230}},\ \bibinfo {pages} {4373} (\bibinfo {year} {2021})},\ \Eprint
  {http://arxiv.org/abs/2111.02504} {arXiv:2111.02504 [hep-ph]} \BibitemShut
  {NoStop}%
\bibitem [{\citenamefont {Denton}\ \emph {et~al.}(2019)\citenamefont {Denton},
  \citenamefont {Farzan},\ and\ \citenamefont {Shoemaker}}]{Denton:2018dqq}%
  \BibitemOpen
  \bibfield  {author} {\bibinfo {author} {\bibfnamefont {P.~B.}\ \bibnamefont
  {Denton}}, \bibinfo {author} {\bibfnamefont {Y.}~\bibnamefont {Farzan}}, \
  and\ \bibinfo {author} {\bibfnamefont {I.~M.}\ \bibnamefont {Shoemaker}},\
  }\href {\doibase 10.1103/PhysRevD.99.035003} {\bibfield  {journal} {\bibinfo
  {journal} {Phys. Rev. D}\ }\textbf {\bibinfo {volume} {99}},\ \bibinfo
  {pages} {035003} (\bibinfo {year} {2019})},\ \Eprint
  {http://arxiv.org/abs/1811.01310} {arXiv:1811.01310 [hep-ph]} \BibitemShut
  {NoStop}%
\bibitem [{\citenamefont {Bertuzzo}\ \emph {et~al.}(2018)\citenamefont
  {Bertuzzo}, \citenamefont {Jana}, \citenamefont {Machado},\ and\
  \citenamefont {Zukanovich~Funchal}}]{Bertuzzo:2018itn}%
  \BibitemOpen
  \bibfield  {author} {\bibinfo {author} {\bibfnamefont {E.}~\bibnamefont
  {Bertuzzo}}, \bibinfo {author} {\bibfnamefont {S.}~\bibnamefont {Jana}},
  \bibinfo {author} {\bibfnamefont {P.~A.~N.}\ \bibnamefont {Machado}}, \ and\
  \bibinfo {author} {\bibfnamefont {R.}~\bibnamefont {Zukanovich~Funchal}},\
  }\href {\doibase 10.1103/PhysRevLett.121.241801} {\bibfield  {journal}
  {\bibinfo  {journal} {Phys. Rev. Lett.}\ }\textbf {\bibinfo {volume} {121}},\
  \bibinfo {pages} {241801} (\bibinfo {year} {2018})},\ \Eprint
  {http://arxiv.org/abs/1807.09877} {arXiv:1807.09877 [hep-ph]} \BibitemShut
  {NoStop}%
\bibitem [{\citenamefont {Bertuzzo}\ \emph {et~al.}(2019)\citenamefont
  {Bertuzzo}, \citenamefont {Jana}, \citenamefont {Machado},\ and\
  \citenamefont {Zukanovich~Funchal}}]{Bertuzzo:2018ftf}%
  \BibitemOpen
  \bibfield  {author} {\bibinfo {author} {\bibfnamefont {E.}~\bibnamefont
  {Bertuzzo}}, \bibinfo {author} {\bibfnamefont {S.}~\bibnamefont {Jana}},
  \bibinfo {author} {\bibfnamefont {P.~A.~N.}\ \bibnamefont {Machado}}, \ and\
  \bibinfo {author} {\bibfnamefont {R.}~\bibnamefont {Zukanovich~Funchal}},\
  }\href {\doibase 10.1016/j.physletb.2019.02.023} {\bibfield  {journal}
  {\bibinfo  {journal} {Phys. Lett. B}\ }\textbf {\bibinfo {volume} {791}},\
  \bibinfo {pages} {210} (\bibinfo {year} {2019})},\ \Eprint
  {http://arxiv.org/abs/1808.02500} {arXiv:1808.02500 [hep-ph]} \BibitemShut
  {NoStop}%
\bibitem [{\citenamefont {Ballett}\ \emph
  {et~al.}(2019{\natexlab{a}})\citenamefont {Ballett}, \citenamefont
  {Pascoli},\ and\ \citenamefont {Ross-Lonergan}}]{Ballett:2018ynz}%
  \BibitemOpen
  \bibfield  {author} {\bibinfo {author} {\bibfnamefont {P.}~\bibnamefont
  {Ballett}}, \bibinfo {author} {\bibfnamefont {S.}~\bibnamefont {Pascoli}}, \
  and\ \bibinfo {author} {\bibfnamefont {M.}~\bibnamefont {Ross-Lonergan}},\
  }\href {\doibase 10.1103/PhysRevD.99.071701} {\bibfield  {journal} {\bibinfo
  {journal} {Phys. Rev. D}\ }\textbf {\bibinfo {volume} {99}},\ \bibinfo
  {pages} {071701} (\bibinfo {year} {2019}{\natexlab{a}})},\ \Eprint
  {http://arxiv.org/abs/1808.02915} {arXiv:1808.02915 [hep-ph]} \BibitemShut
  {NoStop}%
\bibitem [{\citenamefont {Arg\"uelles}\ \emph
  {et~al.}(2019{\natexlab{a}})\citenamefont {Arg\"uelles}, \citenamefont
  {Hostert},\ and\ \citenamefont {Tsai}}]{Arguelles:2018mtc}%
  \BibitemOpen
  \bibfield  {author} {\bibinfo {author} {\bibfnamefont {C.~A.}\ \bibnamefont
  {Arg\"uelles}}, \bibinfo {author} {\bibfnamefont {M.}~\bibnamefont
  {Hostert}}, \ and\ \bibinfo {author} {\bibfnamefont {Y.-D.}\ \bibnamefont
  {Tsai}},\ }\href {\doibase 10.1103/PhysRevLett.123.261801} {\bibfield
  {journal} {\bibinfo  {journal} {Phys. Rev. Lett.}\ }\textbf {\bibinfo
  {volume} {123}},\ \bibinfo {pages} {261801} (\bibinfo {year}
  {2019}{\natexlab{a}})},\ \Eprint {http://arxiv.org/abs/1812.08768}
  {arXiv:1812.08768 [hep-ph]} \BibitemShut {NoStop}%
\bibitem [{\citenamefont {Ballett}\ \emph {et~al.}(2020)\citenamefont
  {Ballett}, \citenamefont {Hostert},\ and\ \citenamefont
  {Pascoli}}]{Ballett:2019pyw}%
  \BibitemOpen
  \bibfield  {author} {\bibinfo {author} {\bibfnamefont {P.}~\bibnamefont
  {Ballett}}, \bibinfo {author} {\bibfnamefont {M.}~\bibnamefont {Hostert}}, \
  and\ \bibinfo {author} {\bibfnamefont {S.}~\bibnamefont {Pascoli}},\ }\href
  {\doibase 10.1103/PhysRevD.101.115025} {\bibfield  {journal} {\bibinfo
  {journal} {Phys. Rev. D}\ }\textbf {\bibinfo {volume} {101}},\ \bibinfo
  {pages} {115025} (\bibinfo {year} {2020})},\ \Eprint
  {http://arxiv.org/abs/1903.07589} {arXiv:1903.07589 [hep-ph]} \BibitemShut
  {NoStop}%
\bibitem [{\citenamefont {Ballett}\ \emph
  {et~al.}(2019{\natexlab{b}})\citenamefont {Ballett}, \citenamefont
  {Hostert},\ and\ \citenamefont {Pascoli}}]{Ballett:2019cqp}%
  \BibitemOpen
  \bibfield  {author} {\bibinfo {author} {\bibfnamefont {P.}~\bibnamefont
  {Ballett}}, \bibinfo {author} {\bibfnamefont {M.}~\bibnamefont {Hostert}}, \
  and\ \bibinfo {author} {\bibfnamefont {S.}~\bibnamefont {Pascoli}},\ }\href
  {\doibase 10.1103/PhysRevD.99.091701} {\bibfield  {journal} {\bibinfo
  {journal} {Phys. Rev. D}\ }\textbf {\bibinfo {volume} {99}},\ \bibinfo
  {pages} {091701} (\bibinfo {year} {2019}{\natexlab{b}})},\ \Eprint
  {http://arxiv.org/abs/1903.07590} {arXiv:1903.07590 [hep-ph]} \BibitemShut
  {NoStop}%
\bibitem [{\citenamefont {Abdullahi}\ \emph {et~al.}(2021)\citenamefont
  {Abdullahi}, \citenamefont {Hostert},\ and\ \citenamefont
  {Pascoli}}]{Abdullahi:2020nyr}%
  \BibitemOpen
  \bibfield  {author} {\bibinfo {author} {\bibfnamefont {A.}~\bibnamefont
  {Abdullahi}}, \bibinfo {author} {\bibfnamefont {M.}~\bibnamefont {Hostert}},
  \ and\ \bibinfo {author} {\bibfnamefont {S.}~\bibnamefont {Pascoli}},\ }\href
  {\doibase 10.1016/j.physletb.2021.136531} {\bibfield  {journal} {\bibinfo
  {journal} {Phys. Lett. B}\ }\textbf {\bibinfo {volume} {820}},\ \bibinfo
  {pages} {136531} (\bibinfo {year} {2021})},\ \Eprint
  {http://arxiv.org/abs/2007.11813} {arXiv:2007.11813 [hep-ph]} \BibitemShut
  {NoStop}%
\bibitem [{\citenamefont {Datta}\ \emph {et~al.}(2020)\citenamefont {Datta},
  \citenamefont {Kamali},\ and\ \citenamefont {Marfatia}}]{Datta:2020auq}%
  \BibitemOpen
  \bibfield  {author} {\bibinfo {author} {\bibfnamefont {A.}~\bibnamefont
  {Datta}}, \bibinfo {author} {\bibfnamefont {S.}~\bibnamefont {Kamali}}, \
  and\ \bibinfo {author} {\bibfnamefont {D.}~\bibnamefont {Marfatia}},\ }\href
  {\doibase 10.1016/j.physletb.2020.135579} {\bibfield  {journal} {\bibinfo
  {journal} {Phys. Lett. B}\ }\textbf {\bibinfo {volume} {807}},\ \bibinfo
  {pages} {135579} (\bibinfo {year} {2020})},\ \Eprint
  {http://arxiv.org/abs/2005.08920} {arXiv:2005.08920 [hep-ph]} \BibitemShut
  {NoStop}%
\bibitem [{\citenamefont {Dutta}\ \emph {et~al.}(2020)\citenamefont {Dutta},
  \citenamefont {Ghosh},\ and\ \citenamefont {Li}}]{Dutta:2020scq}%
  \BibitemOpen
  \bibfield  {author} {\bibinfo {author} {\bibfnamefont {B.}~\bibnamefont
  {Dutta}}, \bibinfo {author} {\bibfnamefont {S.}~\bibnamefont {Ghosh}}, \ and\
  \bibinfo {author} {\bibfnamefont {T.}~\bibnamefont {Li}},\ }\href {\doibase
  10.1103/PhysRevD.102.055017} {\bibfield  {journal} {\bibinfo  {journal}
  {Phys. Rev. D}\ }\textbf {\bibinfo {volume} {102}},\ \bibinfo {pages}
  {055017} (\bibinfo {year} {2020})},\ \Eprint
  {http://arxiv.org/abs/2006.01319} {arXiv:2006.01319 [hep-ph]} \BibitemShut
  {NoStop}%
\bibitem [{\citenamefont {Abdallah}\ \emph {et~al.}(2020)\citenamefont
  {Abdallah}, \citenamefont {Gandhi},\ and\ \citenamefont
  {Roy}}]{Abdallah:2020biq}%
  \BibitemOpen
  \bibfield  {author} {\bibinfo {author} {\bibfnamefont {W.}~\bibnamefont
  {Abdallah}}, \bibinfo {author} {\bibfnamefont {R.}~\bibnamefont {Gandhi}}, \
  and\ \bibinfo {author} {\bibfnamefont {S.}~\bibnamefont {Roy}},\ }\href
  {\doibase 10.1007/JHEP12(2020)188} {\bibfield  {journal} {\bibinfo  {journal}
  {JHEP}\ }\textbf {\bibinfo {volume} {12}},\ \bibinfo {pages} {188} (\bibinfo
  {year} {2020})},\ \Eprint {http://arxiv.org/abs/2006.01948} {arXiv:2006.01948
  [hep-ph]} \BibitemShut {NoStop}%
\bibitem [{\citenamefont {Abdallah}\ \emph {et~al.}(2021)\citenamefont
  {Abdallah}, \citenamefont {Gandhi},\ and\ \citenamefont
  {Roy}}]{Abdallah:2020vgg}%
  \BibitemOpen
  \bibfield  {author} {\bibinfo {author} {\bibfnamefont {W.}~\bibnamefont
  {Abdallah}}, \bibinfo {author} {\bibfnamefont {R.}~\bibnamefont {Gandhi}}, \
  and\ \bibinfo {author} {\bibfnamefont {S.}~\bibnamefont {Roy}},\ }\href
  {\doibase 10.1103/PhysRevD.104.055028} {\bibfield  {journal} {\bibinfo
  {journal} {Phys. Rev. D}\ }\textbf {\bibinfo {volume} {104}},\ \bibinfo
  {pages} {055028} (\bibinfo {year} {2021})},\ \Eprint
  {http://arxiv.org/abs/2010.06159} {arXiv:2010.06159 [hep-ph]} \BibitemShut
  {NoStop}%
\bibitem [{\citenamefont {Hammad}\ \emph {et~al.}(2022)\citenamefont {Hammad},
  \citenamefont {Rashed},\ and\ \citenamefont {Moretti}}]{Hammad:2021mpl}%
  \BibitemOpen
  \bibfield  {author} {\bibinfo {author} {\bibfnamefont {A.}~\bibnamefont
  {Hammad}}, \bibinfo {author} {\bibfnamefont {A.}~\bibnamefont {Rashed}}, \
  and\ \bibinfo {author} {\bibfnamefont {S.}~\bibnamefont {Moretti}},\ }\href
  {\doibase 10.1016/j.physletb.2022.136945} {\bibfield  {journal} {\bibinfo
  {journal} {Phys. Lett. B}\ }\textbf {\bibinfo {volume} {827}},\ \bibinfo
  {pages} {136945} (\bibinfo {year} {2022})},\ \Eprint
  {http://arxiv.org/abs/2110.08651} {arXiv:2110.08651 [hep-ph]} \BibitemShut
  {NoStop}%
\bibitem [{\citenamefont {Dutta}\ \emph {et~al.}(2021)\citenamefont {Dutta},
  \citenamefont {Kim}, \citenamefont {Thompson}, \citenamefont {Thornton},\
  and\ \citenamefont {Van~de Water}}]{Dutta:2021cip}%
  \BibitemOpen
  \bibfield  {author} {\bibinfo {author} {\bibfnamefont {B.}~\bibnamefont
  {Dutta}}, \bibinfo {author} {\bibfnamefont {D.}~\bibnamefont {Kim}}, \bibinfo
  {author} {\bibfnamefont {A.}~\bibnamefont {Thompson}}, \bibinfo {author}
  {\bibfnamefont {R.~T.}\ \bibnamefont {Thornton}}, \ and\ \bibinfo {author}
  {\bibfnamefont {R.~G.}\ \bibnamefont {Van~de Water}},\ }\href@noop {} {\
  (\bibinfo {year} {2021})},\ \Eprint {http://arxiv.org/abs/2110.11944}
  {arXiv:2110.11944 [hep-ph]} \BibitemShut {NoStop}%
\bibitem [{\citenamefont {Fischer}\ \emph {et~al.}(2020)\citenamefont
  {Fischer}, \citenamefont {Hern\'andez-Cabezudo},\ and\ \citenamefont
  {Schwetz}}]{Fischer:2019fbw}%
  \BibitemOpen
  \bibfield  {author} {\bibinfo {author} {\bibfnamefont {O.}~\bibnamefont
  {Fischer}}, \bibinfo {author} {\bibfnamefont {A.}~\bibnamefont
  {Hern\'andez-Cabezudo}}, \ and\ \bibinfo {author} {\bibfnamefont
  {T.}~\bibnamefont {Schwetz}},\ }\href {\doibase 10.1103/PhysRevD.101.075045}
  {\bibfield  {journal} {\bibinfo  {journal} {Phys. Rev. D}\ }\textbf {\bibinfo
  {volume} {101}},\ \bibinfo {pages} {075045} (\bibinfo {year} {2020})},\
  \Eprint {http://arxiv.org/abs/1909.09561} {arXiv:1909.09561 [hep-ph]}
  \BibitemShut {NoStop}%
\bibitem [{\citenamefont {Chang}\ \emph {et~al.}(2021)\citenamefont {Chang},
  \citenamefont {Chen}, \citenamefont {Ho},\ and\ \citenamefont
  {Tseng}}]{Chang:2021myh}%
  \BibitemOpen
  \bibfield  {author} {\bibinfo {author} {\bibfnamefont {C.-H.~V.}\
  \bibnamefont {Chang}}, \bibinfo {author} {\bibfnamefont {C.-R.}\ \bibnamefont
  {Chen}}, \bibinfo {author} {\bibfnamefont {S.-Y.}\ \bibnamefont {Ho}}, \ and\
  \bibinfo {author} {\bibfnamefont {S.-Y.}\ \bibnamefont {Tseng}},\ }\href
  {\doibase 10.1103/PhysRevD.104.015030} {\bibfield  {journal} {\bibinfo
  {journal} {Phys. Rev. D}\ }\textbf {\bibinfo {volume} {104}},\ \bibinfo
  {pages} {015030} (\bibinfo {year} {2021})},\ \Eprint
  {http://arxiv.org/abs/2102.05012} {arXiv:2102.05012 [hep-ph]} \BibitemShut
  {NoStop}%
\bibitem [{\citenamefont {Abdallah}\ \emph {et~al.}(2022)\citenamefont
  {Abdallah}, \citenamefont {Gandhi},\ and\ \citenamefont
  {Roy}}]{Abdallah:2022grs}%
  \BibitemOpen
  \bibfield  {author} {\bibinfo {author} {\bibfnamefont {W.}~\bibnamefont
  {Abdallah}}, \bibinfo {author} {\bibfnamefont {R.}~\bibnamefont {Gandhi}}, \
  and\ \bibinfo {author} {\bibfnamefont {S.}~\bibnamefont {Roy}},\ }\href@noop
  {} {\  (\bibinfo {year} {2022})},\ \Eprint {http://arxiv.org/abs/2202.09373}
  {arXiv:2202.09373 [hep-ph]} \BibitemShut {NoStop}%
\bibitem [{\citenamefont {Abe}\ \emph {et~al.}(2019{\natexlab{a}})\citenamefont
  {Abe} \emph {et~al.}}]{Abe:2019kgx}%
  \BibitemOpen
  \bibfield  {author} {\bibinfo {author} {\bibfnamefont {K.}~\bibnamefont
  {Abe}} \emph {et~al.} (\bibinfo {collaboration} {T2K}),\ }\href {\doibase
  10.1103/PhysRevD.100.052006} {\bibfield  {journal} {\bibinfo  {journal}
  {Phys. Rev. D}\ }\textbf {\bibinfo {volume} {100}},\ \bibinfo {pages}
  {052006} (\bibinfo {year} {2019}{\natexlab{a}})},\ \Eprint
  {http://arxiv.org/abs/1902.07598} {arXiv:1902.07598 [hep-ex]} \BibitemShut
  {NoStop}%
\bibitem [{\citenamefont {Buchmuller}\ \emph {et~al.}(1991)\citenamefont
  {Buchmuller}, \citenamefont {Greub},\ and\ \citenamefont
  {Minkowski}}]{Buchmuller:1991ce}%
  \BibitemOpen
  \bibfield  {author} {\bibinfo {author} {\bibfnamefont {W.}~\bibnamefont
  {Buchmuller}}, \bibinfo {author} {\bibfnamefont {C.}~\bibnamefont {Greub}}, \
  and\ \bibinfo {author} {\bibfnamefont {P.}~\bibnamefont {Minkowski}},\ }\href
  {\doibase 10.1016/0370-2693(91)90952-M} {\bibfield  {journal} {\bibinfo
  {journal} {Phys. Lett. B}\ }\textbf {\bibinfo {volume} {267}},\ \bibinfo
  {pages} {395} (\bibinfo {year} {1991})}\BibitemShut {NoStop}%
\bibitem [{\citenamefont {Khalil}(2008)}]{Khalil:2006yi}%
  \BibitemOpen
  \bibfield  {author} {\bibinfo {author} {\bibfnamefont {S.}~\bibnamefont
  {Khalil}},\ }\href {\doibase 10.1088/0954-3899/35/5/055001} {\bibfield
  {journal} {\bibinfo  {journal} {J. Phys. G}\ }\textbf {\bibinfo {volume}
  {35}},\ \bibinfo {pages} {055001} (\bibinfo {year} {2008})},\ \Eprint
  {http://arxiv.org/abs/hep-ph/0611205} {arXiv:hep-ph/0611205} \BibitemShut
  {NoStop}%
\bibitem [{\citenamefont {Fileviez~Perez}\ \emph {et~al.}(2009)\citenamefont
  {Fileviez~Perez}, \citenamefont {Han},\ and\ \citenamefont
  {Li}}]{Perez:2009mu}%
  \BibitemOpen
  \bibfield  {author} {\bibinfo {author} {\bibfnamefont {P.}~\bibnamefont
  {Fileviez~Perez}}, \bibinfo {author} {\bibfnamefont {T.}~\bibnamefont {Han}},
  \ and\ \bibinfo {author} {\bibfnamefont {T.}~\bibnamefont {Li}},\ }\href
  {\doibase 10.1103/PhysRevD.80.073015} {\bibfield  {journal} {\bibinfo
  {journal} {Phys. Rev. D}\ }\textbf {\bibinfo {volume} {80}},\ \bibinfo
  {pages} {073015} (\bibinfo {year} {2009})},\ \Eprint
  {http://arxiv.org/abs/0907.4186} {arXiv:0907.4186 [hep-ph]} \BibitemShut
  {NoStop}%
\bibitem [{\citenamefont {Khalil}(2010)}]{Khalil:2010iu}%
  \BibitemOpen
  \bibfield  {author} {\bibinfo {author} {\bibfnamefont {S.}~\bibnamefont
  {Khalil}},\ }\href {\doibase 10.1103/PhysRevD.82.077702} {\bibfield
  {journal} {\bibinfo  {journal} {Phys. Rev. D}\ }\textbf {\bibinfo {volume}
  {82}},\ \bibinfo {pages} {077702} (\bibinfo {year} {2010})},\ \Eprint
  {http://arxiv.org/abs/1004.0013} {arXiv:1004.0013 [hep-ph]} \BibitemShut
  {NoStop}%
\bibitem [{\citenamefont {Harnik}\ \emph {et~al.}(2012)\citenamefont {Harnik},
  \citenamefont {Kopp},\ and\ \citenamefont {Machado}}]{Harnik:2012ni}%
  \BibitemOpen
  \bibfield  {author} {\bibinfo {author} {\bibfnamefont {R.}~\bibnamefont
  {Harnik}}, \bibinfo {author} {\bibfnamefont {J.}~\bibnamefont {Kopp}}, \ and\
  \bibinfo {author} {\bibfnamefont {P.~A.~N.}\ \bibnamefont {Machado}},\ }\href
  {\doibase 10.1088/1475-7516/2012/07/026} {\bibfield  {journal} {\bibinfo
  {journal} {JCAP}\ }\textbf {\bibinfo {volume} {07}},\ \bibinfo {pages} {026}
  (\bibinfo {year} {2012})},\ \Eprint {http://arxiv.org/abs/1202.6073}
  {arXiv:1202.6073 [hep-ph]} \BibitemShut {NoStop}%
\bibitem [{\citenamefont {Batell}\ \emph {et~al.}(2016)\citenamefont {Batell},
  \citenamefont {Pospelov},\ and\ \citenamefont {Shuve}}]{Batell:2016zod}%
  \BibitemOpen
  \bibfield  {author} {\bibinfo {author} {\bibfnamefont {B.}~\bibnamefont
  {Batell}}, \bibinfo {author} {\bibfnamefont {M.}~\bibnamefont {Pospelov}}, \
  and\ \bibinfo {author} {\bibfnamefont {B.}~\bibnamefont {Shuve}},\ }\href
  {\doibase 10.1007/JHEP08(2016)052} {\bibfield  {journal} {\bibinfo  {journal}
  {JHEP}\ }\textbf {\bibinfo {volume} {08}},\ \bibinfo {pages} {052} (\bibinfo
  {year} {2016})},\ \Eprint {http://arxiv.org/abs/1604.06099} {arXiv:1604.06099
  [hep-ph]} \BibitemShut {NoStop}%
\bibitem [{\citenamefont {Dib}\ \emph {et~al.}(2014)\citenamefont {Dib},
  \citenamefont {Moreno},\ and\ \citenamefont {Neill}}]{Dib:2014fua}%
  \BibitemOpen
  \bibfield  {author} {\bibinfo {author} {\bibfnamefont {C.~O.}\ \bibnamefont
  {Dib}}, \bibinfo {author} {\bibfnamefont {G.~R.}\ \bibnamefont {Moreno}}, \
  and\ \bibinfo {author} {\bibfnamefont {N.~A.}\ \bibnamefont {Neill}},\ }\href
  {\doibase 10.1103/PhysRevD.90.113003} {\bibfield  {journal} {\bibinfo
  {journal} {Phys. Rev. D}\ }\textbf {\bibinfo {volume} {90}},\ \bibinfo
  {pages} {113003} (\bibinfo {year} {2014})},\ \Eprint
  {http://arxiv.org/abs/1409.1868} {arXiv:1409.1868 [hep-ph]} \BibitemShut
  {NoStop}%
\bibitem [{\citenamefont {De~Romeri}\ \emph {et~al.}(2017)\citenamefont
  {De~Romeri}, \citenamefont {Fernandez-Martinez}, \citenamefont {Gehrlein},
  \citenamefont {Machado},\ and\ \citenamefont {Niro}}]{DeRomeri:2017oxa}%
  \BibitemOpen
  \bibfield  {author} {\bibinfo {author} {\bibfnamefont {V.}~\bibnamefont
  {De~Romeri}}, \bibinfo {author} {\bibfnamefont {E.}~\bibnamefont
  {Fernandez-Martinez}}, \bibinfo {author} {\bibfnamefont {J.}~\bibnamefont
  {Gehrlein}}, \bibinfo {author} {\bibfnamefont {P.~A.~N.}\ \bibnamefont
  {Machado}}, \ and\ \bibinfo {author} {\bibfnamefont {V.}~\bibnamefont
  {Niro}},\ }\href {\doibase 10.1007/JHEP10(2017)169} {\bibfield  {journal}
  {\bibinfo  {journal} {JHEP}\ }\textbf {\bibinfo {volume} {10}},\ \bibinfo
  {pages} {169} (\bibinfo {year} {2017})},\ \Eprint
  {http://arxiv.org/abs/1707.08606} {arXiv:1707.08606 [hep-ph]} \BibitemShut
  {NoStop}%
\bibitem [{\citenamefont {Camargo}\ \emph {et~al.}(2019)\citenamefont
  {Camargo}, \citenamefont {Campos}, \citenamefont {de~Melo},\ and\
  \citenamefont {Queiroz}}]{Camargo:2019ukv}%
  \BibitemOpen
  \bibfield  {author} {\bibinfo {author} {\bibfnamefont {D.~A.}\ \bibnamefont
  {Camargo}}, \bibinfo {author} {\bibfnamefont {M.~D.}\ \bibnamefont {Campos}},
  \bibinfo {author} {\bibfnamefont {T.~B.}\ \bibnamefont {de~Melo}}, \ and\
  \bibinfo {author} {\bibfnamefont {F.~S.}\ \bibnamefont {Queiroz}},\ }\href
  {\doibase 10.1016/j.physletb.2019.06.020} {\bibfield  {journal} {\bibinfo
  {journal} {Phys. Lett. B}\ }\textbf {\bibinfo {volume} {795}},\ \bibinfo
  {pages} {319} (\bibinfo {year} {2019})},\ \Eprint
  {http://arxiv.org/abs/1901.05476} {arXiv:1901.05476 [hep-ph]} \BibitemShut
  {NoStop}%
\bibitem [{\citenamefont {Arg\"uelles}\ \emph {et~al.}(2020)\citenamefont
  {Arg\"uelles}, \citenamefont {Coloma}, \citenamefont {Hern\'andez},\ and\
  \citenamefont {Mu\~noz}}]{Arguelles:2019ziu}%
  \BibitemOpen
  \bibfield  {author} {\bibinfo {author} {\bibfnamefont {C.}~\bibnamefont
  {Arg\"uelles}}, \bibinfo {author} {\bibfnamefont {P.}~\bibnamefont {Coloma}},
  \bibinfo {author} {\bibfnamefont {P.}~\bibnamefont {Hern\'andez}}, \ and\
  \bibinfo {author} {\bibfnamefont {V.}~\bibnamefont {Mu\~noz}},\ }\href
  {\doibase 10.1007/JHEP02(2020)190} {\bibfield  {journal} {\bibinfo  {journal}
  {JHEP}\ }\textbf {\bibinfo {volume} {02}},\ \bibinfo {pages} {190} (\bibinfo
  {year} {2020})},\ \Eprint {http://arxiv.org/abs/1910.12839} {arXiv:1910.12839
  [hep-ph]} \BibitemShut {NoStop}%
\bibitem [{\citenamefont {Pospelov}(2011)}]{Pospelov:2011ha}%
  \BibitemOpen
  \bibfield  {author} {\bibinfo {author} {\bibfnamefont {M.}~\bibnamefont
  {Pospelov}},\ }\href {\doibase 10.1103/PhysRevD.84.085008} {\bibfield
  {journal} {\bibinfo  {journal} {Phys. Rev. D}\ }\textbf {\bibinfo {volume}
  {84}},\ \bibinfo {pages} {085008} (\bibinfo {year} {2011})},\ \Eprint
  {http://arxiv.org/abs/1103.3261} {arXiv:1103.3261 [hep-ph]} \BibitemShut
  {NoStop}%
\bibitem [{\citenamefont {Baek}\ \emph {et~al.}(2015)\citenamefont {Baek},
  \citenamefont {Okada},\ and\ \citenamefont {Yagyu}}]{Baek:2015mna}%
  \BibitemOpen
  \bibfield  {author} {\bibinfo {author} {\bibfnamefont {S.}~\bibnamefont
  {Baek}}, \bibinfo {author} {\bibfnamefont {H.}~\bibnamefont {Okada}}, \ and\
  \bibinfo {author} {\bibfnamefont {K.}~\bibnamefont {Yagyu}},\ }\href
  {\doibase 10.1007/JHEP04(2015)049} {\bibfield  {journal} {\bibinfo  {journal}
  {JHEP}\ }\textbf {\bibinfo {volume} {04}},\ \bibinfo {pages} {049} (\bibinfo
  {year} {2015})},\ \Eprint {http://arxiv.org/abs/1501.01530} {arXiv:1501.01530
  [hep-ph]} \BibitemShut {NoStop}%
\bibitem [{\citenamefont {Kamada}\ and\ \citenamefont
  {Yu}(2015)}]{Kamada:2015era}%
  \BibitemOpen
  \bibfield  {author} {\bibinfo {author} {\bibfnamefont {A.}~\bibnamefont
  {Kamada}}\ and\ \bibinfo {author} {\bibfnamefont {H.-B.}\ \bibnamefont
  {Yu}},\ }\href {\doibase 10.1103/PhysRevD.92.113004} {\bibfield  {journal}
  {\bibinfo  {journal} {Phys. Rev. D}\ }\textbf {\bibinfo {volume} {92}},\
  \bibinfo {pages} {113004} (\bibinfo {year} {2015})},\ \Eprint
  {http://arxiv.org/abs/1504.00711} {arXiv:1504.00711 [hep-ph]} \BibitemShut
  {NoStop}%
\bibitem [{\citenamefont {Baek}\ \emph {et~al.}(2013)\citenamefont {Baek},
  \citenamefont {Ko},\ and\ \citenamefont {Park}}]{Baek:2013qwa}%
  \BibitemOpen
  \bibfield  {author} {\bibinfo {author} {\bibfnamefont {S.}~\bibnamefont
  {Baek}}, \bibinfo {author} {\bibfnamefont {P.}~\bibnamefont {Ko}}, \ and\
  \bibinfo {author} {\bibfnamefont {W.-I.}\ \bibnamefont {Park}},\ }\href
  {\doibase 10.1007/JHEP07(2013)013} {\bibfield  {journal} {\bibinfo  {journal}
  {JHEP}\ }\textbf {\bibinfo {volume} {07}},\ \bibinfo {pages} {013} (\bibinfo
  {year} {2013})},\ \Eprint {http://arxiv.org/abs/1303.4280} {arXiv:1303.4280
  [hep-ph]} \BibitemShut {NoStop}%
\bibitem [{\citenamefont {Okada}\ and\ \citenamefont
  {Yagyu}(2014)}]{Okada:2014nsa}%
  \BibitemOpen
  \bibfield  {author} {\bibinfo {author} {\bibfnamefont {H.}~\bibnamefont
  {Okada}}\ and\ \bibinfo {author} {\bibfnamefont {K.}~\bibnamefont {Yagyu}},\
  }\href {\doibase 10.1103/PhysRevD.90.035019} {\bibfield  {journal} {\bibinfo
  {journal} {Phys. Rev. D}\ }\textbf {\bibinfo {volume} {90}},\ \bibinfo
  {pages} {035019} (\bibinfo {year} {2014})},\ \Eprint
  {http://arxiv.org/abs/1405.2368} {arXiv:1405.2368 [hep-ph]} \BibitemShut
  {NoStop}%
\bibitem [{\citenamefont {Ko}\ and\ \citenamefont {Tang}(2014)}]{Ko:2014bka}%
  \BibitemOpen
  \bibfield  {author} {\bibinfo {author} {\bibfnamefont {P.}~\bibnamefont
  {Ko}}\ and\ \bibinfo {author} {\bibfnamefont {Y.}~\bibnamefont {Tang}},\
  }\href {\doibase 10.1016/j.physletb.2014.10.035} {\bibfield  {journal}
  {\bibinfo  {journal} {Phys. Lett. B}\ }\textbf {\bibinfo {volume} {739}},\
  \bibinfo {pages} {62} (\bibinfo {year} {2014})},\ \Eprint
  {http://arxiv.org/abs/1404.0236} {arXiv:1404.0236 [hep-ph]} \BibitemShut
  {NoStop}%
\bibitem [{\citenamefont {Diaz}\ \emph {et~al.}(2017)\citenamefont {Diaz},
  \citenamefont {Mantilla},\ and\ \citenamefont {Martinez}}]{Diaz:2017edh}%
  \BibitemOpen
  \bibfield  {author} {\bibinfo {author} {\bibfnamefont {C.~E.}\ \bibnamefont
  {Diaz}}, \bibinfo {author} {\bibfnamefont {S.~F.}\ \bibnamefont {Mantilla}},
  \ and\ \bibinfo {author} {\bibfnamefont {R.}~\bibnamefont {Martinez}},\
  }\href@noop {} {\  (\bibinfo {year} {2017})},\ \Eprint
  {http://arxiv.org/abs/1712.05433} {arXiv:1712.05433 [hep-ph]} \BibitemShut
  {NoStop}%
\bibitem [{\citenamefont {Nomura}\ and\ \citenamefont
  {Okada}(2019)}]{Nomura:2018ibs}%
  \BibitemOpen
  \bibfield  {author} {\bibinfo {author} {\bibfnamefont {T.}~\bibnamefont
  {Nomura}}\ and\ \bibinfo {author} {\bibfnamefont {H.}~\bibnamefont {Okada}},\
  }\href {\doibase 10.1103/PhysRevD.99.055033} {\bibfield  {journal} {\bibinfo
  {journal} {Phys. Rev. D}\ }\textbf {\bibinfo {volume} {99}},\ \bibinfo
  {pages} {055033} (\bibinfo {year} {2019})},\ \Eprint
  {http://arxiv.org/abs/1806.07182} {arXiv:1806.07182 [hep-ph]} \BibitemShut
  {NoStop}%
\bibitem [{\citenamefont {Hagedorn}\ \emph {et~al.}(2018)\citenamefont
  {Hagedorn}, \citenamefont {Herrero-Garc\'\i{}a}, \citenamefont {Molinaro},\
  and\ \citenamefont {Schmidt}}]{Hagedorn:2018spx}%
  \BibitemOpen
  \bibfield  {author} {\bibinfo {author} {\bibfnamefont {C.}~\bibnamefont
  {Hagedorn}}, \bibinfo {author} {\bibfnamefont {J.}~\bibnamefont
  {Herrero-Garc\'\i{}a}}, \bibinfo {author} {\bibfnamefont {E.}~\bibnamefont
  {Molinaro}}, \ and\ \bibinfo {author} {\bibfnamefont {M.~A.}\ \bibnamefont
  {Schmidt}},\ }\href {\doibase 10.1007/JHEP11(2018)103} {\bibfield  {journal}
  {\bibinfo  {journal} {JHEP}\ }\textbf {\bibinfo {volume} {11}},\ \bibinfo
  {pages} {103} (\bibinfo {year} {2018})},\ \Eprint
  {http://arxiv.org/abs/1804.04117} {arXiv:1804.04117 [hep-ph]} \BibitemShut
  {NoStop}%
\bibitem [{\citenamefont {Shakya}\ and\ \citenamefont
  {Wells}(2019)}]{Shakya:2018qzg}%
  \BibitemOpen
  \bibfield  {author} {\bibinfo {author} {\bibfnamefont {B.}~\bibnamefont
  {Shakya}}\ and\ \bibinfo {author} {\bibfnamefont {J.~D.}\ \bibnamefont
  {Wells}},\ }\href {\doibase 10.1007/JHEP02(2019)174} {\bibfield  {journal}
  {\bibinfo  {journal} {JHEP}\ }\textbf {\bibinfo {volume} {02}},\ \bibinfo
  {pages} {174} (\bibinfo {year} {2019})},\ \Eprint
  {http://arxiv.org/abs/1801.02640} {arXiv:1801.02640 [hep-ph]} \BibitemShut
  {NoStop}%
\bibitem [{\citenamefont {De~Jager}\ \emph {et~al.}(1974)\citenamefont
  {De~Jager}, \citenamefont {De~Vries},\ and\ \citenamefont
  {De~Vries}}]{DeJager:1974liz}%
  \BibitemOpen
  \bibfield  {author} {\bibinfo {author} {\bibfnamefont {C.~W.}\ \bibnamefont
  {De~Jager}}, \bibinfo {author} {\bibfnamefont {H.}~\bibnamefont {De~Vries}},
  \ and\ \bibinfo {author} {\bibfnamefont {C.}~\bibnamefont {De~Vries}},\
  }\href {\doibase 10.1016/S0092-640X(74)80002-1} {\bibfield  {journal}
  {\bibinfo  {journal} {Atom. Data Nucl. Data Tabl.}\ }\textbf {\bibinfo
  {volume} {14}},\ \bibinfo {pages} {479} (\bibinfo {year} {1974})},\ \bibinfo
  {note} {[Erratum: Atom.Data Nucl.Data Tabl. 16, 580--580 (1975)]}\BibitemShut
  {NoStop}%
\bibitem [{\citenamefont {Abratenko}\ \emph
  {et~al.}(2021{\natexlab{a}})\citenamefont {Abratenko} \emph
  {et~al.}}]{MicroBooNE:2021rmx}%
  \BibitemOpen
  \bibfield  {author} {\bibinfo {author} {\bibfnamefont {P.}~\bibnamefont
  {Abratenko}} \emph {et~al.} (\bibinfo {collaboration} {MicroBooNE}),\
  }\href@noop {} {\  (\bibinfo {year} {2021}{\natexlab{a}})},\ \Eprint
  {http://arxiv.org/abs/2110.14054} {arXiv:2110.14054 [hep-ex]} \BibitemShut
  {NoStop}%
\bibitem [{\citenamefont {Abratenko}\ \emph
  {et~al.}(2021{\natexlab{b}})\citenamefont {Abratenko} \emph
  {et~al.}}]{MicroBooNE:2021nxr}%
  \BibitemOpen
  \bibfield  {author} {\bibinfo {author} {\bibfnamefont {P.}~\bibnamefont
  {Abratenko}} \emph {et~al.} (\bibinfo {collaboration} {MicroBooNE}),\
  }\href@noop {} {\  (\bibinfo {year} {2021}{\natexlab{b}})},\ \Eprint
  {http://arxiv.org/abs/2110.13978} {arXiv:2110.13978 [hep-ex]} \BibitemShut
  {NoStop}%
\bibitem [{\citenamefont {Abratenko}\ \emph
  {et~al.}(2021{\natexlab{c}})\citenamefont {Abratenko} \emph
  {et~al.}}]{MicroBooNE:2021jwr}%
  \BibitemOpen
  \bibfield  {author} {\bibinfo {author} {\bibfnamefont {P.}~\bibnamefont
  {Abratenko}} \emph {et~al.} (\bibinfo {collaboration} {MicroBooNE}),\
  }\href@noop {} {\  (\bibinfo {year} {2021}{\natexlab{c}})},\ \Eprint
  {http://arxiv.org/abs/2110.14080} {arXiv:2110.14080 [hep-ex]} \BibitemShut
  {NoStop}%
\bibitem [{\citenamefont {Abratenko}\ \emph
  {et~al.}(2021{\natexlab{d}})\citenamefont {Abratenko} \emph
  {et~al.}}]{MicroBooNE:2021sne}%
  \BibitemOpen
  \bibfield  {author} {\bibinfo {author} {\bibfnamefont {P.}~\bibnamefont
  {Abratenko}} \emph {et~al.} (\bibinfo {collaboration} {MicroBooNE}),\
  }\href@noop {} {\  (\bibinfo {year} {2021}{\natexlab{d}})},\ \Eprint
  {http://arxiv.org/abs/2110.14065} {arXiv:2110.14065 [hep-ex]} \BibitemShut
  {NoStop}%
\bibitem [{\citenamefont {Arg\"uelles}\ \emph {et~al.}(2021)\citenamefont
  {Arg\"uelles}, \citenamefont {Esteban}, \citenamefont {Hostert},
  \citenamefont {Kelly}, \citenamefont {Kopp}, \citenamefont {Machado},
  \citenamefont {Martinez-Soler},\ and\ \citenamefont
  {Perez-Gonzalez}}]{Arguelles:2021meu}%
  \BibitemOpen
  \bibfield  {author} {\bibinfo {author} {\bibfnamefont {C.~A.}\ \bibnamefont
  {Arg\"uelles}}, \bibinfo {author} {\bibfnamefont {I.}~\bibnamefont
  {Esteban}}, \bibinfo {author} {\bibfnamefont {M.}~\bibnamefont {Hostert}},
  \bibinfo {author} {\bibfnamefont {K.~J.}\ \bibnamefont {Kelly}}, \bibinfo
  {author} {\bibfnamefont {J.}~\bibnamefont {Kopp}}, \bibinfo {author}
  {\bibfnamefont {P.~A.~N.}\ \bibnamefont {Machado}}, \bibinfo {author}
  {\bibfnamefont {I.}~\bibnamefont {Martinez-Soler}}, \ and\ \bibinfo {author}
  {\bibfnamefont {Y.~F.}\ \bibnamefont {Perez-Gonzalez}},\ }\href@noop {} {\
  (\bibinfo {year} {2021})},\ \Eprint {http://arxiv.org/abs/2111.10359}
  {arXiv:2111.10359 [hep-ph]} \BibitemShut {NoStop}%
\bibitem [{\citenamefont {Denton}(2021)}]{Denton:2021czb}%
  \BibitemOpen
  \bibfield  {author} {\bibinfo {author} {\bibfnamefont {P.~B.}\ \bibnamefont
  {Denton}},\ }\href@noop {} {\  (\bibinfo {year} {2021})},\ \Eprint
  {http://arxiv.org/abs/2111.05793} {arXiv:2111.05793 [hep-ph]} \BibitemShut
  {NoStop}%
\bibitem [{\citenamefont {Brdar}\ \emph {et~al.}(2021)\citenamefont {Brdar},
  \citenamefont {Fischer},\ and\ \citenamefont {Smirnov}}]{Brdar:2020tle}%
  \BibitemOpen
  \bibfield  {author} {\bibinfo {author} {\bibfnamefont {V.}~\bibnamefont
  {Brdar}}, \bibinfo {author} {\bibfnamefont {O.}~\bibnamefont {Fischer}}, \
  and\ \bibinfo {author} {\bibfnamefont {A.~Y.}\ \bibnamefont {Smirnov}},\
  }\href {\doibase 10.1103/PhysRevD.103.075008} {\bibfield  {journal} {\bibinfo
   {journal} {Phys. Rev. D}\ }\textbf {\bibinfo {volume} {103}},\ \bibinfo
  {pages} {075008} (\bibinfo {year} {2021})},\ \Eprint
  {http://arxiv.org/abs/2007.14411} {arXiv:2007.14411 [hep-ph]} \BibitemShut
  {NoStop}%
\bibitem [{\citenamefont {Abe}\ \emph {et~al.}(2011)\citenamefont {Abe} \emph
  {et~al.}}]{T2K:2011qtm}%
  \BibitemOpen
  \bibfield  {author} {\bibinfo {author} {\bibfnamefont {K.}~\bibnamefont
  {Abe}} \emph {et~al.} (\bibinfo {collaboration} {T2K}),\ }\href {\doibase
  10.1016/j.nima.2011.06.067} {\bibfield  {journal} {\bibinfo  {journal} {Nucl.
  Instrum. Meth. A}\ }\textbf {\bibinfo {volume} {659}},\ \bibinfo {pages}
  {106} (\bibinfo {year} {2011})},\ \Eprint {http://arxiv.org/abs/1106.1238}
  {arXiv:1106.1238 [physics.ins-det]} \BibitemShut {NoStop}%
\bibitem [{\citenamefont {Abe}\ \emph {et~al.}(2020)\citenamefont {Abe} \emph
  {et~al.}}]{T2K:2020lrr}%
  \BibitemOpen
  \bibfield  {author} {\bibinfo {author} {\bibfnamefont {K.}~\bibnamefont
  {Abe}} \emph {et~al.} (\bibinfo {collaboration} {T2K}),\ }\href {\doibase
  10.1007/JHEP10(2020)114} {\bibfield  {journal} {\bibinfo  {journal} {JHEP}\
  }\textbf {\bibinfo {volume} {10}},\ \bibinfo {pages} {114} (\bibinfo {year}
  {2020})},\ \Eprint {http://arxiv.org/abs/2002.11986} {arXiv:2002.11986
  [hep-ex]} \BibitemShut {NoStop}%
\bibitem [{\citenamefont {Assylbekov}\ \emph {et~al.}(2012)\citenamefont
  {Assylbekov} \emph {et~al.}}]{Assylbekov:2011sh}%
  \BibitemOpen
  \bibfield  {author} {\bibinfo {author} {\bibfnamefont {S.}~\bibnamefont
  {Assylbekov}} \emph {et~al.},\ }\href {\doibase 10.1016/j.nima.2012.05.028}
  {\bibfield  {journal} {\bibinfo  {journal} {Nucl. Instrum. Meth. A}\ }\textbf
  {\bibinfo {volume} {686}},\ \bibinfo {pages} {48} (\bibinfo {year} {2012})},\
  \Eprint {http://arxiv.org/abs/1111.5030} {arXiv:1111.5030 [physics.ins-det]}
  \BibitemShut {NoStop}%
\bibitem [{\citenamefont {Gilje}(2014)}]{Gilje:2014cwd}%
  \BibitemOpen
  \bibfield  {author} {\bibinfo {author} {\bibfnamefont {K.}~\bibnamefont
  {Gilje}},\ }\emph {\bibinfo {title} {{Neutral Current $\pi^0$ Production Rate
  Measurement On-Water Using the $\pi^0$ Detector in the Near Detector of the
  T2K Experiment}}},\ \href@noop {} {Ph.D. thesis},\ \bibinfo  {school} {SUNY,
  Stony Brook} (\bibinfo {year} {2014})\BibitemShut {NoStop}%
\bibitem [{\citenamefont {Lamoureux}(2018)}]{Lamoureux:2018owo}%
  \BibitemOpen
  \bibfield  {author} {\bibinfo {author} {\bibfnamefont {M.}~\bibnamefont
  {Lamoureux}},\ }\emph {\bibinfo {title} {{Recherche de neutrinos lourds avec
  l'exp\'erience T2K}}},\ \href@noop {} {Ph.D. thesis},\ \bibinfo  {school}
  {Saclay} (\bibinfo {year} {2018})\BibitemShut {NoStop}%
\bibitem [{\citenamefont {Arg\"uelles}\ \emph {et~al.}(2022)\citenamefont
  {Arg\"uelles}, \citenamefont {Foppiani},\ and\ \citenamefont
  {Hostert}}]{Arguelles:2021dqn}%
  \BibitemOpen
  \bibfield  {author} {\bibinfo {author} {\bibfnamefont {C.~A.}\ \bibnamefont
  {Arg\"uelles}}, \bibinfo {author} {\bibfnamefont {N.}~\bibnamefont
  {Foppiani}}, \ and\ \bibinfo {author} {\bibfnamefont {M.}~\bibnamefont
  {Hostert}},\ }\href {\doibase 10.1103/PhysRevD.105.095006} {\bibfield
  {journal} {\bibinfo  {journal} {Phys. Rev. D}\ }\textbf {\bibinfo {volume}
  {105}},\ \bibinfo {pages} {095006} (\bibinfo {year} {2022})},\ \Eprint
  {http://arxiv.org/abs/2109.03831} {arXiv:2109.03831 [hep-ph]} \BibitemShut
  {NoStop}%
\bibitem [{\citenamefont {Abdullahi}\ \emph
  {et~al.}(2022{\natexlab{b}})\citenamefont {Abdullahi}, \citenamefont
  {Hoefken~Zink}, \citenamefont {Hostert}, \citenamefont {Massaro},\ and\
  \citenamefont {Pascoli}}]{Abdullahi:2022cdw}%
  \BibitemOpen
  \bibfield  {author} {\bibinfo {author} {\bibfnamefont {A.~M.}\ \bibnamefont
  {Abdullahi}}, \bibinfo {author} {\bibfnamefont {J.}~\bibnamefont
  {Hoefken~Zink}}, \bibinfo {author} {\bibfnamefont {M.}~\bibnamefont
  {Hostert}}, \bibinfo {author} {\bibfnamefont {D.}~\bibnamefont {Massaro}}, \
  and\ \bibinfo {author} {\bibfnamefont {S.}~\bibnamefont {Pascoli}},\
  }\href@noop {} {\  (\bibinfo {year} {2022}{\natexlab{b}})},\ \Eprint
  {http://arxiv.org/abs/2207.04137} {arXiv:2207.04137 [hep-ph]} \BibitemShut
  {NoStop}%
\bibitem [{\citenamefont {Abe}\ \emph {et~al.}(2019{\natexlab{b}})\citenamefont
  {Abe} \emph {et~al.}}]{T2K:2019bbb}%
  \BibitemOpen
  \bibfield  {author} {\bibinfo {author} {\bibfnamefont {K.}~\bibnamefont
  {Abe}} \emph {et~al.} (\bibinfo {collaboration} {T2K}),\ }\href@noop {} {\
  (\bibinfo {year} {2019}{\natexlab{b}})},\ \Eprint
  {http://arxiv.org/abs/1901.03750} {arXiv:1901.03750 [physics.ins-det]}
  \BibitemShut {NoStop}%
\bibitem [{\citenamefont {Abe}\ \emph {et~al.}(2016)\citenamefont {Abe} \emph
  {et~al.}}]{Abe:2016tii}%
  \BibitemOpen
  \bibfield  {author} {\bibinfo {author} {\bibfnamefont {K.}~\bibnamefont
  {Abe}} \emph {et~al.},\ }\href@noop {} {\  (\bibinfo {year} {2016})},\
  \Eprint {http://arxiv.org/abs/1609.04111} {arXiv:1609.04111 [hep-ex]}
  \BibitemShut {NoStop}%
\bibitem [{\citenamefont {King}\ \emph {et~al.}(2021)\citenamefont {King},
  \citenamefont {Christodoulou}, \citenamefont {Lamoureux},\ and\ \citenamefont
  {Experiment}}]{king_sophie_2021_5543856}%
  \BibitemOpen
  \bibfield  {author} {\bibinfo {author} {\bibfnamefont {S.}~\bibnamefont
  {King}}, \bibinfo {author} {\bibfnamefont {G.}~\bibnamefont {Christodoulou}},
  \bibinfo {author} {\bibfnamefont {M.}~\bibnamefont {Lamoureux}}, \ and\
  \bibinfo {author} {\bibfnamefont {T.~T.}\ \bibnamefont {Experiment}},\ }\href
  {\doibase 10.5281/zenodo.5543856} {\enquote {\bibinfo {title} {{Data release
  for the "Measurement of the charged- current electron (anti-)neutrino
  inclusive cross- sections at the T2K off-axis near detector ND280"}},}\ }
  (\bibinfo {year} {2021})\BibitemShut {NoStop}%
\bibitem [{\citenamefont {Buckley}\ \emph {et~al.}(2010)\citenamefont
  {Buckley}, \citenamefont {Hoeth}, \citenamefont {Lacker}, \citenamefont
  {Schulz},\ and\ \citenamefont {von Seggern}}]{Buckley:2009bj}%
  \BibitemOpen
  \bibfield  {author} {\bibinfo {author} {\bibfnamefont {A.}~\bibnamefont
  {Buckley}}, \bibinfo {author} {\bibfnamefont {H.}~\bibnamefont {Hoeth}},
  \bibinfo {author} {\bibfnamefont {H.}~\bibnamefont {Lacker}}, \bibinfo
  {author} {\bibfnamefont {H.}~\bibnamefont {Schulz}}, \ and\ \bibinfo {author}
  {\bibfnamefont {J.~E.}\ \bibnamefont {von Seggern}},\ }\href {\doibase
  10.1140/epjc/s10052-009-1196-7} {\bibfield  {journal} {\bibinfo  {journal}
  {Eur. Phys. J. C}\ }\textbf {\bibinfo {volume} {65}},\ \bibinfo {pages} {331}
  (\bibinfo {year} {2010})},\ \Eprint {http://arxiv.org/abs/0907.2973}
  {arXiv:0907.2973 [hep-ph]} \BibitemShut {NoStop}%
\bibitem [{\citenamefont {Krishnamoorthy}\ \emph {et~al.}(2021)\citenamefont
  {Krishnamoorthy}, \citenamefont {Schulz}, \citenamefont {Ju}, \citenamefont
  {Wang}, \citenamefont {Leyffer}, \citenamefont {Marshall}, \citenamefont
  {Mrenna}, \citenamefont {M\"uller},\ and\ \citenamefont
  {Kowalkowski}}]{Krishnamoorthy:2021nwv}%
  \BibitemOpen
  \bibfield  {author} {\bibinfo {author} {\bibfnamefont {M.}~\bibnamefont
  {Krishnamoorthy}}, \bibinfo {author} {\bibfnamefont {H.}~\bibnamefont
  {Schulz}}, \bibinfo {author} {\bibfnamefont {X.}~\bibnamefont {Ju}}, \bibinfo
  {author} {\bibfnamefont {W.}~\bibnamefont {Wang}}, \bibinfo {author}
  {\bibfnamefont {S.}~\bibnamefont {Leyffer}}, \bibinfo {author} {\bibfnamefont
  {Z.}~\bibnamefont {Marshall}}, \bibinfo {author} {\bibfnamefont
  {S.}~\bibnamefont {Mrenna}}, \bibinfo {author} {\bibfnamefont
  {J.}~\bibnamefont {M\"uller}}, \ and\ \bibinfo {author} {\bibfnamefont
  {J.~B.}\ \bibnamefont {Kowalkowski}},\ }\href {\doibase
  10.1051/epjconf/202125103060} {\bibfield  {journal} {\bibinfo  {journal} {EPJ
  Web Conf.}\ }\textbf {\bibinfo {volume} {251}},\ \bibinfo {pages} {03060}
  (\bibinfo {year} {2021})},\ \Eprint {http://arxiv.org/abs/2103.05748}
  {arXiv:2103.05748 [hep-ex]} \BibitemShut {NoStop}%
\bibitem [{\citenamefont {Cerde\~no}\ \emph {et~al.}(2018)\citenamefont
  {Cerde\~no}, \citenamefont {Cheek}, \citenamefont {Reid},\ and\ \citenamefont
  {Schulz}}]{Cerdeno:2018bty}%
  \BibitemOpen
  \bibfield  {author} {\bibinfo {author} {\bibfnamefont {D.~G.}\ \bibnamefont
  {Cerde\~no}}, \bibinfo {author} {\bibfnamefont {A.}~\bibnamefont {Cheek}},
  \bibinfo {author} {\bibfnamefont {E.}~\bibnamefont {Reid}}, \ and\ \bibinfo
  {author} {\bibfnamefont {H.}~\bibnamefont {Schulz}},\ }\href {\doibase
  10.1088/1475-7516/2018/08/011} {\bibfield  {journal} {\bibinfo  {journal}
  {JCAP}\ }\textbf {\bibinfo {volume} {08}},\ \bibinfo {pages} {011} (\bibinfo
  {year} {2018})},\ \Eprint {http://arxiv.org/abs/1802.03174} {arXiv:1802.03174
  [hep-ph]} \BibitemShut {NoStop}%
\bibitem [{\citenamefont {Aartsen}\ \emph {et~al.}(2019)\citenamefont {Aartsen}
  \emph {et~al.}}]{IceCube:2019lxi}%
  \BibitemOpen
  \bibfield  {author} {\bibinfo {author} {\bibfnamefont {M.~G.}\ \bibnamefont
  {Aartsen}} \emph {et~al.} (\bibinfo {collaboration} {IceCube}),\ }\href
  {\doibase 10.1088/1475-7516/2019/10/048} {\bibfield  {journal} {\bibinfo
  {journal} {JCAP}\ }\textbf {\bibinfo {volume} {10}},\ \bibinfo {pages} {048}
  (\bibinfo {year} {2019})},\ \Eprint {http://arxiv.org/abs/1909.01530}
  {arXiv:1909.01530 [hep-ex]} \BibitemShut {NoStop}%
\bibitem [{\citenamefont {{Peter Lepage}}(1978)}]{PETERLEPAGE1978192}%
  \BibitemOpen
  \bibfield  {author} {\bibinfo {author} {\bibfnamefont {G.}~\bibnamefont
  {{Peter Lepage}}},\ }\href {\doibase
  https://doi.org/10.1016/0021-9991(78)90004-9} {\bibfield  {journal} {\bibinfo
   {journal} {Journal of Computational Physics}\ }\textbf {\bibinfo {volume}
  {27}},\ \bibinfo {pages} {192} (\bibinfo {year} {1978})}\BibitemShut
  {NoStop}%
\bibitem [{\citenamefont {Lepage}(2021)}]{Lepage:2020tgj}%
  \BibitemOpen
  \bibfield  {author} {\bibinfo {author} {\bibfnamefont {G.~P.}\ \bibnamefont
  {Lepage}},\ }\href {\doibase 10.1016/j.jcp.2021.110386} {\bibfield  {journal}
  {\bibinfo  {journal} {J. Comput. Phys.}\ }\textbf {\bibinfo {volume} {439}},\
  \bibinfo {pages} {110386} (\bibinfo {year} {2021})},\ \Eprint
  {http://arxiv.org/abs/2009.05112} {arXiv:2009.05112 [physics.comp-ph]}
  \BibitemShut {NoStop}%
\bibitem [{\citenamefont {Lepage}(2022)}]{peter_lepage_2022_5893494}%
  \BibitemOpen
  \bibfield  {author} {\bibinfo {author} {\bibfnamefont {P.}~\bibnamefont
  {Lepage}},\ }\href {\doibase 10.5281/zenodo.5893494} {\enquote {\bibinfo
  {title} {gplepage/vegas: vegas version 5.1.1},}\ } (\bibinfo {year}
  {2022})\BibitemShut {NoStop}%
\bibitem [{\citenamefont {Ilten}\ \emph {et~al.}(2018)\citenamefont {Ilten},
  \citenamefont {Soreq}, \citenamefont {Williams},\ and\ \citenamefont
  {Xue}}]{Ilten:2018crw}%
  \BibitemOpen
  \bibfield  {author} {\bibinfo {author} {\bibfnamefont {P.}~\bibnamefont
  {Ilten}}, \bibinfo {author} {\bibfnamefont {Y.}~\bibnamefont {Soreq}},
  \bibinfo {author} {\bibfnamefont {M.}~\bibnamefont {Williams}}, \ and\
  \bibinfo {author} {\bibfnamefont {W.}~\bibnamefont {Xue}},\ }\href {\doibase
  10.1007/JHEP06(2018)004} {\bibfield  {journal} {\bibinfo  {journal} {JHEP}\
  }\textbf {\bibinfo {volume} {06}},\ \bibinfo {pages} {004} (\bibinfo {year}
  {2018})},\ \Eprint {http://arxiv.org/abs/1801.04847} {arXiv:1801.04847
  [hep-ph]} \BibitemShut {NoStop}%
\bibitem [{\citenamefont {Arg\"uelles}\ \emph
  {et~al.}(2019{\natexlab{b}})\citenamefont {Arg\"uelles}, \citenamefont
  {Schneider},\ and\ \citenamefont {Yuan}}]{Arguelles:2019izp}%
  \BibitemOpen
  \bibfield  {author} {\bibinfo {author} {\bibfnamefont {C.~A.}\ \bibnamefont
  {Arg\"uelles}}, \bibinfo {author} {\bibfnamefont {A.}~\bibnamefont
  {Schneider}}, \ and\ \bibinfo {author} {\bibfnamefont {T.}~\bibnamefont
  {Yuan}},\ }\href {\doibase 10.1007/JHEP06(2019)030} {\bibfield  {journal}
  {\bibinfo  {journal} {JHEP}\ }\textbf {\bibinfo {volume} {06}},\ \bibinfo
  {pages} {030} (\bibinfo {year} {2019}{\natexlab{b}})},\ \Eprint
  {http://arxiv.org/abs/1901.04645} {arXiv:1901.04645 [physics.data-an]}
  \BibitemShut {NoStop}%
\bibitem [{\citenamefont {Asaka}\ \emph {et~al.}(2013)\citenamefont {Asaka},
  \citenamefont {Eijima},\ and\ \citenamefont {Watanabe}}]{Asaka:2012bb}%
  \BibitemOpen
  \bibfield  {author} {\bibinfo {author} {\bibfnamefont {T.}~\bibnamefont
  {Asaka}}, \bibinfo {author} {\bibfnamefont {S.}~\bibnamefont {Eijima}}, \
  and\ \bibinfo {author} {\bibfnamefont {A.}~\bibnamefont {Watanabe}},\ }\href
  {\doibase 10.1007/JHEP03(2013)125} {\bibfield  {journal} {\bibinfo  {journal}
  {JHEP}\ }\textbf {\bibinfo {volume} {03}},\ \bibinfo {pages} {125} (\bibinfo
  {year} {2013})},\ \Eprint {http://arxiv.org/abs/1212.1062} {arXiv:1212.1062
  [hep-ph]} \BibitemShut {NoStop}%
\bibitem [{\citenamefont {de~Gouv\^ea}\ \emph {et~al.}(2019)\citenamefont
  {de~Gouv\^ea}, \citenamefont {Fox}, \citenamefont {Harnik}, \citenamefont
  {Kelly},\ and\ \citenamefont {Zhang}}]{deGouvea:2018cfv}%
  \BibitemOpen
  \bibfield  {author} {\bibinfo {author} {\bibfnamefont {A.}~\bibnamefont
  {de~Gouv\^ea}}, \bibinfo {author} {\bibfnamefont {P.~J.}\ \bibnamefont
  {Fox}}, \bibinfo {author} {\bibfnamefont {R.}~\bibnamefont {Harnik}},
  \bibinfo {author} {\bibfnamefont {K.~J.}\ \bibnamefont {Kelly}}, \ and\
  \bibinfo {author} {\bibfnamefont {Y.}~\bibnamefont {Zhang}},\ }\href
  {\doibase 10.1007/JHEP01(2019)001} {\bibfield  {journal} {\bibinfo  {journal}
  {JHEP}\ }\textbf {\bibinfo {volume} {01}},\ \bibinfo {pages} {001} (\bibinfo
  {year} {2019})},\ \Eprint {http://arxiv.org/abs/1809.06388} {arXiv:1809.06388
  [hep-ph]} \BibitemShut {NoStop}%
\bibitem [{\citenamefont {Isaacson}\ \emph {et~al.}(2021)\citenamefont
  {Isaacson}, \citenamefont {Jay}, \citenamefont {Lovato}, \citenamefont
  {Machado},\ and\ \citenamefont {Rocco}}]{Isaacson:2020wlx}%
  \BibitemOpen
  \bibfield  {author} {\bibinfo {author} {\bibfnamefont {J.}~\bibnamefont
  {Isaacson}}, \bibinfo {author} {\bibfnamefont {W.~I.}\ \bibnamefont {Jay}},
  \bibinfo {author} {\bibfnamefont {A.}~\bibnamefont {Lovato}}, \bibinfo
  {author} {\bibfnamefont {P.~A.~N.}\ \bibnamefont {Machado}}, \ and\ \bibinfo
  {author} {\bibfnamefont {N.}~\bibnamefont {Rocco}},\ }\href {\doibase
  10.1103/PhysRevC.103.015502} {\bibfield  {journal} {\bibinfo  {journal}
  {Phys. Rev. C}\ }\textbf {\bibinfo {volume} {103}},\ \bibinfo {pages}
  {015502} (\bibinfo {year} {2021})},\ \Eprint
  {http://arxiv.org/abs/2007.15570} {arXiv:2007.15570 [hep-ph]} \BibitemShut
  {NoStop}%
\bibitem [{\citenamefont {Isaacson}\ \emph {et~al.}(2022)\citenamefont
  {Isaacson}, \citenamefont {H\"oche}, \citenamefont {Gutierrez~Lopez},\ and\
  \citenamefont {Rocco}}]{Isaacson:2021xty}%
  \BibitemOpen
  \bibfield  {author} {\bibinfo {author} {\bibfnamefont {J.}~\bibnamefont
  {Isaacson}}, \bibinfo {author} {\bibfnamefont {S.}~\bibnamefont {H\"oche}},
  \bibinfo {author} {\bibfnamefont {D.}~\bibnamefont {Gutierrez~Lopez}}, \ and\
  \bibinfo {author} {\bibfnamefont {N.}~\bibnamefont {Rocco}},\ }\href
  {\doibase 10.1103/PhysRevD.105.096006} {\bibfield  {journal} {\bibinfo
  {journal} {Phys. Rev. D}\ }\textbf {\bibinfo {volume} {105}},\ \bibinfo
  {pages} {096006} (\bibinfo {year} {2022})},\ \Eprint
  {http://arxiv.org/abs/2110.15319} {arXiv:2110.15319 [hep-ph]} \BibitemShut
  {NoStop}%
\bibitem [{\citenamefont {Campbell}\ \emph {et~al.}(2022)\citenamefont
  {Campbell} \emph {et~al.}}]{Campbell:2022qmc}%
  \BibitemOpen
  \bibfield  {author} {\bibinfo {author} {\bibfnamefont {J.~M.}\ \bibnamefont
  {Campbell}} \emph {et~al.},\ }in\ \href@noop {} {\emph {\bibinfo {booktitle}
  {{2022 Snowmass Summer Study}}}}\ (\bibinfo {year} {2022})\ \Eprint
  {http://arxiv.org/abs/2203.11110} {arXiv:2203.11110 [hep-ph]} \BibitemShut
  {NoStop}%
\bibitem [{\citenamefont {Antonello}\ \emph {et~al.}(2015)\citenamefont
  {Antonello} \emph {et~al.}}]{MicroBooNE:2015bmn}%
  \BibitemOpen
  \bibfield  {author} {\bibinfo {author} {\bibfnamefont {M.}~\bibnamefont
  {Antonello}} \emph {et~al.} (\bibinfo {collaboration} {MicroBooNE, LAr1-ND,
  ICARUS-WA104}),\ }\href@noop {} {\  (\bibinfo {year} {2015})},\ \Eprint
  {http://arxiv.org/abs/1503.01520} {arXiv:1503.01520 [physics.ins-det]}
  \BibitemShut {NoStop}%
\bibitem [{\citenamefont {Machado}\ \emph {et~al.}(2019)\citenamefont
  {Machado}, \citenamefont {Palamara},\ and\ \citenamefont
  {Schmitz}}]{Machado:2019oxb}%
  \BibitemOpen
  \bibfield  {author} {\bibinfo {author} {\bibfnamefont {P.~A.}\ \bibnamefont
  {Machado}}, \bibinfo {author} {\bibfnamefont {O.}~\bibnamefont {Palamara}}, \
  and\ \bibinfo {author} {\bibfnamefont {D.~W.}\ \bibnamefont {Schmitz}},\
  }\href {\doibase 10.1146/annurev-nucl-101917-020949} {\bibfield  {journal}
  {\bibinfo  {journal} {Ann. Rev. Nucl. Part. Sci.}\ }\textbf {\bibinfo
  {volume} {69}},\ \bibinfo {pages} {363} (\bibinfo {year} {2019})},\ \Eprint
  {http://arxiv.org/abs/1903.04608} {arXiv:1903.04608 [hep-ex]} \BibitemShut
  {NoStop}%
\bibitem [{\citenamefont {Adrian-Martinez}\ \emph {et~al.}(2016)\citenamefont
  {Adrian-Martinez} \emph {et~al.}}]{KM3Net:2016zxf}%
  \BibitemOpen
  \bibfield  {author} {\bibinfo {author} {\bibfnamefont {S.}~\bibnamefont
  {Adrian-Martinez}} \emph {et~al.} (\bibinfo {collaboration} {KM3Net}),\
  }\href {\doibase 10.1088/0954-3899/43/8/084001} {\bibfield  {journal}
  {\bibinfo  {journal} {J. Phys. G}\ }\textbf {\bibinfo {volume} {43}},\
  \bibinfo {pages} {084001} (\bibinfo {year} {2016})},\ \Eprint
  {http://arxiv.org/abs/1601.07459} {arXiv:1601.07459 [astro-ph.IM]}
  \BibitemShut {NoStop}%
\bibitem [{\citenamefont {Abi}\ \emph {et~al.}(2020)\citenamefont {Abi} \emph
  {et~al.}}]{DUNE:2020ypp}%
  \BibitemOpen
  \bibfield  {author} {\bibinfo {author} {\bibfnamefont {B.}~\bibnamefont
  {Abi}} \emph {et~al.} (\bibinfo {collaboration} {DUNE}),\ }\href@noop {} {\
  (\bibinfo {year} {2020})},\ \Eprint {http://arxiv.org/abs/2002.03005}
  {arXiv:2002.03005 [hep-ex]} \BibitemShut {NoStop}%
\bibitem [{\citenamefont {Abe}\ \emph {et~al.}(2018)\citenamefont {Abe} \emph
  {et~al.}}]{Hyper-Kamiokande:2018ofw}%
  \BibitemOpen
  \bibfield  {author} {\bibinfo {author} {\bibfnamefont {K.}~\bibnamefont
  {Abe}} \emph {et~al.} (\bibinfo {collaboration} {Hyper-Kamiokande}),\
  }\href@noop {} {\  (\bibinfo {year} {2018})},\ \Eprint
  {http://arxiv.org/abs/1805.04163} {arXiv:1805.04163 [physics.ins-det]}
  \BibitemShut {NoStop}%
\end{thebibliography}%
\end{document}